\pdfoutput=1
\documentclass[11pt]{article}
\usepackage[margin=1.1in]{geometry}
\usepackage{amsmath,amssymb,amsthm,bm}
\usepackage{booktabs}
\usepackage{graphicx}
\usepackage{needspace}
\usepackage[section]{placeins}
\usepackage[authoryear,round]{natbib}
\usepackage[colorlinks=true,linkcolor=blue,citecolor=blue]{hyperref}

\makeatletter
\renewenvironment{proof}[1][\proofname]{\par\pushQED{\qed}\normalfont
  \topsep6\p@\@plus6\p@\relax\trivlist\item[\hskip\labelsep\bfseries#1\@addpunct{.}]\ignorespaces}
  {\popQED\endtrivlist\@endpefalse}
\makeatother

\newtheorem{theorem}{Theorem}
\newtheorem{proposition}{Proposition}
\newtheorem{lemma}[proposition]{Lemma}

\theoremstyle{definition}

\newtheorem{conjecture}[proposition]{Conjecture}

\DeclareMathOperator{\vecl}{vecl}
\DeclareMathOperator{\tr}{tr}
\DeclareMathOperator{\diag}{diag}
\DeclareMathOperator{\Diag}{Diag}
\newcommand{\Cor}{\operatorname{Cor}}
\newcommand{\R}{\mathbb{R}}
\newcommand{\Cn}{\mathcal{C}_n}
\newcommand{\Cbar}{\overline{\mathcal{C}}_n}
\newcommand{\Sn}{\mathbb{S}_n}

\title{Box--Cox Transformations of Correlation Matrices\thanks{We thank Yuichi Kitamura and seminar participants at Yale University for helpful comments.}}
\author{Ilya Archakov\\ York University
  \and
  Peter Reinhard Hansen\\ University of North Carolina at Chapel Hill}
\date{\today}

\begin{document}
\maketitle

\begin{abstract}
We consider Box--Cox transformations of correlation matrices,
$\Phi_\alpha(C)=\vecl\{(C^{\alpha}-I)/\alpha\}$, with Jacobian $J_\alpha$. The family nests the raw correlations at
$\alpha=1$, the generalised Fisher
transformation (GFT), $C\mapsto\vecl(\log C)$, at $\alpha=0$, and the negated off-diagonal concentration coordinates at $\alpha=-1$. The inverse
problem is a strictly convex program for every $\alpha$, so every member is injective. The map is onto $\R^{d}$,
$d=n(n-1)/2$, if and only if $\alpha\le0$, such that $\alpha=0$ defines the boundary for unrestricted parametrizations.
For $\alpha>0$ the image is bounded.
The conditioning obeys the bound $\kappa(J_\alpha)\le\kappa(C)^{|1-\alpha|}$, where
$\kappa(C)=\lambda_{\max}(C)/\lambda_{\min}(C)$, and on
the surjective range the exponent is smallest at $\alpha=0$.
Under Gaussian sampling with $n\geq3$, the asymptotic covariance of the transformed sample correlations is the identity at independence for every member,
but only $\alpha=0$ retains the identity to first order in the correlations, whereas the leading
joint third cumulant at independence vanishes only at $\alpha=\tfrac13$.
Finally, applying the transformation to the diagonal rescaling of $C$ that makes the result
hollow adds a new family to the existing correlation parametrizations. For every $\alpha$ the rescaling solves a strictly convex problem
and the inverse is explicit, and the scaled coordinates are unrestricted exactly when $\alpha=0$,
where they recover the known global normal coordinates, centred at independence, for the
efficient information metric of the correlations.
\end{abstract}



\clearpage
\section{Introduction}

A correlation matrix is constrained by positive definiteness and by its unit
diagonal. The $n\times n$ positive definite correlation matrices form the interior of the elliptope, a convex set of dimension $d=n(n-1)/2$, and positive definiteness cannot be characterised by separate bounds on the individual correlations. Estimation, simulation and modelling all become easier if the constraint can be
removed by a change of variables onto $\R^{d}$.

\citet{ArchakovHansen:Correlation} showed that the matrix logarithm does exactly that. The map
\[
  C\mapsto\gamma=\vecl(\log C)
\]
is a bijection from the correlation matrices onto $\R^{d}$. It is called the generalised
Fisher transformation (GFT), because it simplifies to the transformation of
\citet{Fisher:1921} when $n=2$. Its finite-sample behaviour is studied in \citet{ArchakovHansen:2026},
and it has been used for canonical representations of block matrices, for generating random
correlation matrices, and for unconstrained dynamic correlation models.

The Box--Cox transformation of \citet{BoxCox:1964}, $x\mapsto(x^{\alpha}-1)/\alpha$, is a one-parameter family of power transformations that embeds the logarithm as the case $\alpha=0$. Applied spectrally to positive definite correlation matrices, it gives
\begin{equation}\label{eq:family}
  \Phi_\alpha(C)=\vecl\Bigl(\frac{C^{\alpha}-I}{\alpha}\Bigr),\qquad
  \Phi_0(C)=\vecl(\log C),
\end{equation}
which includes three familiar objects. The first is the usual correlation vector, $\Phi_1(C)=\vecl(C)$, the second 
is the negated off diagonal of the concentration matrix, $\Phi_{-1}(C)=-\vecl(C^{-1})$, and the third is the GFT, $\Phi_0$. 
Spectral Box--Cox transformations of covariance matrices have been used in
volatility modelling by \citet{Weigand:2014}, where the object of interest is the covariance
matrix itself. The correlation problem is different
precisely because the diagonal of $f_\alpha(C)$ is pinned down by the requirement $\diag C=\iota$, where $\iota$ is the vector of ones. Inverting $\Phi_\alpha$ therefore means solving
$n$ equations, $\diag e_\alpha(G)=\iota$, for the $n$ unknown diagonal entries of $G$, the
symmetric matrix with off diagonal $\gamma$, where $e_\alpha=f_\alpha^{-1}$. This diagonal recovery problem, which we call the \emph{diagonal problem}, is the reason that surjectivity is nontrivial.

In this paper we analyze what makes GFT ($\alpha=0$) unique in the family of Box--Cox transformed correlation matrices, and identify other values of $\alpha$ with unique properties.
Some of the properties we identify for $\alpha=0$ are shared with other members of the family, others hold
for the GFT alone, and only the GFT has them all. The diagonal problem is a strictly convex program for every $\alpha$,
so every member is injective with a nonsingular Jacobian. The family splits at $\alpha=0$, because 
$\Phi_\alpha$ maps onto $\R^{d}$ if and only if $\alpha\le0$, while for $\alpha>0$ the image
is bounded. We say that $\Phi_\alpha$ is \emph{surjective} when it maps onto all of $\R^{d}$, that
is, when every $\gamma\in\R^{d}$ is the image of some correlation matrix, and we call
$\alpha\le0$ the \emph{surjective range}. Moreover the image is convex for $\alpha\in[\tfrac12,1]$, with an explicit inverse at
$\alpha=\tfrac12$, and it is nonconvex for $\alpha>1$ when $n\ge3$ and near $\alpha=\tfrac14$ when
$n\ge4$. The lack of convexity matters, because an average of two correlation forecasts, taken in the coordinates $\Phi_\alpha$, need not be the image of a valid correlation matrix, whereas for $\alpha\in[\tfrac12,1]$ every convex combination of points of the image lies in the image. We prove the conditioning
bound $\kappa(J_\alpha(C))\le\kappa(C)^{|1-\alpha|}$, which is attained for $n\ge4$. On the surjective range the exponent
is smallest at $\alpha=0$, i.e.\ GFT.

We also show that
the asymptotic covariance of $\Phi_\alpha(\hat C)$ under Gaussian sampling expands around the identity as
$I_d+\alpha B_\Delta+O(\|\Delta\|^{2})$, so that $\alpha=0$ is the unique member whose
asymptotic Cram\'er--Rao bound is free of the correlation matrix to first order when $n\ge3$. For
$n=2$ this is the variance stabilising property of Fisher's $z$. For every $n\ge3$ the GFT is also exactly variance stabilising along the equicorrelation line, on a subspace of dimension $1+n(n-3)/2$. The variance result can also
be stated in terms of the information matrix. Concentrating the variances out of the Gaussian information matrix
leaves the efficient information for the correlations, which only for $\alpha=0$ is the identity
at independence with vanishing first derivatives there, and in the geometric language of the
quotient of the Fisher--Rao geometry by diagonal rescaling \citep{DavidGu:2019,ThanwerdasPennec:2021}
the GFT agrees with normal coordinates to second order. Third moments behave differently. At
independence every coordinate is symmetric for every $\alpha$, and the leading joint third
cumulant on a triangle of distinct pairs is proportional to $(3\alpha-1)/2$, so it vanishes at
$\alpha=\tfrac13$ in every dimension $n\ge3$. We also derive the
leading bias of the coordinates near independence. On the surjective range it is smallest at
$\alpha=0$, although $\alpha=0$ is not the least biased member of the whole family.

The stabilisation along the equicorrelation line gives a statistic for pair-specific departures from equicorrelation whose limit under Gaussian sampling is $\chi^{2}$ without an estimated covariance. Separability of a Kronecker structure $C_A\otimes C_B$ is a linear restriction in GFT coordinates, as in \citet{HafnerLintonTang:2020}, so a restricted estimate requires no constrained optimisation. In simulations with Gaussian data, parametric bootstrap critical values, evaluated by the warp-speed method and checked with conventional bootstrap tests in the most distorted designs, bring the rejection rates of the GFT Wald test close to the nominal level, whereas the Wald test in raw correlations with the same calibration over-rejects in the larger designs (Section~\ref{sec:tests}).

Finally, applying
$f_\alpha$ to a diagonal rescaling of $C$ chosen to make the result hollow produces a second
family in which the inverse is explicit and the forward map is a convex scaling problem. Its
image is bounded unless $\alpha=0$, where every correlation matrix is $\Cor(e^{H})$ for exactly one
hollow symmetric $H$, which in the geometric language says that the exponential map of the
quotient is a global diffeomorphism. This logarithmic case is known. The uniqueness of the
hollowing rescaling was conjectured by \citet[Conjecture 3.1]{Thanwerdas:2024}, and it follows
from the decomposition in \citet[Section 1.2.1]{TumpachLarotonda:2024} and is stated in
\citet[Theorem 2.39]{BissonPennec:2025}. What is new is that one strictly convex scaling problem
works for every Box--Cox index, gives an explicit inverse, and identifies exactly when the scaled
coordinates are unrestricted.

Surjectivity for $\alpha\le0$ can also be derived from the mixed parametrization
theorem of \citet{Zwiernik:2025}, who placed the bijection of
\citet{ArchakovHansen:Correlation} in a general framework of entropic covariance models. In these
models a convex generator $F$ on the positive definite matrices defines the coordinates
$L=\nabla F(\Sigma)$, and Zwiernik shows that a positive definite matrix is determined by any
chosen subset of its own entries together with the remaining entries of $L$, provided that the
gradient of $F$ diverges at the boundary of the cone, which is to say that $F$ is essentially
smooth. Fixing the diagonal of $\Sigma$ to ones and taking the
off diagonal of $L$ gives a parametrization of correlation matrices, which recovers
the GFT for $F=\tr(\Sigma\log\Sigma-\Sigma)$ and gives the off diagonal of the inverse for
$F=-\log\det\Sigma$. That a positive definite matrix is determined by its diagonal together with
the off diagonal of its inverse, the two being variation independent, is the mixed
parametrization of the Gaussian exponential family \citep{BarndorffNielsen:1978}, of which
covariance selection \citep{Dempster:1972} is the case with zeros in the inverse. For
correlation matrices it is stated in \citet[Corollary 5.8]{Zwiernik:2025}, who proposes it as an
algebraic alternative to the logarithm. The Box--Cox family corresponds to the generators
$F_\alpha=\tr\Psi^{*}_\alpha$ of Section~\ref{sec:dual}, which are essentially smooth exactly
when $\alpha\le0$.
What is new here is the classification of the family. We give a direct proof of the dichotomy,
which demonstrates the mechanism and why it fails for $\alpha>0$, and we establish the results that
separate $\alpha=0$ from the rest of the surjective range. These include the cost of the
algebraic alternative at $\alpha=-1$, which is an extra factor of $\kappa(C)$ in the worst-case
conditioning and an asymptotic covariance that depends on $C$ at first order.

Section~\ref{sec:family} introduces the family. Section~\ref{sec:fibre} shows that the diagonal problem is a convex program for every $\alpha$, and identifies its dual as
a penalised maximisation over the elliptope, which at $\alpha=0$ is von Neumann entropy
maximisation. Section~\ref{sec:dichotomy} proves
the dichotomy and the convexity results for the image. Section~\ref{sec:special} collects the
properties that isolate $\alpha=0$ within the surjective range, which concern conditioning,
variance stabilisation and the efficient information matrix, and uses them in tests of
separability and of departures from equicorrelation. Section~\ref{sec:scaled} treats the
scaled family and the related parametrizations in the literature, and Section~\ref{sec:conclusion}
concludes. Appendix~\ref{app:proofs} contains the proofs of the results in the main text,
Appendix~\ref{app:supp} collects supplementary results on the diagonal problem, the image, the
conditioning for $n=3$, the second order covariance and bias, the information geometry and the
scaling problem, Appendix~\ref{sec:algo} collects algorithms for the inverse for general
$\alpha$, and Appendix~\ref{app:num} gives the computational details.

\section{The Box--Cox family for correlation matrices}\label{sec:family}

Let $\Sn$ denote the symmetric $n\times n$ matrices, $\Sn^{0}\subset\Sn$ the hollow symmetric
matrices, that is, those with zero diagonal, and $\Cn\subset\Sn$ the positive definite
correlation matrices, which form the relative interior of the elliptope $\Cbar$ inside the
affine space of unit-diagonal symmetric matrices. Let $\vecl:\Sn\to\R^{d}$ be the
strict lower triangle vectorisation, $d=n(n-1)/2$, and $\iota$ the vector of ones. We write
$\alpha$ for the index of the Box--Cox family, which \citet{BoxCox:1964} denote by $\lambda$,
because we reserve $\lambda$ for eigenvalues, with $\lambda_k(A)$ the $k$-th largest eigenvalue
of a symmetric matrix $A$. For a positive definite $A$ the condition number is
$\kappa(A)=\lambda_{\max}(A)/\lambda_{\min}(A)$, which is $\|A\|\|A^{-1}\|$ in the spectral norm.
For a positive definite $\Sigma$, $\Cor(\Sigma)=\Diag(\Sigma)^{-1/2}\Sigma\Diag(\Sigma)^{-1/2}$ is the
corresponding correlation matrix, where $\Diag(\Sigma)$ is the diagonal matrix with the same
diagonal as $\Sigma$. For $\alpha\in\R$ write
\[
  f_\alpha(x)=\frac{x^{\alpha}-1}{\alpha},\qquad f_0(x)=\log x,\qquad x>0,
\]
and let $e_\alpha=f_\alpha^{-1}$ be the $\alpha$-exponential
\[
  e_\alpha(y)=(1+\alpha y)^{1/\alpha},\qquad e_0(y)=e^{y},\qquad 1+\alpha y>0 .
\]
Both are applied to symmetric matrices spectrally, so that $g(U\diag(a)U')=U\diag(g(a))U'$. The
object of study is $\Phi_\alpha(C)=\vecl f_\alpha(C)$ as in (\ref{eq:family}). Since $I$ is
diagonal, $\Phi_\alpha(C)=\alpha^{-1}\vecl(C^{\alpha})$ for $\alpha\ne0$, and the affine
normalisation matters only because it makes $\alpha\mapsto\Phi_\alpha$ continuous at
$\alpha=0$.

For $n=2$ the family collapses to a scalar transform of $\rho$,
\begin{equation}\label{eq:n2}
  \Phi_\alpha(\rho)=\frac{(1+\rho)^{\alpha}-(1-\rho)^{\alpha}}{2\alpha},
\end{equation}
odd and strictly increasing for every $\alpha$, equal to $\operatorname{artanh}\rho$ at
$\alpha=0$, to $\rho$ at $\alpha=1$ and to $\rho/(1-\rho^{2})$ at $\alpha=-1$.

Two symmetries hold for the whole family and are the symmetries of the elliptope itself. For any
permutation matrix $P$ and signature matrix $D=\diag(\pm1)$ we have
$(PCP')^{\alpha}=PC^{\alpha}P'$ and $(DCD)^{\alpha}=DC^{\alpha}D$, so $\Phi_\alpha$ is
permutation equivariant and sign equivariant, $\gamma_{ij}\mapsto s_is_j\gamma_{ij}$.

For equicorrelation matrices the transformation has a closed form, which is useful for
interpreting several of the later results. Let $C_\rho$ be equicorrelated with parameter
$\rho\in(-\tfrac{1}{n-1},1)$. It has the eigenvalue $1+(n-1)\rho$ on $\iota$ and $1-\rho$ on the
orthogonal complement of $\iota$, so
$C_\rho^{\alpha}=(1-\rho)^{\alpha}I+n^{-1}[\{1+(n-1)\rho\}^{\alpha}-(1-\rho)^{\alpha}]\iota\iota'$,
and every coordinate of $\Phi_\alpha(C_\rho)$ equals
\begin{equation}\label{eq:equi}
  a_{\alpha,n}(\rho)=\frac{\{1+(n-1)\rho\}^{\alpha}-(1-\rho)^{\alpha}}{n\alpha},
  \qquad
  a_{0,n}(\rho)=\frac{1}{n} \log\frac{1+(n-1)\rho}{1-\rho}.
\end{equation}

The expression for $a_{0,n}$ is given in \citet{ArchakovHansenLuo-RandomCorr:2024}, together
with its inverse $\rho=(1-e^{-na})/\{1+(n-1)e^{-na}\}$.
At $n=2$, (\ref{eq:equi}) simplifies to (\ref{eq:n2}). Spectral powers shift the index of the family,
\begin{equation}\label{eq:power} 
  f_\alpha(A^{t})=t\,f_{\alpha t}(A),
\end{equation}
so a power path $t\mapsto A^{t}$ is a straight line through the origin in the $\alpha$
coordinates only when $\alpha=0$, and inversion is the case $t=-1$. Scaling acts by
$\vecl f_\alpha(aA)=a^{\alpha}\vecl f_\alpha(A)$ for $a>0$. Applied to a covariance matrix, the
off diagonal coordinates are therefore unaffected by a common change of scale only when
$\alpha=0$, since $\log(a\Sigma)=(\log a)I+\log\Sigma$.

We adopt the following terminology from matrix analysis. A \emph{diffeomorphism} is a bijection
that is differentiable in both directions, and a differentiable bijection need not have a
differentiable inverse.\footnote{The map $x\mapsto x^{3}$ on $\R$ is a differentiable bijection,
but its inverse $x^{1/3}$ is not differentiable at the origin, where the derivative of $x^{3}$
vanishes.} This is why injectivity and
nonsingularity of the Jacobian are established separately below. A map is \emph{real analytic}
if it is locally the sum of a convergent power series, which is strictly stronger than being
infinitely differentiable. Given a scalar $g$ and $A=U\diag(a)U'$, the derivative of $A\mapsto
g(A)$ is described by the first divided differences
$g^{[1]}(a_k,a_\ell)=\{g(a_k)-g(a_\ell)\}/(a_k-a_\ell)$, with $g'(a_k)$ on the diagonal, through
the Daleckii--Krein formula $Dg(A)[H]=U\{\Lambda^{[1]}\circ(U'HU)\}U'$, where $\Lambda^{[1]}$
is the matrix with entries $g^{[1]}(a_k,a_\ell)$ and $\circ$ is the elementwise product. For
a specific $g$ we write $g^{[1]}$ with that function's name, as in $e_\alpha^{[1]}$ or
$f_\alpha^{[1]}$.

\section{The diagonal problem}\label{sec:fibre}

\subsection{Convexity}\label{sec:convexity}

Given $\gamma\in\R^{d}$, the inverse problem is to find $C\in\Cn$ with $\Phi_\alpha(C)=\gamma$. Writing
$G=f_\alpha(C)$, so that $\gamma=\vecl(G)$ as in the GFT case $G=\log C$, the off diagonal of
$G$ is prescribed and the diagonal is free. Let $G(\delta)\in\Sn$ be the symmetric matrix with
$\vecl G(\delta)=\gamma$ and $\diag G(\delta)=\delta$, and let
\[
  \mathcal{D}_\alpha=\{\delta\in\R^{n}:I+\alpha G(\delta)\succ0\},
\]
with $\mathcal{D}_0=\R^{n}$. On $\mathcal{D}_\alpha$ set $C(\delta)=e_\alpha(G(\delta))\succ0$.
The \emph{diagonal problem} is to find $\delta\in\mathcal{D}_\alpha$ with
$\diag C(\delta)=\iota$, and its solution $\delta^{*}$ gives the inverse as
$\Phi_\alpha^{-1}(\gamma)=C(\delta^{*})$. We solve it by minimising the objective function
\begin{equation}\label{eq:L}
  L_\alpha(\delta)=\tr\Psi_\alpha\bigl(G(\delta)\bigr)-\iota'\delta,
  \qquad
  \Psi_\alpha(y)=\frac{(1+\alpha y)^{(1+\alpha)/\alpha}}{1+\alpha},
\end{equation}
with $\Psi_0(y)=e^{y}$ and $\Psi_{-1}(y)=-\log(1-y)$. These are the antiderivatives of
$e_\alpha$ at the two values where the closed form fails, since $e_0(y)=e^{y}$ and
$e_{-1}(y)=(1-y)^{-1}$ integrates to $-\log(1-y)$. Only $\Psi_\alpha'$ enters the first
order condition, so $\Psi_\alpha$ is in any case determined only up to an additive constant. The
point of the Box--Cox coordinate is that $\Psi_\alpha'=e_\alpha$ and
$\Psi_\alpha''=e_\alpha'=(1+\alpha y)^{1/\alpha-1}>0$ on the whole domain, for every sign of
$\alpha$. Up to sign, $\Psi_\alpha$ is the Cressie--Read criterion function of the generalized
empirical likelihood estimators in \citet[Theorem 2.2]{NeweySmith:2004}, with the Cressie--Read
index equal to $\alpha$. The case $\alpha=0$ corresponds to the exponential tilting estimator of
\citet{KitamuraStutzer:1997}, $\alpha=-1$ to empirical likelihood
\citep{Owen:1988,Kitamura:2001}, and $\alpha=1$ to the continuously updated estimator.

\begin{proposition}[Strict convexity and injectivity]\label{prop:convex}
For every $\alpha\in\R$ the set $\mathcal{D}_\alpha$ is open and convex. Moreover $L_\alpha$ is
strictly convex on $\mathcal{D}_\alpha$ with
\[
  \nabla L_\alpha(\delta)=\diag\bigl(e_\alpha(G(\delta))\bigr)-\iota,
  \qquad
  \bigl[\nabla^{2}L_\alpha(\delta)\bigr]_{ij}=w_{ij}'\Lambda^{[1]}w_{ij},
\]
where $G(\delta)=U\diag(a)U'$, $w_{ij}=(u_{i1}u_{j1},\dots,u_{in}u_{jn})'$ and
$\Lambda^{[1]}_{k\ell}=e_\alpha^{[1]}(a_k,a_\ell)$. Consequently $L_\alpha$ has at most one
stationary point and $\Phi_\alpha$ is injective.
\end{proposition}

The Jacobian $J_\alpha(C)$ of $\Phi_\alpha$ with respect to the free off diagonal entries of $C$
is symmetric positive definite at every $C\in\Cn$, so $\Phi_\alpha$ is a real analytic
diffeomorphism onto its image $\mathcal{R}_\alpha=\Phi_\alpha(\Cn)$, which is an open subset of
$\R^{d}$ (Lemma~\ref{lem:jac} in Appendix~\ref{app:diag}). The recovered diagonal
$\delta^{*}=\diag f_\alpha(C)$ also has a definite sign, nonpositive for $\alpha<1$ and
nonnegative for $\alpha>1$ (Proposition~\ref{prop:sign}). For $\alpha=0$ this is
\citet[Lemma 2]{ArchakovHansen:Correlation}.

The Jacobian determinant is needed for densities and likelihoods, and it reduces to an
$n\times n$ determinant.

\begin{proposition}\label{lem:detJ}
Let $C=U\diag(\lambda)U'\in\Cn$, and let $\delta^{*}$ solve the diagonal problem for
$\gamma=\Phi_\alpha(C)$. Then
\[
  \det J_\alpha(C)=\det\nabla^{2}L_\alpha(\delta^{*})\,\prod_{k}\lambda_k^{\alpha-1}
  \prod_{k<\ell}f_\alpha^{[1]}(\lambda_k,\lambda_\ell).
\]
\end{proposition}

Proposition~\ref{lem:detJ} gives the density of $\Phi_\alpha(\hat C)$ whenever $\hat C$ has a density.
For $\alpha\le0$ it also turns a Gaussian model for $\Phi_\alpha(C_t)$ into a likelihood for the
correlations that can be compared across $\alpha$, so that $\alpha$ can be estimated as in
\citet{BoxCox:1964}. For $\alpha>0$ the image $\mathcal{R}_\alpha$ is bounded, so the Gaussian
density must in addition be divided by the probability of $\mathcal{R}_\alpha$, which depends on
$\alpha$ and on the parameters.

\subsection{The extended convex objective and its dual}\label{sec:dual}

The objective $L_\alpha$ is defined on $\mathcal{D}_\alpha$, and for $\alpha>0$ we extend it
to a finite convex function on all of $\R^{n}$, although $\mathcal{D}_\alpha$ is a proper subset
of $\R^{n}$. At $\alpha=0$ there is nothing to extend, since $\mathcal{D}_0=\R^{n}$. For
$\alpha<0$ no finite convex extension to $\R^{n}$ exists. The function $\Psi_\alpha$ has infinite slope at
the boundary of its domain, since $e_\alpha(y)=(1+\alpha y)^{1/\alpha}\to\infty$ as
$1+\alpha y\downarrow0$, so every convex extension is $+\infty$ beyond the boundary point. For
$-1\le\alpha<0$ the value diverges there as well, whereas for $\alpha<-1$ it stays finite, so
that, for example, $\Psi_{-2}(y)=-\sqrt{1-2y}$ extends convexly to the boundary point
$y=\tfrac12$ with value zero. For $\alpha>0$ the slope $e_\alpha$ tends to
zero at the boundary and $\Psi_\alpha$ extends to a finite convex function on $\R$,
\[
  \hat\Psi_\alpha(y)=\frac{(1+\alpha y)_{+}^{(1+\alpha)/\alpha}}{1+\alpha}
  \qquad(\alpha>0),
\]
which is continuously differentiable with $\hat\Psi_\alpha'(y)=\hat e_\alpha(y)=(1+\alpha y)_{+}^{1/\alpha}$
because the exponent $(1+\alpha)/\alpha$ exceeds one. Let
\begin{equation}\label{eq:Lhat}
  \hat L_\alpha(\delta)=\tr\hat\Psi_\alpha\bigl(G(\delta)\bigr)-\iota'\delta,
  \qquad\delta\in\R^{n},
\end{equation}
a finite convex function on $\R^{n}$ that agrees with $L_\alpha$ on $\mathcal{D}_\alpha$. It is
coercive and attains its infimum over $\R^{n}$. If $\gamma\in\mathcal{R}_\alpha$ its minimiser
is the stationary point $\delta^{*}\in\mathcal{D}_\alpha$ of Proposition~\ref{prop:convex}, an
interior solution, and otherwise every minimiser lies outside $\mathcal{D}_\alpha$
(Proposition~\ref{prop:dual} in Appendix~\ref{app:diag}).

For $\alpha<0$ the objective is defined only on $\mathcal{D}_\alpha$, so the diagonal problem is
a minimisation over a proper subset of $\R^{n}$. The proof of Theorem~\ref{thm:main} shows that
the minimum is always attained inside $\mathcal{D}_\alpha$ and never on its boundary, so this
constraint never binds. Only at $\alpha=0$ is the objective finite on all of $\R^{n}$
and its minimiser a solution for every $\gamma$.

Since $(\Psi_\alpha')^{-1}=f_\alpha$, the convex conjugate of $\Psi_\alpha$ is
\begin{equation}\label{eq:conj}
  \Psi_\alpha^{*}(y)=
  \begin{cases}
    \dfrac{1}{\alpha}\Bigl(\dfrac{y^{1+\alpha}}{1+\alpha}-y\Bigr), & \alpha\ne0,-1,\\[8pt]
    y\log y-y, & \alpha=0,\\[4pt]
    y-1-\log y, & \alpha=-1,
  \end{cases}
\end{equation}
the last two being the limits of the first modulo affine functions of $y$. All are strictly
convex on $y>0$, since $(\Psi^{*}_\alpha)''(y)=y^{\alpha-1}>0$.

The inverse map can also be characterised as the solution of a maximisation problem over
correlation matrices. By Fenchel--Rockafellar
duality, see \citet[Thm.~31.1]{Rockafellar:1970}, together with the fact that conjugation
commutes with the spectral construction \citep{Lewis:1995}, the diagonal problem has the dual
\begin{equation}\label{eq:maxent}
  \inf_{\delta\in\R^{n}}\hat L_\alpha(\delta)
  =\sup_{C\in\Cbar}\;\Bigl\{2\gamma'\vecl(C)-\tr\Psi^{*}_\alpha(C)\Bigr\},
\end{equation}
with no gap, the supremum attained, and the maximiser equal to $\Phi_\alpha^{-1}(\gamma)$
whenever $\gamma\in\mathcal{R}_\alpha$. For $\alpha\le0$ the closure of $\Psi_\alpha$ takes
the place of $\hat\Psi_\alpha$.

In the language of Lagrangian duality, the dual is a maximisation over positive semidefinite
matrices subject to the $n$ constraints $C_{ii}=1$, and $\delta$ is the vector of Lagrange
multipliers for these unit-diagonal constraints, with Lagrangian
$\tr\{CG(\delta)\}-\tr\Psi^{*}_\alpha(C)-\iota'\delta$. At an interior maximiser the first order condition in $C$ is
$f_\alpha(C)=G(\delta)$, whose off diagonal is $\Phi_\alpha(C)=\gamma$, so the diagonal problem
determines the multipliers and the corresponding maximiser is the inverse. For $\alpha>0$ the maximiser can be singular, and the cone constraint then adds a complementary slackness term to the first order condition. In the analogy with
generalized empirical likelihood, $\delta$ takes the place of the tilting parameters, and since
$\Psi^{*}_\alpha$ differs from the Cressie--Read divergence by an affine function, whose trace is
constant on the elliptope, the dual is the counterpart of the minimum discrepancy problem in
\citet{NeweySmith:2004}.

The objective in (\ref{eq:maxent}) is a linear tilt by $\gamma$ minus a convex penalty on the
spectrum of $C$. The inverse therefore selects, among all correlation matrices, the one that best
balances the tilt, which pulls $C_{ij}$ up when $\gamma_{ij}$ is positive and down when it is
negative, against a penalty that grows as $C$ approaches singularity. At $\alpha=0$ the penalty
$\tr\Psi^{*}_0(C)=\tr(C\log C)-n$ is, up to an additive constant, the negative of the von Neumann
entropy $-\tr(C\log C)$, so the dual is von Neumann entropy maximisation, the convex program for the inverse GFT in \citet{ArchakovHansen:Variational} and
\citet[Remark 5.6]{Zwiernik:2025}. At
$\alpha=-1$ it is the maximisation of $2\gamma'\vecl(C)+\log\det C$, which has the form of a
Gaussian log likelihood $\log\det K-\tr(SK)$ with $C$ in the role of the concentration matrix $K$
and $S$ the unit-diagonal matrix with off-diagonal entries $-\gamma_{ij}$. Its first order
condition $(C^{-1})_{ij}=-\gamma_{ij}$ is the classical variational characterisation of the
concentration matrix. How steeply the penalty rises as an eigenvalue approaches zero depends on
$\alpha$, and this is what drives the dichotomy of the next section.

\needspace{16\baselineskip}
\section{Unrestricted parametrization and the shape of the image}\label{sec:dichotomy}

\begin{theorem}\label{thm:main}
Let $n\ge2$, $d=n(n-1)/2$, and $\mathcal{R}_\alpha=\Phi_\alpha(\Cn)\subseteq\R^{d}$.
\begin{enumerate}
\item[(i)] For every $\alpha\in\R$, $\Phi_\alpha$ is injective and a real analytic
diffeomorphism onto the open set $\mathcal{R}_\alpha$.
\item[(ii)] $\mathcal{R}_\alpha=\R^{d}$ if and only if $\alpha\le0$. In that case
$\Phi_\alpha$ is a global real analytic diffeomorphism $\Cn\to\R^{d}$.
\item[(iii)] For $\alpha>0$ the image is bounded, with
\begin{equation}\label{eq:bound}
  \sup_{C\in\Cn}\|\Phi_\alpha(C)\|_{\infty}=\frac{1}{\alpha}
  \begin{cases}
    2^{\alpha-1}, & 0<\alpha\le1,\\
    n^{\alpha-1}, & \alpha>1,
  \end{cases}
\end{equation}
where the supremum is not attained.
\end{enumerate}
\end{theorem}

\begin{figure}[!tbp]
\centering
\includegraphics[width=\textwidth]{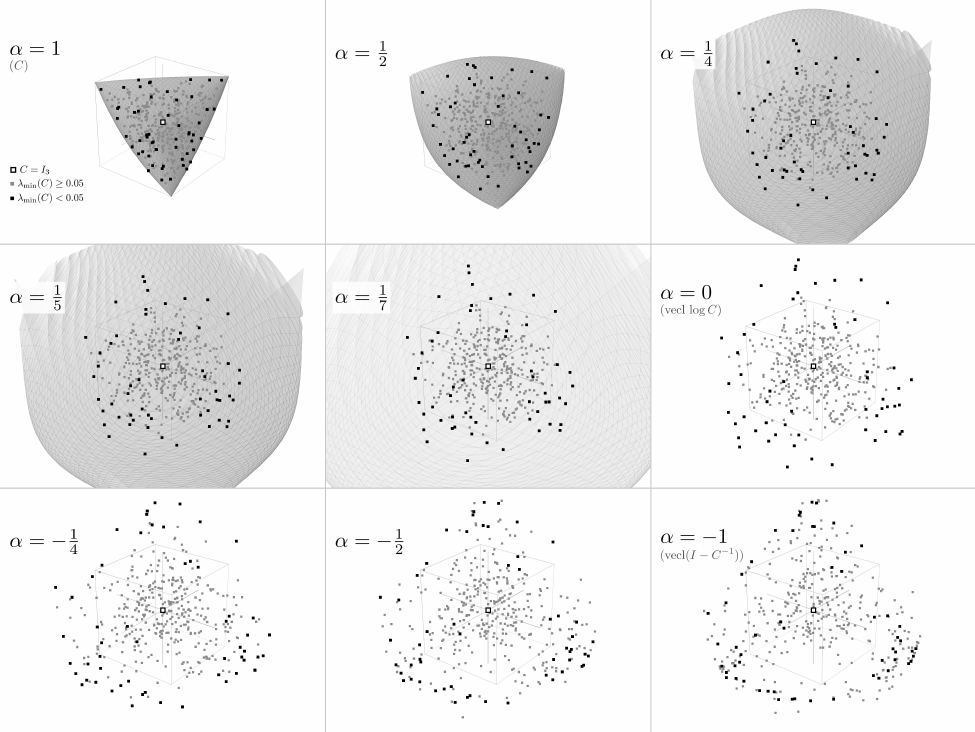}
\caption{The image of the $3\times3$ correlation matrices under $\Phi_\alpha$ for nine values of
$\alpha$. Each panel shows the same 400 correlation matrices, drawn as described in Appendix~\ref{app:num}. The white square is $C=I_3$, and black points have $\lambda_{\min}(C)<0.05$. All panels use the scale of the panel for $\alpha=1$, and points outside it are drawn on its edge. For $\alpha>0$ the grey surface is the image of the singular correlation matrices, which encloses $\mathcal{R}_\alpha$.}
\label{fig:image}
\end{figure}

For $n=2$, (\ref{eq:n2}) gives $\mathcal{R}_\alpha=(-2^{\alpha-1}/\alpha,2^{\alpha-1}/\alpha)$
for $\alpha>0$, in agreement with (\ref{eq:bound}), and $\mathcal{R}_\alpha=\R$ for $\alpha\le0$.
For $0<\alpha\le1$ every coordinate therefore obeys the bivariate bound in every dimension. Figure~\ref{fig:image} shows the image for $n=3$ and nine values of
$\alpha$. As $\alpha$ decreases to zero the bounded image grows without limit, and for
$\alpha\le0$ it is all of $\R^{3}$.

The dual (\ref{eq:maxent}) explains why the family splits at $\alpha=0$. The inverse maximises a
linear tilt minus the penalty $\tr\Psi^{*}_\alpha(C)$ over the elliptope, and the question is
whether a large tilt $\gamma$ can push the maximiser onto the boundary, where $C$ is singular.
By (\ref{eq:conj}) the slope of the penalty in an eigenvalue $y$ is
$(\Psi_\alpha^{*})'(y)=f_\alpha(y)=(y^{\alpha}-1)/\alpha$, which tends to $-\infty$ as
$y\downarrow0$ if and only if $\alpha\le0$. For $\alpha\le0$ the marginal reduction in the
penalty from moving an eigenvalue away from zero is therefore infinite, in the same way that an
Inada condition, infinite marginal utility at zero consumption, rules out zero consumption, and
no tilt, however large, makes a singular matrix optimal. The maximiser is then interior for
every $\gamma$, which is surjectivity. For $\alpha>0$ the slope at zero is finite, a
sufficiently large tilt produces a corner solution on the boundary of the elliptope, and such a
$\gamma$ has no preimage. Proposition~\ref{prop:dual} shows the same thing on the primal side,
where for $\alpha>0$ the minimiser of $\hat L_\alpha$ leaves $\mathcal{D}_\alpha$ exactly when
no correlation matrix maps to $\gamma$.

In convex analysis this property of $\Psi^{*}_\alpha$ is called essential smoothness, or
steepness, and it concerns the slope of the penalty rather than its value. As $y\downarrow0$ the
value $\Psi^{*}_\alpha(y)$ diverges polynomially for $\alpha<-1$, since $y^{1+\alpha}\to\infty$,
and logarithmically at $\alpha=-1$, whereas for $-1<\alpha\le0$ it stays finite while the slope
diverges, and for $\alpha>0$ both stay finite. It is the slope, not the value, that drives
Theorem~\ref{thm:main}.

Part (ii) of Theorem~\ref{thm:main} can also be obtained from \citet[Theorem 5.4]{Zwiernik:2025}.
The generator $F_\alpha(\Sigma)=\tr\Psi^{*}_\alpha(\Sigma)$ is closed, strictly convex and
differentiable on the positive definite cone with $\nabla F_\alpha=f_\alpha$, and by the
preceding paragraph its gradient diverges at the boundary of the cone exactly when $\alpha\le0$.
Zwiernik's theorem then says that $\Sigma\succ0$ is determined by its diagonal together with the
off diagonal of $f_\alpha(\Sigma)$, and the latter ranges over all of $\R^{d}$ for $\alpha\le0$,
because the diagonal of $G=f_\alpha(\Sigma)$ can be chosen to make $I+\alpha G$ diagonally
dominant. Fixing the diagonal of $\Sigma$ to ones gives Theorem~\ref{thm:main}(ii), and at
$\alpha=-1$ this is \citet[Corollary 5.8]{Zwiernik:2025}. Our proof of Theorem~\ref{thm:main} is self contained and
shows where the hypothesis fails for $\alpha>0$, since the generator then has a finite value and
a finite first derivative at the boundary of the cone, and the dual maximiser can sit on the
boundary.

\needspace{14\baselineskip}
At $\alpha=1$ the image is the interior of the elliptope, hence convex. Convexity persists below
$\alpha=1$ down to $\alpha=\tfrac12$, where the inverse is in fact explicit, but not above
$\alpha=1$, and for $n\ge4$ it fails near $\alpha=\tfrac14$. For $\alpha\in[\tfrac12,1]$ set $p=1/\alpha\in[1,2]$, so that $C=A^{p}$ with
$A\equiv C^{\alpha}$.

\begin{theorem}\label{thm:shape}
\begin{enumerate}
\item[(i)] For every $\alpha\in[\tfrac12,1]$ the image $\mathcal{R}_\alpha$ is convex. At
$\alpha=\tfrac12$ the inverse is explicit. Let $B(\gamma)$ be the symmetric matrix with off
diagonal entries $\gamma_{ij}/2$ and diagonal entries
$b_i=\{1-\tfrac14\sum_{j\ne i}\gamma_{ij}^{2}\}^{1/2}$. Then $\gamma\in\mathcal{R}_{1/2}$ if and
only if $B(\gamma)$ is well defined and positive definite, and in that case
$\Phi_{1/2}^{-1}(\gamma)=B(\gamma)^{2}$.
\item[(ii)] For every $n\ge3$ and every $\alpha>1$ the image $\mathcal{R}_\alpha$ is not convex.
\item[(iii)] There is an open interval containing $\tfrac14$ such that for every $n\ge4$ and every
$\alpha$ in the interval the image $\mathcal{R}_\alpha$ is not convex.
\end{enumerate}
\end{theorem}

At $\alpha=2$ and $n=3$ the nonconvexity in Theorem~\ref{thm:shape}(ii) can be seen without a limit. With $a=1/\sqrt2$ and $s=4/5$, the matrices $C_\pm$ with off
diagonal entries $(\pm a,\pm a,s)$ have $\det C_\pm=4/25>0$, and since
$\Phi_2(C)=(x+\tfrac12yz,\,y+\tfrac12xz,\,z+\tfrac12xy)$ for off diagonal entries $(x,y,z)$,
their images have midpoint $(0,0,21/20)$. If that were $\Phi_2(C(x,y,z))$ then the first two
coordinates force $x(1-\tfrac14z^{2})=0$, hence $x=y=0$, and the third then requires $z=21/20$,
contradicting $|z|<1$.

At $\alpha=\tfrac12$ the convexity is also visible directly, because each $b_i$ is a concave
function of $\gamma$, so that $B(t\gamma_1+(1-t)\gamma_2)\succeq tB(\gamma_1)+(1-t)B(\gamma_2)\succ0$ whenever
both endpoints lie in $\mathcal{R}_{1/2}$. The diagonal problem at $\alpha=\tfrac12$ is solved
in closed form by $\delta^{*}_i=2(b_i-1)$, so that at $\alpha=\tfrac12$, as at $\alpha=1$,
the inverse requires no iteration.

The convexity argument does not extend below $\alpha=\tfrac12$, because $x^{p}$ is not operator
convex for $p>2$. The proof of part (iii) shows that the midpoint of the images of a nearly
singular $4\times4$ correlation matrix and of the matrix obtained by changing the sign of one
variable lies outside $\mathcal{R}_\alpha$,
although every coordinate obeys the bivariate bound in (\ref{eq:bound}), so the obstruction comes
from which coordinates can occur together. Appendix~\ref{app:shape} describes the construction,
the numerical evidence for other values of $\alpha$ in $(0,\tfrac12)$, and a conjecture for the
remaining range.

\section{What makes the GFT special}\label{sec:special}

By Theorem~\ref{thm:main}, only the members with $\alpha\le0$ map the correlation matrices onto
all of $\R^{d}$. Within this range, $\alpha=0$ is singled out by several further properties,
and most of them can be traced to the behaviour of $f_\alpha$ as an eigenvalue approaches zero,
that is, to the boundary singularity of $f_\alpha$. The GFT has the \emph{critical} singularity,
the weakest one that is compatible with surjectivity.

\begin{figure}[!tbp]
\centering
\includegraphics[width=\textwidth]{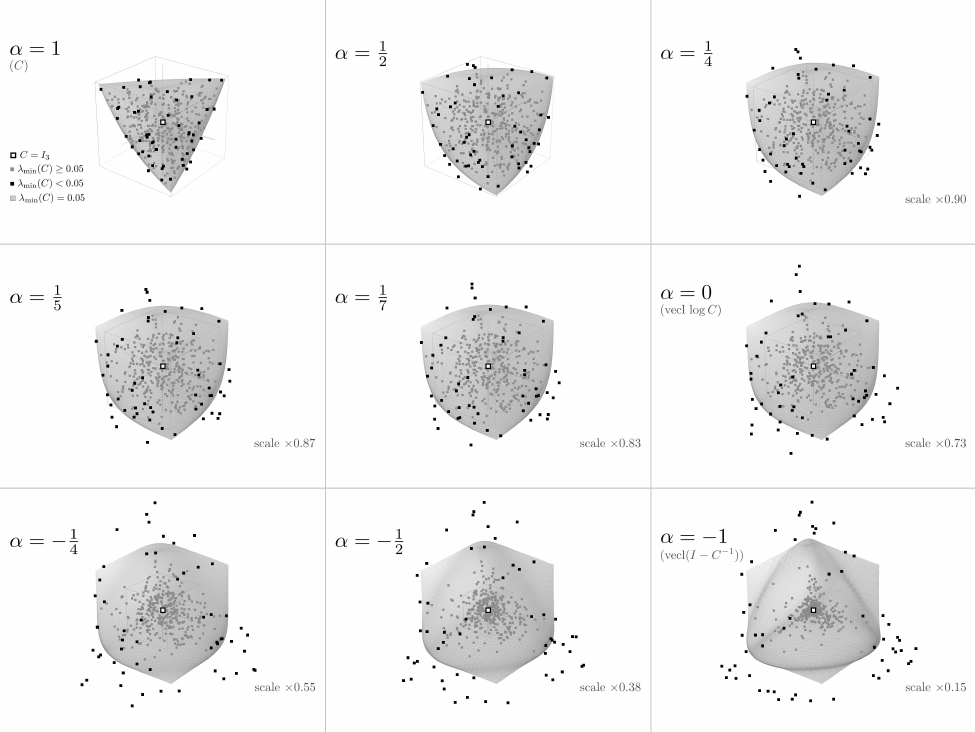}
\caption{The image of the well-conditioned $3\times3$ correlation matrices under $\Phi_\alpha$.
The points are those of Figure~\ref{fig:image}. The grey surface is now the image of the
correlation matrices with $\lambda_{\min}(C)=0.05$, which encloses the image of
$\{C:\lambda_{\min}(C)\ge0.05\}$. Each panel is scaled to fit the surface, and the scale relative
to Figure~\ref{fig:image} is given in its lower right corner when it differs from one.}
\label{fig:hull}
\end{figure}

As $\lambda\downarrow0$,
\begin{equation}\label{eq:sing}
  f_\alpha(\lambda)=\frac{\lambda^{\alpha}-1}{\alpha}\sim
  \begin{cases}
    -|\alpha|^{-1}\lambda^{-|\alpha|}, & \alpha<0,\\
    \log\lambda, & \alpha=0,\\
    -\alpha^{-1}, & \alpha>0,
  \end{cases}
\end{equation}
so the family has three regimes at the boundary of the elliptope, a pole of order $|\alpha|$
for $\alpha<0$, a logarithmic divergence at $\alpha=0$, and a finite limit for $\alpha>0$. Call $|\alpha|$ the
\emph{order} of the boundary singularity for $\alpha\le0$, with order zero meaning the
logarithmic case. Figure~\ref{fig:hull} shows the effect for $n=3$ on the correlation matrices
with $\lambda_{\min}(C)\ge0.05$, whose image is bounded for every $\alpha$. This set is the shrunken elliptope $0.05I+0.95\,\overline{\mathcal{C}}_3$, an affine copy of the elliptope \citep[Theorem 1]{Hansen:CorrelationMatrix}. For $\alpha<0$ the
matrices near the corners of this set, where two eigenvalues equal $0.05$, are pulled out into
spikes, and the spikes lengthen as the order of the singularity increases.

Surjectivity requires that the
singularity be present at all, since by Theorem~\ref{thm:main}, $\mathcal{R}_\alpha=\R^{d}$ exactly
when $f_\alpha(0^{+})=-\infty$, which by $(\Psi^{*}_\alpha)'=f_\alpha$ is the infinite slope of
the dual penalty discussed after Theorem~\ref{thm:main}. The conditioning bound depends on the order, because on the surjective range
the exponent of Theorem~\ref{thm:cond} below is
\[
  |1-\alpha|=1+|\alpha|,\qquad \alpha\le0,
\]
one plus the order of the singularity. A stronger singularity adds nothing to surjectivity once
the divergence is present, and adds one power of $\kappa(C)$ to the bound per unit of order. This
is the sense in which the GFT is the critical member.

\needspace{12\baselineskip}
\subsection{Conditioning}

\begin{theorem}\label{thm:cond}
For every $\alpha\in\R$ and every $C\in\Cn$,
\begin{equation}\label{eq:cond}
  \kappa\bigl(J_\alpha(C)\bigr)\le\kappa(C)^{|1-\alpha|}.
\end{equation}
For $n\ge4$ the bound is sharp, in the sense that for every $\alpha$ and every $\kappa_0>1$
there is a $C\in\Cn$ with $\kappa(C)=\kappa_0$ and $\kappa(J_\alpha(C))=\kappa(C)^{|1-\alpha|}$.
\end{theorem}

The exponent $|1-\alpha|$ is minimised over the surjective range $\alpha\le0$ at $\alpha=0$,
where it equals one. Among the members of the family that map $\Cn$ onto $\R^{d}$, the GFT
therefore has the smallest worst-case amplification of the condition number, and for $n\ge4$
this worst case is attained rather than merely bounded. Each unit of $\alpha$ below zero adds
one further factor of $\kappa(C)$ in the worst case.

The bound need not be approached at a given $C$, and how closely it is approached depends on $C$ and not only on $\kappa(C)$. At $n=2$ the Jacobian is scalar and $\kappa(J_\alpha)\equiv1$ however ill conditioned
$C$ is, so the conditioning criterion separates the members only for $n\ge3$. At $n=3$ the
exponent is asymptotically sharp for every $\alpha<1$, but for $\alpha\ne1$ equality holds only at
$C=I$, and for
$\alpha>1$ the condition number of $J_\alpha$ is bounded on $\mathcal{C}_3$
(Proposition~\ref{prop:n3} in Appendix~\ref{app:cond}). For $\alpha\ne1$, exact attainment away from
$C=I$ therefore requires $n\ge4$.

\subsection{Variance stabilisation}\label{sec:varstab}

For $n=2$ the asymptotic variance of $\hat\rho$ is $(1-\rho^{2})^{2}/T$, so $\Phi_\alpha$ is
variance stabilising if and only if $h_\alpha(\rho)=\Phi_\alpha'(\rho)(1-\rho^{2})$ is
constant. From (\ref{eq:n2}),
\[
  h_\alpha(\rho)=\tfrac12\bigl[(1+\rho)^{\alpha}(1-\rho)+(1-\rho)^{\alpha}(1+\rho)\bigr]
  =1+\Bigl(\frac{\alpha(\alpha-1)}{2}-\alpha\Bigr)\rho^{2}+O(\rho^{4}).
\]
The quadratic term vanishes at $\alpha=0$ and $\alpha=3$, but $h_3(\rho)=1-\rho^{4}$, so only
$\alpha=0$ gives $h_\alpha\equiv1$, recovering Fisher's $z$. The same differential equation
governs the leading skewness. By the delta method, the third cumulant of $\Phi_\alpha(\hat\rho)$ is
\[
  \frac{(1-\rho^{2})^{3}\Phi_\alpha'(\rho)^{2}}{T^{2}}
  \bigl\{3(1-\rho^{2})\Phi_\alpha''(\rho)-6\rho\,\Phi_\alpha'(\rho)\bigr\}+O(T^{-3}),
\]
which vanishes for every $\rho$ if and only if $\Phi_\alpha'$ is proportional to $(1-\rho^{2})^{-1}$.
Fisher's $z$ is therefore the only member that both stabilises the variance and removes the
leading skewness, a point emphasised by \citet{Hotelling:1953}.

For $n>2$, \citet{ArchakovHansen:2026} find in simulations that the GFT coordinates are nearly uncorrelated
and that their covariance changes little with $C$, also far from independence. We consider the
corresponding local problem at the identity, which again has a unique solution. Let $\hat C$ be the sample correlation matrix from $T$ independent draws of
$N(0,C)$, so that $\sqrt{T}\{\vecl(\hat C)-\vecl(C)\}\Rightarrow N(0,V(C))$. By the delta method
\[
  \sqrt{T}\{\Phi_\alpha(\hat C)-\Phi_\alpha(C)\}\Rightarrow N\bigl(0,V_\alpha(C)\bigr),
  \qquad V_\alpha(C)=J_\alpha(C)V(C)J_\alpha(C)',
\]
and we refer to $V_\alpha(C)$ as the asymptotic covariance of $\hat\gamma_\alpha=\Phi_\alpha(\hat C)$. For a hollow $\Delta$ define the \emph{overlap
matrix} $B_\Delta\in\R^{d\times d}$ by
\begin{equation}\label{eq:overlap}
  (B_\Delta)_{(i,j),(k,l)}=
  \begin{cases}
    \Delta_{uv} & \text{if }\{i,j\}\cap\{k,l\}=\{p\}\text{ with }\{u\}=\{i,j\}\setminus\{p\},\;\{v\}=\{k,l\}\setminus\{p\},\\
    0 & \text{if the pairs are equal or disjoint.}
  \end{cases}
\end{equation}

\begin{theorem}[Local variance stabilisation]\label{thm:varstab}
Let $n\ge3$. As $\Delta\to0$ through hollow symmetric matrices,
\begin{equation}\label{eq:varstab}
  V_\alpha(I+\Delta)=I_d+\alpha B_\Delta+O(\|\Delta\|^{2}).
\end{equation}
Hence $\alpha=0$ is the unique member of the family whose asymptotic covariance is free of $C$
to first order at the identity.
\end{theorem}

For $\alpha=0$ and $\alpha=1$ the expansion (\ref{eq:varstab}) is in
\citet[Corollary 1]{ArchakovHansen:2026}, and Theorem~\ref{thm:varstab} extends it to the whole
family. Their result holds under elliptical sampling, where the asymptotic covariance of the
sample correlations is $(1+\kappa)V(C)$, with $\kappa$ the kurtosis parameter
\citep{BrowneShapiro:1986}. Since $J_\alpha(C)$ does not depend on the sampling distribution,
Theorem~\ref{thm:varstab} also holds under elliptical sampling, with $V_\alpha$ replaced by
$(1+\kappa)V_\alpha$.

The proof of Theorem~\ref{thm:varstab} uses only $f_\alpha(1)=0$, $f_\alpha'(1)=1$ and
$f_\alpha''(1)=\alpha-1$, together with the smoothness of $f_\alpha$ near one. For any spectral
transformation $f$ that is three times continuously differentiable in a neighbourhood of one, with
$f(1)=0$ and $f'(1)=1$, the same argument gives
\[
  V_f(I+\Delta)=I_d+\{1+f''(1)\}B_\Delta+O(\|\Delta\|^{2}),
\]
so first-order stabilisation at independence is a property of the curvature of $f$ at one, and
it requires $f''(1)=-1$. It is not a consequence of the behaviour of $f$ near zero, which governs
surjectivity and conditioning.

Since $\hat C$ is the profiled maximum likelihood estimator of $C$ in the centred Gaussian
covariance model with unrestricted marginal variances,
$V_\alpha(C)$ is the inverse of the efficient information for $\gamma_\alpha$ after the
variances have been profiled out, so $\mathcal{I}_\alpha(C)=V_\alpha(C)^{-1}$ is the Fisher
information in the coordinates $\gamma_\alpha$ and $V_\alpha$ is the \emph{asymptotic}
Cram\'er--Rao bound per observation. It is asymptotic in the sense that the intrinsic bound of
\citet{SmithCRB:2005} carries additional bias and curvature corrections which (\ref{eq:varstab})
does not describe. In this language Theorem~\ref{thm:varstab} reads
$\mathcal{I}_\alpha(I)=I_d$ and $D\mathcal{I}_\alpha(I)[\Delta]=-\alpha B_\Delta$, so every
member is orthonormal at independence and only $\alpha=0$ has an information metric whose first
derivatives vanish there. Variance stabilisation, information stabilisation and Cram\'er--Rao
stabilisation are three formulations of one local property and should not be counted separately.
The efficient information for the raw correlations has a closed form. For $x=\vecl(X)$ with $X$
hollow symmetric,
\[
  x'V(C)^{-1}x=\tfrac12\tr\{(C^{-1}X)^{2}\}-\diag(C^{-1}X)'(I+C\circ C^{-1})^{-1}\diag(C^{-1}X),
\]
which is the quadratic form in the test of \citet{Jennrich:1970}.

Fisher information transforms with the coordinates. All members describe the same correlation
matrix, so no member has a smaller bound in an absolute sense, and the members differ only in
whether the bound varies with $C$, to first order at independence. The stabilisation property also concerns the joint covariance
matrix. Since $B_\Delta$ has zero diagonal by (\ref{eq:overlap}), the \emph{marginal} asymptotic
variance of each coordinate is stable to first order for every $\alpha$, and what is constant to
first order only at $\alpha=0$ is the orientation and eccentricity of the efficient confidence
ellipsoid.

The restriction to $n\ge3$ is needed for the uniqueness claim and not for
(\ref{eq:varstab}) itself. When $n=2$ we have $d=1$, there are no overlapping pairs, so
$B_\Delta=0$ and the first order term vanishes for every $\alpha$. The two results are
therefore complementary. The $n=2$ statement is exact and global, while
Theorem~\ref{thm:varstab} is local and has content precisely when $n\ge3$. In both cases $\alpha=0$ is the unique stabilising member. In other words, at $\alpha=0$ the root $\sqrt T\{\Phi_0(\hat C)-\Phi_0(C)\}$ is asymptotically pivotal to first order at independence, whereas for $n=2$ Fisher's $z$ makes it exactly asymptotically pivotal.

Under elliptical sampling the first order term in (\ref{eq:varstab}) vanishes at $\alpha=0$, and the second order term has a closed form, by which each GFT variance falls below $1+\kappa$ by $(1+\kappa)/12$ times the sum of the squared correlations of variables $i$ and $j$ with the other variables (Proposition~\ref{prop:secondorder} in Appendix~\ref{app:sampling}).

Along the equicorrelation line part of the exact $n=2$ result holds in every dimension. Let $C_\rho$ be equicorrelated, with eigenvalues $\lambda_1=1+(n-1)\rho$ and $\lambda_2=1-\rho$. The permutations of the variables act on the pairs and split $\R^{d}$ into three invariant subspaces, the constant vectors, the vectors with $\gamma_{ij}=a_i+a_j$ for some $a\in\R^{n}$ with $\iota'a=0$, and the orthogonal complement of the two, which consists of the vectors $\vecl(X)$ with $X$ hollow and $X\iota=0$. Their dimensions are $1$, $n-1$ and $n(n-3)/2$. We call them the common, node and pair-specific subspaces, and write $\Pi_0$, $\Pi_1$ and $\Pi_2$ for the orthogonal projections onto them.

\begin{proposition}\label{prop:equivar}
Let $n\ge3$ and $\rho\in(-\tfrac{1}{n-1},1)$. Under Gaussian sampling $V_\alpha(C_\rho)=\omega_0\Pi_0+\omega_1\Pi_1+\omega_2\Pi_2$, where
\[
  \omega_0=\Bigl\{\frac{(n-1)\lambda_1^{\alpha}\lambda_2+\lambda_1\lambda_2^{\alpha}}{n}\Bigr\}^{2},\qquad
  \omega_2=\lambda_2^{2\alpha},
\]
and $\omega_1$ is given in (\ref{eq:omega1}) in Appendix~\ref{app:sampling}. In particular $V_0(C_\rho)=\Pi_0+\omega_1\Pi_1+\Pi_2$ for every $\rho$, and $\alpha=0$ is the only member for which $\omega_0$ does not depend on $\rho$. For $n\ge4$ it is also the only member for which $\omega_2$ does not depend on $\rho$.
\end{proposition}

Under elliptical sampling the three eigenvalues are multiplied by $1+\kappa$. Along the equicorrelation line the GFT is therefore exactly stabilised on a subspace of dimension $1+n(n-3)/2$, which is most of $\R^{d}$ when $n$ is large. Only the node effects, which tilt the common eigendirection $\iota$, are not stabilised, and $\omega_1\to2/n$ as $\rho\to1$. By (\ref{eq:equi}) the common coordinate is $a_{0,n}(\rho)=\tfrac{2}{n}z_I(\rho)$, where $z_I(\rho)=\tfrac12\log[\{1+(n-1)\rho\}/(1-\rho)]$ is the transformation that \citet[Chapter~VII]{Fisher:1925} applied to the intraclass correlation. Under the intraclass model with equal variances, the analysis of variance estimate satisfies $z_I(\hat\rho)-z_I(\rho)=\tfrac12\log F$, where $F$ has an $F$ distribution that does not depend on $\rho$ \citep[Chapter~VII]{Fisher:1925}. On the common direction the GFT thus extends Fisher's intraclass $z$ to the unrestricted sample correlation matrix. For $n=3$ the pair-specific subspace is empty, but at $\rho=0.8$ the proposition gives $\omega_0=1$ and $\omega_1\approx0.81$, so two GFT coordinates have the asymptotic correlation $0.07$, against $0.54$ for two raw or two Fisher transformed correlations. This is in line with the nearly circular contours of the GFT coordinates in Figure~1 of \citet{ArchakovHansen:2026}, which are computed with $T=8$.

Two consequences are useful for inference. First, $\Phi_0(C_\rho)$ is a constant vector, so $\Pi_2\Phi_0(C_\rho)=0$, and under Gaussian sampling $T\|\Pi_2\Phi_0(\hat C)\|^{2}$ converges in distribution to $\chi^{2}_{n(n-3)/2}$ under every equicorrelated $C_\rho$, without an estimated covariance. The statistic tests pair-specific departures from equicorrelation. It does not test the node component, so it does not test all the restrictions implied by equicorrelation. For $\alpha\ne0$ the limit of the corresponding statistic is $\lambda_2^{2\alpha}\chi^{2}_{n(n-3)/2}$, which depends on $\rho$. Under elliptical sampling the limit of the GFT statistic is $(1+\kappa)\chi^{2}_{n(n-3)/2}$, so the statistic is divided by a consistent estimate of $1+\kappa$. Second, the argument applies within blocks. If $C$ is block equicorrelated and block $k$ has $m_k\ge4$ variables and within-block correlation $\rho_k$, then the vectors $\vecl(X)$ with $X$ hollow, supported on block $k$ and with zero row sums form an eigenspace of $V_\alpha(C)$ of dimension $m_k(m_k-3)/2$ with eigenvalue $(1-\rho_k)^{2\alpha}$ under Gaussian sampling, which is one at $\alpha=0$. Under elliptical sampling the eigenvalue is multiplied by $1+\kappa$.

For the bias and the skewness we return to Gaussian sampling. Near independence the leading bias of $\Phi_\alpha(\hat C)$ is $c_n(\alpha)\vecl(\Delta)/T$ to first order in $\Delta$, with a coefficient $c_n(\alpha)$ that is quadratic in $\alpha$ (Proposition~\ref{prop:bias} in Appendix~\ref{app:sampling}). On the surjective range it is positive and decreasing in $\alpha$, so the GFT is the least biased member with $\alpha\le0$, but it is not the least biased member of the whole family, since $c_n(\alpha)$ vanishes at a value of $\alpha$ between $0.38$ and $\tfrac12$. The scaled family of Section~\ref{sec:scaled} changes this comparison (Proposition~\ref{prop:scaledbias}).

The leading skewness of the coordinates at independence has a simple form. For distinct
indices, write $\hat\gamma_{ij}$ for the coordinate of $\hat\gamma=\Phi_\alpha(\hat C)$ that
corresponds to the pair $\{i,j\}$.

\begin{proposition}\label{prop:skew}
Under Gaussian sampling the following hold.
\begin{enumerate}
\item[(i)] Let $n\ge3$ and $C=I$. Every coordinate of $\hat\gamma=\Phi_\alpha(\hat C)$ has mean zero. The
joint third cumulant of three coordinates is zero for every $T$ unless the three pairs are the
sides $\{i,j\}$, $\{j,k\}$ and $\{k,i\}$ of a triangle with distinct $i,j,k$. For such a
triangle,
\[
  \kappa_3(\hat\gamma_{ij},\hat\gamma_{jk},\hat\gamma_{ki})
  =\mathbb{E}(\hat\gamma_{ij}\hat\gamma_{jk}\hat\gamma_{ki})
  =\frac{3\alpha-1}{2T^{2}}+O(T^{-3}).
\]
The result holds whether the mean of the data is known or estimated. For $\alpha\le0$ we assume
$T>n+12\max(1,-\alpha)$, which ensures that the moments used in the proof are finite.
\item[(ii)] Let $n\ge2$ and $C=I+\Delta$. As $\Delta\to0$ through hollow symmetric matrices, the leading third cumulant of $\hat\gamma_{ij}$ given by the delta method is $\{3\alpha(\alpha-3)\Delta_{ij}+O(\|\Delta\|^{2})\}/T^{2}$.
\end{enumerate}
\end{proposition}

The leading triangle cumulant is $T^{-2}$ for $\hat C$ itself and $-\tfrac12T^{-2}$ for the GFT.
It does not depend on $n$, and it vanishes only at $\alpha=\tfrac13$. At independence every coordinate is symmetric for every $\alpha$. Near independence, by part (ii), only $\alpha=0$ and the non-surjective $\alpha=3$ remove the leading marginal skewness to first order, in every dimension. Far from independence the marginal skewness of the GFT coordinates is also close to zero in simulations, as for Fisher's $z$, whereas that of the raw correlations is large \citep[Figure~3]{ArchakovHansen:2026}. Part (ii) is the local counterpart of the fact that Fisher's $z$ removes the leading skewness for $n=2$, and for the GFT the leading skewness near independence lies in the joint distribution. The GFT stabilises the joint
covariance to first order around $I$, whereas $\alpha=\tfrac13$ cancels the leading joint third
cumulant at $I$, much as the cube root of \citet{WilsonHilferty:1931} removes the leading
skewness of a chi-square variable, and as the Box--Cox power $\tfrac13$ removes the leading skewness of realized volatility when volatility is constant \citep{GoncalvesMeddahi:2011}. For a Wishart matrix with identity scale, the same second
order expansion shows that $f_{1/3}$ removes every leading third cumulant, but this does not
hold for a general scale matrix. For $n=2$ there are no triangles, and the skewness is governed
by the formula given above, under which Fisher's $z$ is the unique member without leading
skewness.

The bias and skewness results extend to Gaussian scale mixtures, a subclass of the elliptical distributions that contains the multivariate $t$. Let $x_t=s_ty_t$, with $y_t\sim N(0,C)$, a scale $s_t>0$ independent of $y_t$, and the pairs $(s_t,y_t)$ independent and identically distributed over $t$. Let $w_t=s_t^{2}$, assume $\mathbb{E}(w^{3})<\infty$, and let $m_4=\mathbb{E}(w^{2})/(\mathbb{E}w)^{2}$ and $m_6=\mathbb{E}(w^{3})/(\mathbb{E}w)^{3}$, so that $m_4=1+\kappa$. For $\alpha\le0$ we also assume that the moments of the coordinates used in the expansions are finite. Given the scales, $\hat C$ is the correlation matrix of a weighted Gaussian sample, whose second and third order cumulants are of order $m_4/T$ and $m_6/T^{2}$. The linear part of each sample correlation is symmetric for every scale, so the leading bias and the asymptotic covariance are those of the Gaussian case with $T$ replaced by $T/m_4$, and so is the leading marginal third cumulant, to first order at independence,
\[
  \kappa_3(\hat\gamma_{ij})=\frac{m_4^{2}\{3\alpha(\alpha-3)\Delta_{ij}+O(\|\Delta\|^{2})\}}{T^{2}}+o(T^{-2}).
\]
In particular the root of $c_n(\alpha)$ and the removal of the leading marginal skewness at $\alpha=0$ do not depend on the mixing distribution. The joint third cumulants do not follow from a change of sample size, because the linear parts contribute a term in $m_6$ and the curvature of the transformation a term in $m_4^{2}$. At independence the triangle cumulant is
\[
  \kappa_3(\hat\gamma_{ij},\hat\gamma_{jk},\hat\gamma_{ki})=\frac{m_6+\tfrac32(\alpha-1)m_4^{2}}{T^{2}}+o(T^{-2}),
\]
which vanishes at $\alpha=1-\tfrac23m_6/m_4^{2}$. This equals $\tfrac13$ only when the scale is constant, and it is zero for the multivariate $t$ distribution with ten degrees of freedom, for which $m_4=\tfrac43$ and $m_6=\tfrac83$. A similar argument applies to realized correlation matrices computed from intraday returns, when the instantaneous correlation matrix is constant within the day and the returns are Gaussian given a common volatility path that is independent of the Gaussian innovations, with no drift and no leverage. Then $m_4$ and $m_6$ are ratios of integrated powers of volatility, and the zero has the form of the power $1-\tfrac23\sigma_2\sigma_6/\sigma_4^{2}$ that removes the leading skewness of realized volatility standardised by its true variance in \citet{GoncalvesMeddahi:2011}, where $\sigma_q$ is the integrated $q$-th power of volatility. Their analysis excludes drift and leverage in the same way, and the corresponding power for the studentised statistic is different. GFT coordinates of empirical realized correlation matrices are studied in \citet[Section~3.5]{ArchakovHansen:2026}.

\subsection{The efficient information matrix for correlations}\label{sec:geom}

Theorem~\ref{thm:varstab} has a coordinate-free formulation, which also explains why $n=2$ is
exceptional. Concentrating the variances out of the Gaussian information matrix leaves the
efficient information for the correlations. By Theorem~\ref{thm:scaled} below, every $C\in\Cn$
is $\Cor(e^{H})$ for exactly one hollow symmetric $H$, and in the geometric literature the
coordinates $\vecl(H)$ are the normal coordinates at $I$ of the quotient of the Fisher--Rao
geometry by diagonal rescaling \citep{DavidGu:2019,ThanwerdasPennec:2021}. Appendix~\ref{app:geom}
gives the details.

\begin{proposition}\label{prop:normal}
Let $H$ be hollow symmetric.
Then for every $\alpha\in\R$,
\begin{equation}\label{eq:normal}
  \Phi_\alpha\{\Cor(e^{H})\}=\vecl(H)+\frac{\alpha}{2}\vecl(H^{2})+O(\|H\|^{3}).
\end{equation}
In the coordinates $\vecl(H)$ of $\Cor(e^{H})$ the efficient information matrix for the
correlations is the identity at independence and has vanishing first derivatives there. For $n\ge3$ the vector $\vecl(H^{2})$ is not identically zero, so $\alpha=0$ is the
unique member of the family agreeing with these coordinates through second order. For $n=2$
every member agrees, because $H^{2}$ is then a multiple of the identity. At $\alpha=0$ the
remainder in (\ref{eq:normal}) is $O(\|H\|^{4})$, so the GFT agrees with these coordinates
through third order.
\end{proposition}

Proposition~\ref{prop:normal} is the coordinate-free content of Theorem~\ref{thm:varstab}. The
reason $V_0(I+\Delta)=I_d+O(\|\Delta\|^{2})$ is that the GFT coordinate and the coordinate
$\vecl(H)$ of $\Cor(e^{H})$ coincide to second order, while every other member introduces
the quadratic distortion $\tfrac{\alpha}{2}\vecl(H^{2})$. It also identifies the source of the
$n=2$ degeneracy. There the distortion vanishes identically, all members agree with
$\vecl(H)$ to second order, and the local argument cannot separate them, which is exactly why
the exact computation at $n=2$ is needed. For $n=2$ the comparison of orders
is degenerate, because $\Cor(e^{H})$ then has off diagonal $\tanh(H_{12})$, so $\Phi_0$
recovers $H$ exactly, and $\alpha=3$, the other root of the quadratic coefficient of $h_\alpha$ in
the variance stabilisation subsection, also agrees beyond third order.

\subsection{Characterisation of the GFT}

The results so far identify $\alpha=0$ by four properties. It is the boundary point of the
surjective range $\alpha\le0$ (Theorem~\ref{thm:main}). On that range it minimises the
worst-case conditioning exponent $|1-\alpha|$, which is attained for $n\ge4$
(Theorem~\ref{thm:cond}), whereas for $n=2$ the Jacobian is scalar and the criterion does not
separate the members. It is the only member whose diagonal problem is both unconstrained and
always solvable, since for $\alpha<0$ the problem is confined to the proper subset
$\mathcal{D}_\alpha$ of $\R^{n}$, which calls for a feasible starting point and a feasibility
line search, and for $\alpha>0$ the extended objective has minimisers that are not solutions
when $\gamma\notin\mathcal{R}_\alpha$ (Section~\ref{sec:dual}). Finally, for $n\ge3$ it is the
only member whose asymptotic covariance is free of $C$ to first order at independence
(Theorem~\ref{thm:varstab}), equivalently the only member that agrees with the normal
coordinates through second order (Proposition~\ref{prop:normal}), and along the
equicorrelation line its asymptotic covariance is constant on the common and pair-specific
subspaces (Proposition~\ref{prop:equivar}). For $n=2$ it is the unique exact variance
stabiliser.

The first two properties isolate $\alpha=0$ on their own.
Surjectivity confines $\alpha$ to $(-\infty,0]$ by Theorem~\ref{thm:main}. A worst-case exponent
of at most one, that is amplification at most linear in $\kappa(C)$ in the worst case, confines
it to $\{\alpha:|1-\alpha|\le1\}=[0,2]$ by Theorem~\ref{thm:cond}, where the sharpness part of
that theorem is what excludes the other values once $n\ge4$. The first constraint is
topological and the second numerical, and
\[
  (-\infty,0]\cap[0,2]=\{0\}.
\]

The value $\alpha=1$ is unimprovable on
conditioning but is not a parametrisation of $\Cn$ by $\R^{d}$, and its diagonal problem has
spurious minimisers outside $\mathcal{D}_1$. The value $\alpha=-1$ is a bijection and gives
the cleanest sign class, since nonnegative coordinates at $\alpha=-1$ are the
$\mathrm{MTP}_2$ property of \citet{KarlinRinott:1980}, but costs an extra factor of
$\kappa(C)$ and its diagonal problem is confined to $\mathcal{D}_{-1}$.

A further property of $\alpha=0$ reflects the familiar fact that the logarithm turns products into sums. By (\ref{eq:power}) power paths are straight lines through the origin only at $\alpha=0$, and by (\ref{eq:kron}) in Section~\ref{sec:tests} Kronecker products are additive in the coordinates only at $\alpha=0$. Together they make the logarithm the unique member of the family that linearises multiplicative structure.

The same ranking appears when the inverse is computed, which Appendix~\ref{sec:algo} treats in
detail. At $\alpha=0$ the inverse is computed by the fixed point and Newton method of
\citet{ArchakovHansen:Variational}. For other values of $\alpha$, damped Newton on the strictly
convex diagonal problem works, with a feasibility line search for $\alpha<0$, and for
$\alpha>0$ minimising the extended objective $\hat L_\alpha$ also decides whether
$\gamma\in\mathcal{R}_\alpha$, by Proposition~\ref{prop:dual}. The Hessian $H$ of the diagonal
problem satisfies $\kappa(H)\le\kappa(C)^{|1-\alpha|}$ at every feasible $\delta$, by the
argument in the proof of Theorem~\ref{thm:cond}, so its worst-case exponent is again smallest at
$\alpha=0$ on the surjective range, and for $\alpha<-1$ the solvers deteriorate in practice as
$\kappa(C)^{|1-\alpha|}$ grows.

\subsection{Testing separability and departures from equicorrelation}\label{sec:tests}

Hypotheses that are linear in GFT coordinates lead to Wald statistics with a simple restricted estimate, and Proposition~\ref{prop:equivar} gives a statistic that needs no estimated covariance. We illustrate both with two structures that are common in applications.

Separable correlation structures, such as the correlation of a panel across units and over time, or of
returns across assets and horizons, are often modelled by a Kronecker product $A\otimes B$.
For positive definite $A$ and $B$,
\begin{equation}\label{eq:kron}
  f_\alpha(A\otimes B)=f_\alpha(A)\otimes I+I\otimes f_\alpha(B)+\alpha f_\alpha(A)\otimes f_\alpha(B),
\end{equation}
so the interaction term disappears exactly at $\alpha=0$. In GFT coordinates a separable
structure is therefore additive, $\log(A\otimes B)=\log A\otimes I+I\otimes\log B$, in the same
way that the logarithm makes a Cobb--Douglas specification additive. This log-linear representation is used by \citet{HafnerLintonTang:2020}, who discuss identifying the off diagonal entries of the logarithms of the factors and recovering their diagonals with the parametrization of \citet{ArchakovHansen:Correlation}, and who propose a specification test. The following proposition states the characterisation in GFT coordinates, which we then use to compare the GFT with raw correlations. Let $n=pq$, and index the variables by pairs $(a,b)$ with $a\le p$ and $b\le q$, in the order of $C_A\otimes C_B$.

\begin{proposition}\label{prop:kron}
Let $S_{p,q}\subset\R^{d}$ consist of the vectors $\gamma$ with $\gamma_{(a,b),(a',b')}=0$ whenever $a\ne a'$ and $b\ne b'$, with $\gamma_{(a,b),(a',b)}$ the same for every $b$, and with $\gamma_{(a,b),(a,b')}$ the same for every $a$. Then $C=C_A\otimes C_B$ for some $C_A\in\mathcal{C}_p$ and $C_B\in\mathcal{C}_q$ if and only if $\Phi_0(C)\in S_{p,q}$, and in that case the common values are the coordinates of $\Phi_0(C_A)$ and $\Phi_0(C_B)$. The subspace $S_{p,q}$ has dimension $p(p-1)/2+q(q-1)/2$.
\end{proposition}

For $\alpha\ne0$ the interaction term in (\ref{eq:kron}) makes the restriction nonlinear. Let $R$ be a matrix whose rows form an orthonormal basis of the orthogonal complement of $S_{p,q}$, so that separability is $R\Phi_0(C)=0$, with $r=d-p(p-1)/2-q(q-1)/2$ restrictions. With $\hat\gamma=\Phi_0(\hat C)$ the Wald statistic is
\[
  W_0=T\hat\gamma'R'\{RV_0(\hat C)R'\}^{-1}R\hat\gamma.
\]
The restricted estimate requires no constrained nonlinear optimisation. Averaging the coordinates of $\hat\gamma$ over the repeated positions gives its orthogonal projection onto $S_{p,q}$, the unweighted minimum distance estimate in GFT coordinates, and the restricted estimate of $C$ is $\tilde C=\tilde C_A\otimes\tilde C_B$, where $\tilde C_A$ and $\tilde C_B$ invert the GFT in dimensions $p$ and $q$.

For the comparison in raw correlations let $\hat c=\vecl(\hat C)$, and let $h(c)$ collect the $r$ restrictions $c_{(a,b),(a',b)}-c_{(a,1),(a',1)}$ for $a'<a$ and $b\ge2$, the restrictions $c_{(a,b),(a,b')}-c_{(1,b),(1,b')}$ for $b'<b$ and $a\ge2$, and the restrictions $c_{(a,b),(a',b')}-c_{(a,1),(a',1)}c_{(1,b),(1,b')}$ for $a'<a$ and $b\ne b'$, which anchor the factors at the first block index. Then
\[
  W_1=T h(\hat c)'\bigl\{Dh(\hat c)V(\hat C)Dh(\hat c)'\bigr\}^{-1}h(\hat c),
\]
where $V(\hat C)$ is the Gaussian asymptotic covariance of the sample correlations. Wald statistics change under nonlinear rewritings of the same null hypothesis, so $W_1$ is one natural choice among several. Under Gaussian sampling $W_0$ and $W_1$ are asymptotically $\chi^{2}_{r}$. For pair-specific departures from equicorrelation we use $W_2=T\|\Pi_2\Phi_0(\hat C)\|^{2}$, which by Proposition~\ref{prop:equivar} is asymptotically $\chi^{2}_{n(n-3)/2}$ under every equicorrelated $C_\rho$. It does not test the node component. In all cases the sample correlations are computed with estimated means.

Each statistic also admits a parametric bootstrap under the null hypothesis, which draws Gaussian samples with correlation matrix $\tilde C$, or, for $W_2$, with the equicorrelation matrix whose parameter is the average sample correlation. We evaluate the bootstrap tests by the warp-speed method of \citet{GiacominiPolitisWhite:2013}. Each Monte Carlo replication draws one bootstrap sample, and the critical value is the $95\%$ quantile of the bootstrap statistics pooled over the replications. This approximates the rejection rate of a conventional bootstrap test, which uses a separate critical value for each data set. In the four cells with the largest distortions, $4\times5$ and $5\times8$ with $T=50$ and $T=100$, we also computed conventional bootstrap tests with $199$ bootstrap samples for each data set. Their rejection rates are between $2.9\%$ and $5.5\%$ for the GFT test and between $6.1\%$ and $21.4\%$ for the raw-correlation test, within $1.3$ percentage points of the warp-speed rates on the same data.

Table~\ref{tab:sim} reports rejection rates at the 5\% level for Gaussian data. In the separability designs $C_A$ is equicorrelated and $C_B$ has the first order autoregressive form $\rho^{|b-b'|}$, with parameters $(0.5,0.7)$ for $3\times4$ and $(0.4,0.6)$ for $4\times5$ and $5\times8$. Appendix~\ref{app:num} gives the details. The $\chi^{2}$ approximation is poor when $r$ is large relative to $T$, as for Wald tests of a simple hypothesis in \citet[Section~5.4]{ArchakovHansen:2026}, where the GFT test approaches its nominal level much faster than tests based on raw or Fisher transformed correlations, because the plug-in covariance is far more accurate in GFT coordinates (Section~5.3 there). At $T=100$ the GFT test rejects in $14.6\%$, $45.3\%$ and $100\%$ of the samples in the three designs, and the raw-correlation test in $15.2\%$, $73.6\%$ and $100\%$. With the warp-speed bootstrap critical values the rejection rates of the GFT test are between $4.7\%$ and $5.6\%$ in every case except $5\times8$ with $T=50$, where the rate is $3.6\%$, and $2.9\%$ with the conventional bootstrap, so the bootstrap test is conservative in that design. With the same calibration the raw-correlation test rejects in $7.8\%$ of the samples for $4\times5$ at $T=50$, and in $20.8\%$ and $15.1\%$ for $5\times8$ at $T=50$ and $T=100$. For $W_2$ the rejection rate against the $\chi^{2}$ distribution changes little over the three values of $\rho$, for example $41.0\%$, $41.7\%$ and $41.7\%$ at $n=20$ and $T=100$ for $\rho=0$, $0.5$ and $0.9$, and the warp-speed bootstrap gives rejection rates between $4.3\%$ and $5.5\%$ in all cells. The corresponding statistics at $\alpha=\tfrac12$ and $\alpha=-1$ have the limits $\lambda_2^{2\alpha}\chi^{2}_{n(n-3)/2}$, so against the unscaled $\chi^{2}$ critical value they reject in at most $0.1\%$ and in at least $74.0\%$ of the samples when $\rho>0$. This shows the nuisance scale that a test at these values of $\alpha$ has to estimate, and which disappears at $\alpha=0$.

For power we use the alternatives $\Cor(C_A\otimes C_B+\tau E)$, with $E$ a fixed positive semidefinite matrix, and critical values from separate simulations under the null hypothesis. The comparison is therefore of size-adjusted power, not of the power of the bootstrap tests above. In the two designs and for this perturbation the size-adjusted power of the GFT test is higher than that of the raw-correlation test, $0.59$ against $0.43$ for $3\times4$ at $T=100$ and $\tau=0.08$, and $0.46$ against $0.37$ for $4\times5$ at $T=250$ and $\tau=0.05$, and the two are about equal for the smallest values of $\tau$. These are illustrations, not a general ranking. Under elliptical sampling the statistics must also be divided by an estimate of $1+\kappa$, or the bootstrap must draw from an elliptical law, which we do not pursue here.

\begin{table}[!tbp]
\centering\small
\caption{Size of the tests of separability and of pair-specific departures from equicorrelation}\label{tab:sim}
\begin{tabular*}{\textwidth}{@{\extracolsep{\fill}}lrrcccc@{}}
\toprule
\multicolumn{7}{@{}l}{\emph{Panel A. Separability}}\\[2pt]
& & & \multicolumn{2}{c}{$W_0$ (GFT)} & \multicolumn{2}{c}{$W_1$ (raw correlations)}\\
\cmidrule(lr){4-5}\cmidrule(l){6-7}
$p\times q$ & $r$ & $T$ & $\chi^{2}$ & Bootstrap & $\chi^{2}$ & Bootstrap\\
\midrule
$3\times4$ & 57 & 50 & 0.329 & 0.050 & 0.301 & 0.060\\
 & & 100 & 0.146 & 0.050 & 0.152 & 0.051\\
 & & 250 & 0.080 & 0.048 & 0.088 & 0.048\\
 & & 1000 & 0.052 & 0.048 & 0.057 & 0.048\\[3pt]
$4\times5$ & 174 & 50 & 0.911 & 0.048 & 0.977 & 0.078\\
 & & 100 & 0.453 & 0.050 & 0.736 & 0.068\\
 & & 250 & 0.148 & 0.047 & 0.285 & 0.050\\
 & & 1000 & 0.074 & 0.056 & 0.095 & 0.056\\[3pt]
$5\times8$ & 742 & 50 & 1.000 & 0.036 & 1.000 & 0.208\\
 & & 100 & 1.000 & 0.055 & 1.000 & 0.151\\
 & & 250 & 0.704 & 0.052 & 0.959 & 0.075\\
 & & 1000 & 0.139 & 0.054 & 0.266 & 0.058\\
\midrule
\end{tabular*}
\begin{tabular*}{\textwidth}{@{\extracolsep{\fill}}rrrcccccc@{}}
\multicolumn{9}{@{}l}{\emph{Panel B. Pair-specific departures from equicorrelation}}\\[2pt]
& & & \multicolumn{3}{c}{$\chi^{2}$} & \multicolumn{3}{c}{Bootstrap}\\
\cmidrule(lr){4-6}\cmidrule(l){7-9}
$n$ & df & $T$ & $\rho=0$ & $\rho=0.5$ & $\rho=0.9$ & $\rho=0$ & $\rho=0.5$ & $\rho=0.9$\\
\midrule
5 & 5 & 50 & 0.073 & 0.071 & 0.071 & 0.047 & 0.043 & 0.045\\
 & & 100 & 0.062 & 0.061 & 0.060 & 0.051 & 0.051 & 0.050\\
 & & 400 & 0.052 & 0.052 & 0.053 & 0.047 & 0.047 & 0.049\\[3pt]
8 & 20 & 50 & 0.134 & 0.137 & 0.136 & 0.049 & 0.051 & 0.052\\
 & & 100 & 0.083 & 0.085 & 0.084 & 0.051 & 0.053 & 0.048\\
 & & 400 & 0.059 & 0.060 & 0.061 & 0.052 & 0.049 & 0.051\\[3pt]
12 & 54 & 50 & 0.295 & 0.301 & 0.303 & 0.051 & 0.053 & 0.055\\
 & & 100 & 0.141 & 0.141 & 0.135 & 0.054 & 0.051 & 0.051\\
 & & 400 & 0.065 & 0.065 & 0.064 & 0.054 & 0.053 & 0.053\\[3pt]
20 & 170 & 50 & 0.877 & 0.885 & 0.882 & 0.052 & 0.051 & 0.048\\
 & & 100 & 0.410 & 0.417 & 0.417 & 0.051 & 0.051 & 0.050\\
 & & 400 & 0.094 & 0.099 & 0.097 & 0.048 & 0.048 & 0.049\\
\bottomrule
\end{tabular*}
\par\medskip
\begin{minipage}{\textwidth}\footnotesize
\emph{Note:} Rejection rates at the $5\%$ level under the null hypothesis for Gaussian data, against the $\chi^{2}$ critical value and with parametric bootstrap critical values under the restricted estimate, evaluated by the warp-speed method. Sample correlations with estimated means. Panel A: test of $C=C_A\otimes C_B$ with $r$ restrictions, by the Wald statistic $W_0$ in GFT coordinates and the Wald statistic $W_1$ in raw correlations. $C_A$ is equicorrelated and $C_B$ is first order autoregressive, with parameters $(0.5,0.7)$ for $3\times4$ and $(0.4,0.6)$ for $4\times5$ and $5\times8$. Panel B: the statistic $W_2=T\|\Pi_2\Phi_0(\hat C)\|^{2}$ with $n(n-3)/2$ degrees of freedom, at the equicorrelation matrix with parameter $\rho$. $10{,}000$ replications, $2{,}500$ for $5\times8$. Warp-speed method: \citet{GiacominiPolitisWhite:2013}.
\end{minipage}
\end{table}

\section{Scaled transformations}\label{sec:scaled}

The family (\ref{eq:family}) applies $f_\alpha$ to the correlation matrix $C$, but it can also be
applied to a rescaled matrix $DCD$, where $D$ is a diagonal
matrix with positive diagonal elements, since such a rescaling does not change the correlations.
We refer to $DCD$ as a representative of $C$. The idea of transforming a representative goes back to \citet{Thanwerdas:2024}, who, building on
\citet{MarshallOlkin:1968} and \citet{JohnsonReams:2009}, applies the logarithm to the
representative with unit row sums. \citet{TumpachLarotonda:2024} and \citet{BissonPennec:2025}
use the representative whose logarithm is hollow, and the balanced correlation parametrization of
\citet{HansenTong:BFT} applies the average $\tfrac12\{f_1(A)+f_{-1}(A)\}=\tfrac12(A-A^{-1})$ to the
representative $A$ with
$\diag A=\diag A^{-1}$. Here we apply $f_\alpha$ to the representative for which
$f_\alpha(DCD)$ is hollow, and read off the off diagonal. This exchanges the roles of the two
directions, so that the inverse becomes explicit and the forward map requires the solution of a
scaling problem. Section~\ref{sec:related} shows that the balanced parametrization combines the
members $\alpha=\pm1$ of the scaled family.

\subsection{The scaled family}

For $C\in\Cn$ and $x\in\R^{n}$ write $X=\Diag(x)$ and $A(x)=e^{X}Ce^{X}$, the representative
of $C$ with scaling $D=e^{X}$. The scaled transformation is
\begin{equation}\label{eq:scaled}
  \tilde\Phi_\alpha(C)=\vecl f_\alpha\bigl(D_\alpha CD_\alpha\bigr),
  \qquad\text{where }\diag f_\alpha(D_\alpha CD_\alpha)=0,
\end{equation}
that is $\diag\{(D_\alpha CD_\alpha)^{\alpha}\}=\iota$ for $\alpha\ne0$ and
$\diag\log(D_\alpha CD_\alpha)=0$ for $\alpha=0$. The first question is whether such a
scaling exists and is unique. It does, for every $\alpha$, because the trace of a power is
convex along diagonal congruences. For $\mu\ne0$ let $p_\mu(x)=\tr A(x)^{\mu}/|\mu|$, and let
$p_0(x)=\tfrac12\|\log A(x)\|_F^{2}$. By Lemma~\ref{lem:congconv} in Appendix~\ref{app:congconv},
$p_\mu$ is strictly convex on $\R^{n}$ for every $\mu$, with gradient
$2\operatorname{sgn}(\mu)\diag\{A(x)^{\mu}\}$ for $\mu\ne0$ and $2\diag\{\log A(x)\}$ for $\mu=0$.

\begin{theorem}\label{thm:scaled}
Let $\alpha\in\R$ and set $\mathcal{H}_\alpha=\{G\in\Sn^{0}:I+\alpha G\succ0\}$ for
$\alpha\ne0$ and $\mathcal{H}_0=\Sn^{0}$.
\begin{enumerate}
\item[(i)] For every $C\in\Cn$ there is exactly one positive diagonal $D_\alpha$ with
$f_\alpha(D_\alpha CD_\alpha)$ hollow. Its logarithm is the unique minimiser of the strictly
convex function $P_\alpha(x)=p_\alpha(x)-2\operatorname{sgn}(\alpha)\iota'x$ for
$\alpha\ne0$ and of $p_0$ for $\alpha=0$.
\item[(ii)] $\tilde\Phi_\alpha$ is a real analytic diffeomorphism of $\Cn$ onto
$\mathcal{H}_\alpha$, identified with its $\vecl$ image, with the explicit inverse
\begin{equation}\label{eq:scaledinv}
  \tilde\Phi_\alpha^{-1}(G)=\Cor\bigl(e_\alpha(G)\bigr)
  =\Cor\bigl\{(I+\alpha G)^{1/\alpha}\bigr\},
  \qquad
  \tilde\Phi_0^{-1}(G)=\Cor(e^{G}).
\end{equation}
\item[(iii)] $\mathcal{H}_\alpha$ is open and convex for every $\alpha$. It is bounded for
$\alpha\ne0$, with $\|G\|_2<(n-1)/|\alpha|$ on $\mathcal{H}_\alpha$, and it is all of
$\Sn^{0}\cong\R^{d}$ exactly when $\alpha=0$.
\end{enumerate}
\end{theorem}

Three members are familiar. At $\alpha=1$ the scaling equation $\diag(DCD)=\iota$ forces
$D=I$, so $\tilde\Phi_1=\Phi_1$ is the raw correlation vector and $\mathcal{H}_1$ is the
elliptope shifted by $-I$. At $\alpha=-1$ the equation $\diag\{(DCD)^{-1}\}=\iota$ reads
$d_i^{2}=K_{ii}$ with $K=C^{-1}$, so $D_{-1}=\Diag(K)^{1/2}$ and
$f_{-1}(D_{-1}CD_{-1})=I-D_{-1}^{-1}KD_{-1}^{-1}$ has off diagonal entries
$-K_{ij}/\sqrt{K_{ii}K_{jj}}=\pi_{ij}$, the full partial correlations. Hence
$\tilde\Phi_{-1}(C)=\vecl(\Pi)$ with $\mathcal{H}_{-1}=\{\Pi\in\Sn^{0}:I-\Pi\succ0\}$, and
(\ref{eq:scaledinv}) is the classical recovery $C=\Cor\{(I-\Pi)^{-1}\}$. At $\alpha=0$
the scaling makes $\log(DCD)$ hollow, and $p_0$ is precisely the objective in (\ref{eq:dcor}).
This member is the inverse of the coordinates in Proposition~\ref{prop:normal}.

The scaled family is symmetric under inversion. If $D$ makes $f_{-\alpha}(DCD)$ hollow, then
$(DCD)^{-1}=D^{-1}C^{-1}D^{-1}$ is a representative of $\Cor(C^{-1})$ for which $f_\alpha$ is hollow,
because $f_\alpha(A^{-1})=-f_{-\alpha}(A)$. Hence
\[
  \tilde\Phi_\alpha\{\Cor(C^{-1})\}=-\tilde\Phi_{-\alpha}(C).
\]
The raw and the partial correlations, $\alpha=\pm1$, are therefore mirror images. For $n=2$
normalized inversion maps $\rho$ to $-\rho$, and every member satisfies
$\tilde\Phi_\alpha\{\Cor(C^{-1})\}=-\tilde\Phi_\alpha(C)$. For $n\ge3$ only $\alpha=0$ does. Indeed,
with $C=\Cor(e^{H})$ we have $\Cor(C^{-1})=\Cor(e^{-H})$, and the expansion of
$\tilde\Phi_\alpha\{\Cor(e^{H})\}$ below shows that the two sides differ by
$\alpha\vecl(H^{2})+O(\|H\|^{3})$. The
GFT does not have this property when $n\ge3$, but by Proposition~\ref{prop:normal} and the
agreement of $\Phi_0$ and $\tilde\Phi_0$ through third order, its departure
$\Phi_0\{\Cor(C^{-1})\}+\Phi_0(C)$ is of fourth order at independence.

At $\alpha=0$, Theorem~\ref{thm:scaled} says that $H\mapsto\Cor(e^{H})$ is a global real
analytic diffeomorphism of $\Sn^{0}$ onto $\Cn$, with inverse $\tilde\Phi_0$, and that the
infimum in (\ref{eq:dcor}) is attained at a unique scaling, the one that makes $\log(DCD)$
hollow. In the geometric language of Appendix~\ref{app:geom}, $t\mapsto\Cor(e^{tH})$ is the
quotient geodesic through $I$ with initial velocity $H$, as the proof of
Proposition~\ref{prop:normal} shows, so $H\mapsto\Cor(e^{H})$ is the exponential map at $I$ of
the quotient Fisher--Rao geometry, $\tilde\Phi_0$ is its inverse, and every $C\in\Cn$ is joined
to $I$ by exactly one quotient geodesic. The ingredients of this interpretation, the horizontal
lift and the projection formula, are in \citet{DavidGu:2019} and \citet{ThanwerdasPennec:2021}.
The diffeomorphism and the uniqueness of the hollowing scaling are known. \citet[Conjecture 3.1]{Thanwerdas:2024} conjectured that for
every $\Sigma\succ0$ there is exactly one positive diagonal $D$ for which $\log(D\Sigma D)$ is
hollow, proved existence, and showed that uniqueness implies uniqueness of the Riemannian
logarithm of the quotient geometry at $I$, which \citet{ThanwerdasPennec:2022} note is not
ensured. The equivalent decomposition $\Sigma=P^{1/2}e^{H}P^{1/2}$, with $P$ positive diagonal and
$H$ hollow, is given by \citet[Section 1.2.1]{TumpachLarotonda:2024}, and
\citet[Theorem 2.39 and Remark 2.40]{BissonPennec:2025} state the uniqueness of the hollowing
scaling and its consequence for the quotient logarithm. Lemma~\ref{lem:congconv} gives a short
proof through strict convexity along diagonal congruences, and Theorem~\ref{thm:scaled} extends
the construction to every Box--Cox index.
The GFT and $\tilde\Phi_0$ are different maps, the off diagonal of $\log C$
against the off diagonal of $\log(DCD)$ for the hollowing scaling, but they agree through third
order at $I$ by Proposition~\ref{prop:normal}, and more generally the argument of
Proposition~\ref{prop:normal} gives
$\tilde\Phi_\alpha\{\Cor(e^{H})\}=\vecl(H)+\tfrac{\alpha}{2}\vecl(H^{2})+O(\|H\|^{3})$
for every $\alpha$, the same quadratic term as for $\Phi_\alpha$, so that, for $n\ge3$,
$\alpha=0$ is also the unique information stabilising member of the scaled family.

The leading bias of the scaled coordinates near independence is given in Proposition~\ref{prop:scaledbias} in Appendix~\ref{app:sampling}. The scaling leaves the coefficient unchanged at $\alpha=0$ and $\alpha=1$, and at $\alpha=-1$ it takes it from $n+\tfrac12$ to $-\tfrac12$, the coefficient of a sample partial correlation. Bias can also be measured without reference to a choice of coordinates. The intrinsic bias of \citet{SmithCRB:2005} is the expected value of the estimate in normal coordinates centred at the true $C$. At $C=I$ these coordinates are $\tilde\Phi_0$. Changing the sign of variable $i$ leaves the law of $\hat C$ unchanged at $C=I$ and changes the sign of row $i$ of $\tilde\Phi_0(\hat C)$, so the intrinsic bias at independence is zero whenever the expectation exists. How it behaves away from independence is an open question.

Theorem~\ref{thm:scaled} is the mirror image of Theorem~\ref{thm:main}. In the family
(\ref{eq:family}) the forward map is explicit, the inverse is the diagonal problem, and the image
is $\R^{d}$ exactly on $\alpha\le0$, with nonconvex images for $\alpha>1$ when $n\ge3$. In the scaled
family the inverse is explicit, the forward map is the scaling problem, which is an
unconstrained strictly convex problem for every $\alpha$, the image is always convex, and it is
$\R^{d}$ exactly at $\alpha=0$. The two families have the same image at $\alpha=0$, where both land on $\R^{d}$ through
different maps, and they coincide at $\alpha=1$, where both are the raw correlations.

Hollowness is one way to fix the representative. As noted above, \citet{Thanwerdas:2024} instead scales so
that $DCD$ has unit row sums, which exists and is unique by \citet{MarshallOlkin:1968}, and
takes $S=\log(DCD)$, which then has zero row sums, because $\iota$ is an eigenvector of $e^{S}=DCD$
with eigenvalue one and hence of $S$ with eigenvalue zero. The image is the linear space of symmetric matrices with zero row
sums, again of dimension $d$, and the inverse is again explicit, $C=\Cor(e^{S})$,
because $e^{S}\iota=\iota$ identifies the scaling. Any normalisation under which each scaling
orbit $\{DCD\}$ meets the chosen section exactly once produces the same pattern, an explicit
inverse and an iterative forward map, and that condition has to be verified for each section,
as Theorem~\ref{thm:scaled}(i) and \citet{MarshallOlkin:1968} do for the two above. It is not
implied by the section being a complement of the diagonal matrices.\footnote{For $n=2$ and
$\alpha=1$, the section $I+g\bigl(\begin{smallmatrix}2&1\\1&2\end{smallmatrix}\bigr)$ is such a
complement, but its members are positive definite only for $g>-\tfrac13$ and have correlation
$g/(1+2g)$, which ranges over $(-1,\tfrac12)$ only, so orbits with $\rho\ge\tfrac12$ never meet
it.}

\subsection{Related parametrizations}\label{sec:related}

The members $\alpha=\pm1$ of the scaled family are explicit in both directions, because
their scaling equations decouple across $i$. The equation $\diag(DCD)=\iota$ forces $D=I$, and
$\diag\{(DCD)^{-1}\}=\iota$ forces $d_i^{2}=K_{ii}$. Their images are bounded. The balanced
correlation parametrization of \citet{HansenTong:BFT} combines them. With
$\Delta=\Diag(K)^{1/4}$ and the balanced representative $A=\Delta C\Delta$, characterised by
$\diag A=\diag A^{-1}$, it is $B(C)=\tfrac12(A-A^{-1})$ with inverse
$C(B)=\Cor(B+\sqrt{I+B^{2}})$. In terms of the scaled family, $B$ is the average
$\tfrac12\{f_1(A)+f_{-1}(A)\}$ of the two Box--Cox transforms at $\alpha=\pm1$, the
balancing condition is hollowness of that average, and the balanced scaling is the geometric
mean $\Delta=(D_1D_{-1})^{1/2}$ of the two scalings above. It is explicit in both directions
with image $\R^{d}$, which neither the GFT nor its scaled counterpart $\tilde\Phi_0$ is. For
its properties, and for the comparison with the ordering-dependent Cholesky and vine
coordinates, which are also explicit in both directions after componentwise transformations,
we refer to that paper. The log-scaling of \citet{Thanwerdas:2024} described above, the
hollow logarithm $\tilde\Phi_0$ of \citet{TumpachLarotonda:2024} and \citet{BissonPennec:2025}, and
the off diagonal of the inverse, which is $\Phi_{-1}$ of the unscaled family, are, to our
knowledge, the other ordering-free constructions in the literature.

Box--Cox transformations are also used in models of realized covariance matrices.
\citet{Weigand:2014} transforms either the eigenvalues of the covariance matrix or the individual
variances, which are then modelled jointly with Fisher's $z$ of each correlation. A scalar
Box--Cox transform has a range that is bounded on one side unless its parameter is zero, so the
transformed variances cannot be exactly Gaussian, and Fisher's $z$ applied entry by entry does not
guarantee a positive definite forecast when $n>2$. The family $\Phi_\alpha$ is a matrix version of
the correlation part of this model. For $\alpha\le0$ every vector in $\R^{d}$ is the image of a
correlation matrix by Theorem~\ref{thm:main}, so a Gaussian model for $\Phi_\alpha(C_t)$ has no
support restriction and every forecast is a valid correlation matrix. Log variances combined with
the GFT therefore give a Box--Cox model of the covariance matrix that is valid for every $n$.

\section{Conclusion}\label{sec:conclusion}

The Box--Cox family identifies the GFT as a member of a one-parameter family that also contains the raw
correlations and the off diagonal of the concentration matrix. All members share a common
structure, in that the inverse is the solution of a strictly convex diagonal problem for every
$\alpha$, so every member is a real analytic diffeomorphism onto an open image. What distinguishes the
members is the behaviour of the scalar function $f_\alpha(x)$ near two points. The first is
$x=0$. As $C$ approaches the boundary of the elliptope its smallest eigenvalue tends to zero, so
the behaviour of $f_\alpha(x)$ as $x\downarrow0$ determines both the image of $\Phi_\alpha$ and
the conditioning of the transformation. The image
is all of $\R^{d}$ exactly for $\alpha\le0$, and for $\alpha>0$ it is bounded and convex on
$[\tfrac12,1]$, whereas for $n\ge4$ it is nonconvex both above $\alpha=1$ and near
$\alpha=\tfrac14$. On the surjective range $\alpha\le0$, that is, for the members
$\Phi_\alpha$ that map onto all of $\R^{d}$, the GFT has
the weakest boundary singularity, which gives the smallest exponent in the conditioning bound and
a diagonal problem that is unconstrained and solvable for every $\gamma$. The second point is
$x=1$. When $C$ is close to independence all its eigenvalues are close to one, so the local
statistical behaviour is determined by the derivatives of $f_\alpha$ at one. Under Gaussian
sampling and for $n\ge3$, the asymptotic covariance is stabilised to first order exactly when
$f''(1)=-1$, which in the Box--Cox family singles out $\alpha=0$. For $n=2$ the first order
dependence vanishes for every member, and the logarithm is distinguished by exact variance
stabilisation. For every $n\ge3$ the GFT is also exactly stabilised along the equicorrelation line, on a subspace of dimension $1+n(n-3)/2$. Neither the boundary behaviour at $x=0$ nor the covariance stabilisation at $x=1$ implies the other. The logarithm is
special because it combines the weakest boundary divergence compatible with surjectivity with the
curvature at independence that stabilises the covariance. Under Gaussian sampling the leading joint third cumulant at independence vanishes only at $\alpha=\tfrac13$, which is analogous to the finding of \citet{GoncalvesMeddahi:2011} that the Box--Cox power $\tfrac13$ removes the leading skewness of realized volatility when volatility is constant. Unlike the properties of the GFT, this value depends on the tails of the data. The exact stabilisation along the equicorrelation line and the linearity of Kronecker separability in GFT coordinates also have uses in inference, which Section~\ref{sec:tests} illustrates with tests of correlation structure.

The scaled family of Section~\ref{sec:scaled} is the mirror image of this picture. There the
inverse is explicit and the forward map is a convex scaling problem, the image is convex for every
$\alpha$ and bounded unless $\alpha=0$, and at $\alpha=0$ it recovers the known result that every
correlation matrix is $\Cor(e^{H})$ for exactly one hollow $H$. The scaled GFT $\tilde\Phi_0$ therefore
extends Fisher's $z$ to the correlation matrix, and its coordinates are global normal coordinates,
centred at independence, for the efficient information metric of the correlations. The explicit inverse makes the scaled family attractive for dynamic models of correlation matrices, in which the coordinates are modelled and the correlation matrix must be recovered at every point in time. The correlation matrix is then obtained in closed form rather than by solving a diagonal problem, which matters most in high dimensions. The scaled coordinates are unrestricted only at $\alpha=0$, and for the other scaled members a dynamic model must respect their bounded image. Both families single out $\alpha=0$, for different reasons, and both place the
GFT and its scaled counterpart among a small number of ordering-free parametrizations of
correlation matrices by $\R^{d}$.

The Box--Cox family is one parametric choice of the scalar function $f$. For any continuous and strictly increasing $f$ the antiderivative of $f^{-1}$ is strictly convex, and the diagonal problem is again a strictly convex program, much as generalized empirical likelihood is defined for any concave criterion function and not only for the Cressie--Read family \citep{NeweySmith:2004}. Other choices of $f$, and hence of the convex objective, may therefore have attractive properties, and we will pursue this in future research.

\section*{Declaration of generative AI and AI-assisted technologies}

During the preparation of this work the authors used Claude Opus 5.5 (Anthropic) and GPT-6 Astra (OpenAI) to draft and edit parts of the text, to check derivations, and to write code for the numerical verifications and simulations. The authors reviewed and edited all output and take full responsibility for the content of the paper.

\appendix
\numberwithin{equation}{section}
\counterwithin{proposition}{section}
\section{Proofs}\label{app:proofs}

\begin{proof}[Proof of Proposition~\ref{prop:convex}]
$\mathcal{D}_\alpha$ is the preimage of the positive definite cone under an affine map, hence
open and convex. Let $G=U\diag(a)U'$. By the Daleckii--Krein formula, see e.g.\ \citet[Ch.~3]{Higham:2008},
$\frac{d}{dt}\tr\Psi_\alpha(G+tH)|_{t=0}=\tr\{e_\alpha(G)H\}$ and
\[
  \frac{d^{2}}{dt^{2}}\tr\Psi_\alpha(G+tH)\Big|_{t=0}
  =\sum_{k,\ell}\Lambda^{[1]}_{k\ell}\bigl[(U'HU)_{k\ell}\bigr]^{2}.
\]
All divided differences of the strictly increasing $e_\alpha$ are positive, so the quadratic
form is positive unless $U'HU=0$, that is unless $H=0$. Taking $H=\diag(v)$ gives the displayed
Hessian and strict convexity. The first order condition is $\diag C(\delta)=\iota$, so any
$C\in\Cn$ with $\Phi_\alpha(C)=\gamma$ corresponds to a stationary point, and strict convexity
forces uniqueness. The matrix $C$ is then recovered from that stationary point as
$C=e_\alpha(G(\delta^{*}))$.
\end{proof}

\begin{proof}[Proof of Proposition~\ref{lem:detJ}]
Let $M=M_\alpha(C)$ be the matrix of $Df_\alpha(C)$ defined in the proof of Lemma~\ref{lem:jac},
with eigenvalues $f_\alpha^{[1]}(\lambda_k,\lambda_\ell)$, $k\le\ell$, and eigenvectors given by
the coordinates of $U(e_ke_\ell'+e_\ell e_k')U'$,
so that $\det M=\prod_{k}\lambda_k^{\alpha-1}\prod_{k<\ell}f_\alpha^{[1]}(\lambda_k,\lambda_\ell)$,
since $f_\alpha'(\lambda)=\lambda^{\alpha-1}$. Partition $M$ into the off diagonal and the
diagonal coordinates, and write $M_{11}$ for the off diagonal block of $M$ and $(M^{-1})_{22}$
for the diagonal block of $M^{-1}$. By the proof of Lemma~\ref{lem:jac}, $J_\alpha(C)=M_{11}$.
Since $e_\alpha\{f_\alpha(C)\}=C$, $M^{-1}$ is the matrix of $De_\alpha(G)$ with
$G=f_\alpha(C)=G(\delta^{*})$, and $(M^{-1})_{22}$ has entries $\partial C(\delta)_{ii}/\partial\delta_j$ at $\delta^{*}$, which by
Proposition~\ref{prop:convex} is $\nabla^{2}L_\alpha(\delta^{*})$, because
$\nabla L_\alpha(\delta)=\diag C(\delta)-\iota$. For a partitioned positive definite matrix, $\det M=\det M_{11}\det(M_{22}-M_{21}M_{11}^{-1}M_{12})$ and
$(M^{-1})_{22}=(M_{22}-M_{21}M_{11}^{-1}M_{12})^{-1}$, so
$\det M_{11}=\det M\cdot\det(M^{-1})_{22}$, which is the result.
\end{proof}

\begin{proof}[Proof of Theorem~\ref{thm:main}]
(i) is Proposition~\ref{prop:convex} and Lemma~\ref{lem:jac}.

(ii) For $\alpha>0$ the image is bounded by part (iii), so it is not $\R^{d}$, and it remains to prove surjectivity for $\alpha\le0$. For $\alpha=0$ we have $\mathcal{D}_0=\R^{n}$, and with $g_1,\dots,g_n$ the eigenvalues of $G(\delta)$ we have $\iota'\delta=\tr G(\delta)=\sum_ig_i$, so that $L_0(\delta)=\sum_{i=1}^{n}(e^{g_i}-g_i)$. Each term is at least one and tends to infinity as $g_i\to\pm\infty$, and $\max_i|g_i|\to\infty$ as $\|\delta\|\to\infty$, so $L_0$ is coercive. Its minimiser exists and satisfies $\diag C(\delta^{*})=\iota$, which is the result of \citet{ArchakovHansen:Correlation}. Now let $\alpha<0$ and write
$M(\delta)=I+\alpha G(\delta)=C(\delta)^{\alpha}$ with eigenvalues $c_1,\dots,c_n>0$ on
$\mathcal{D}_\alpha$. Since $\diag G(\delta)=\delta$ we have
$\iota'\delta=\tr G=\sum_i(c_i-1)/\alpha$, so (\ref{eq:L}) separates as
\[
  L_\alpha(\delta)=\sum_{i=1}^{n}\psi(c_i),
  \qquad
  \psi(c)=\frac{c^{q}}{1+\alpha}-\frac{c-1}{\alpha},
  \qquad q=\frac{1+\alpha}{\alpha},
\]
with $\psi(c)=-\log c+c-1$ when $\alpha=-1$. Then $\psi'(c)=(c^{1/\alpha}-1)/\alpha$ and
$\psi''(c)=\alpha^{-2}c^{1/\alpha-1}>0$, so $\psi$ is strictly convex with a unique minimum at
$c=1$.

\emph{Behaviour at infinity.} As $c\to\infty$, $\psi(c)\to\infty$ for every $\alpha<0$. For
$-1<\alpha<0$ we have $q<0$, so $c^{q}\to0$ while $(c-1)/|\alpha|\to\infty$. For $\alpha=-1$
the logarithm is dominated by the linear term, and for $\alpha<-1$ we have $q\in(0,1)$, so the
sublinear $c^{q}/(1+\alpha)$ cannot overcome the linear growth of $(c-1)/|\alpha|$. On
$\mathcal{D}_\alpha$ the diagonal satisfies $M_{ii}=1+\alpha\delta_i>0$, so
$\delta_i<1/|\alpha|$ is bounded above. If $\|\delta\|\to\infty$, then some
$\delta_i\to-\infty$, hence $M_{ii}\to\infty$, hence $c_1\to\infty$ and $L_\alpha\to\infty$.
Since $\psi$ is bounded below, $L_\alpha$ is coercive on $\mathcal{D}_\alpha$.

\emph{Behaviour at the boundary.} Let $\bar\delta\in\partial\mathcal{D}_\alpha$, so
$M(\bar\delta)\succeq0$ has a zero eigenvalue. For $-1\le\alpha<0$ we have $\psi(0^{+})=+\infty$
and no boundary minimum is possible. For $\alpha<-1$ the value $\psi(0^{+})=1/\alpha$ is
finite but the slope is not, since $\psi'(c)\to-\infty$ as $c\downarrow0$ because
$1/\alpha\in(-1,0)$. Along the feasible ray $\delta(t)=\bar\delta-t\iota$, which gives
$M(\delta(t))=M(\bar\delta)+|\alpha|tI\succ0$ for $t>0$, we get
$\frac{d}{dt}L_\alpha(\delta(t))=|\alpha|\sum_i\psi'(c_i+|\alpha|t)\to-\infty$ as
$t\downarrow0$, so $L_\alpha(\delta(t))<L_\alpha(\bar\delta)$ for small $t>0$ and $\bar\delta$
is not a minimiser.

Hence the infimum of the lower semicontinuous convex extension of $L_\alpha$ to
$\overline{\mathcal{D}_\alpha}$, which takes the value $+\infty$ on the boundary when
$-1\le\alpha<0$ and is finite there when $\alpha<-1$, is attained at an interior $\delta^{*}$,
where $\diag C(\delta^{*})=\iota$. As $\gamma$ was arbitrary, $\mathcal{R}_\alpha=\R^{d}$.

(iii) Since $\Phi_\alpha(C)=\alpha^{-1}\vecl(C^{\alpha})$, it suffices to bound
$|(C^{\alpha})_{ij}|$ for $i\ne j$. Let $\lambda_1\ge\dots\ge\lambda_n>0$ be the eigenvalues of
$C$, with $\sum_k\lambda_k=n$ and $\lambda_1<n$. For $\alpha\in(0,1]$ choose $s\in\{-1,1\}$ with
$s(C^{\alpha})_{ij}=|(C^{\alpha})_{ij}|$, and set $u=(e_i+se_j)/\sqrt2$ and $w=(e_i-se_j)/\sqrt2$.
Then $u'C^{\alpha}u-w'C^{\alpha}w=2|(C^{\alpha})_{ij}|$ and $w'C^{\alpha}w\ge0$, so
$2|(C^{\alpha})_{ij}|\le u'C^{\alpha}u$. As $u$ is a unit vector, $u'C^{\alpha}u$ is an average of the
$\lambda_k^{\alpha}$ with weights $(u'u_k)^{2}$, and concavity of $x^{\alpha}$ gives
$u'C^{\alpha}u\le(u'Cu)^{\alpha}=(1+sC_{ij})^{\alpha}<2^{\alpha}$. Hence
$|(C^{\alpha})_{ij}|<2^{\alpha-1}$. For $\alpha>1$, $|(C^{\alpha})_{ij}|\le\{(C^{\alpha})_{ii}(C^{\alpha})_{jj}\}^{1/2}$
and $(C^{\alpha})_{ii}=\sum_k\lambda_k^{\alpha-1}\lambda_ku_{ik}^{2}\le\lambda_1^{\alpha-1}<n^{\alpha-1}$,
because $\sum_k\lambda_ku_{ik}^{2}=C_{ii}=1$. Dividing by $\alpha$ bounds the coordinates. The bounds are approached but not attained. For $0<\alpha\le1$ let $C$
have a first $2\times2$ block with correlation $\rho\uparrow1$ and the identity elsewhere, so that
$(C^{\alpha})_{12}=\{(1+\rho)^{\alpha}-(1-\rho)^{\alpha}\}/2\to2^{\alpha-1}$. For $\alpha>1$ let $C$ be
equicorrelated with $\rho\uparrow1$, so that by (\ref{eq:equi}) every coordinate tends
to $n^{\alpha-1}/\alpha$.
\end{proof}

The proof of Theorem~\ref{thm:shape}(iii) uses the following lemma, in which $p$ and $q$ are positive numbers.

\begin{lemma}\label{lem:quarter}
Let $p,q>0$ and $\bar v\ge0$ satisfy
\begin{align}
  &16q^{3}+4p^{2}q-2p^{2}>0, \label{eq:qa}\\
  &2p^{2}(2q+\bar v)^{2}+(\bar v^{2}+2p^{2})^{2}\ge1, \label{eq:qb}\\
  &\Delta(\bar v)\equiv2(2q^{2}+p^{2})^{2}-p^{2}(2q+\bar v)^{2}-(\bar v^{2}+2p^{2})^{2}>0. \label{eq:qc}
\end{align}
Then there is no positive semidefinite $3\times3$ matrix $A$ with $A_{12}=-p$,
$A_{13}=-q$, $A_{23}=p$ and $\diag(A^{4})=\iota$.
\end{lemma}

\begin{proof}
Suppose such an $A$ exists. Let $d=\diag(A)$, which is nonnegative because $A\succeq0$, and
let $S=A^{2}$. Direct multiplication gives $S_{11}=d_1^{2}+p^{2}+q^{2}$,
$S_{22}=d_2^{2}+2p^{2}$, $S_{33}=d_3^{2}+p^{2}+q^{2}$, $S_{12}=-p(d_1+d_2+q)$,
$S_{13}=-\{q(d_1+d_3)+p^{2}\}$ and $S_{23}=p(d_2+d_3+q)$. Since $A^{4}=S^{2}$ and $S$ is
symmetric, $(A^{4})_{ii}=\sum_kS_{ik}^{2}$. Hence
\[
  (A^{4})_{11}-(A^{4})_{33}=(d_1-d_3)\bigl\{(d_1+d_3)(d_1^{2}+d_3^{2}+2p^{2}+2q^{2})
  +p^{2}(d_1+d_3+2d_2+2q)\bigr\}.
\]
The term in braces is at least $2p^{2}q>0$, so $(A^{4})_{11}=(A^{4})_{33}$ forces
$d_1=d_3$. Let $u=d_1=d_3$ and $v=d_2$.

Three facts follow. First, $(1,0,1)'$ is an eigenvector of $A$ with eigenvalue $u-q$, so
$u\ge q$. Second, by the Cauchy--Schwarz inequality
$S_{22}=e_2'A^{2}e_2\le\|A^{2}e_2\|=(A^{4})_{22}^{1/2}=1$, so $v^{2}+2p^{2}\le1$ and
$v\le1$. Third, $(A^{4})_{11}=g_1(u,v)$ and $(A^{4})_{22}=g_2(u,v)$ with
\[
  g_1(u,v)=(u^{2}+p^{2}+q^{2})^{2}+p^{2}(u+v+q)^{2}+(2qu+p^{2})^{2},\qquad
  g_2(u,v)=2p^{2}(u+v+q)^{2}+(v^{2}+2p^{2})^{2},
\]
and $g_1(u,v)=g_2(u,v)=1$.

The function $g_2$ is increasing in each argument on $[0,\infty)^{2}$ and strictly increasing
in $v$. Since $u\ge q$, we have $g_2(q,v)\le g_2(u,v)=1\le g_2(q,\bar v)$ by
(\ref{eq:qb}), and therefore $v\le\bar v$. Next, expanding shows that
\[
  (g_1-g_2)(u,v)-(g_1-g_2)(q,v)=(u-q)B(u,v),
\]
with $B(u,v)=(u+q)(u^{2}+3q^{2})+4q^{2}(u+q)+p^{2}u+3p^{2}q-2p^{2}v$. For $u\ge q$ and
$v\le1$ this gives $B(u,v)\ge16q^{3}+4p^{2}q-2p^{2}>0$ by (\ref{eq:qa}), and so
$(g_1-g_2)(u,v)\ge(g_1-g_2)(q,v)$. Finally $(g_1-g_2)(q,v)=\Delta(v)$, and $\Delta$ is
decreasing on $[0,\infty)$, so $\Delta(v)\ge\Delta(\bar v)>0$ by (\ref{eq:qc}). Hence
$g_1(u,v)>g_2(u,v)$, which contradicts $g_1(u,v)=g_2(u,v)=1$.
\end{proof}

\begin{proof}[Proof of Theorem~\ref{thm:shape}]
(i) Since $\Phi_\alpha(C)=\alpha^{-1}\vecl(C^{\alpha})$, a vector $\gamma$ lies in
$\mathcal{R}_\alpha$ if and only if there is $A\succ0$ with $\vecl(A)=\alpha\gamma$ and
$\diag(A^{p})=\iota$, and then $\Phi_\alpha^{-1}(\gamma)=A^{p}$.

For convexity, write $p=1+s$ with $s\in[0,1]$. For $s\in(0,1)$ and $x>0$,
$x^{s}=c_s\int_0^{\infty}t^{s-1}x/(x+t)dt$ with $c_s=\sin(\pi s)/\pi$, see
\citet[Ch.~V]{Bhatia:1997}, and multiplying by $x$ and writing $x^{2}/(x+t)=x-t+t^{2}/(x+t)$
gives, spectrally,
\begin{equation}\label{eq:powint}
  A^{p}=c_s\int_0^{\infty}t^{s-1}\bigl\{A-tI+t^{2}(A+tI)^{-1}\bigr\}dt,
  \qquad A\succ0 .
\end{equation}
Two properties of $A\mapsto A^{p}$ follow. First, $A\mapsto(A+tI)^{-1}$ is convex in the
Loewner order, so by (\ref{eq:powint}) each diagonal entry $(A^{p})_{ii}$ is a convex function of
$A$ on the positive definite cone. Second, for $i\ne j$,
\[
  \frac{\partial(A^{p})_{ii}}{\partial A_{jj}}
  =-c_s\int_0^{\infty}t^{s+1}\bigl\{[(A+tI)^{-1}]_{ij}\bigr\}^{2}dt\le0,
\]
because the first two terms in (\ref{eq:powint}) have no cross derivative and
$\partial(A+tI)^{-1}/\partial A_{jj}=-(A+tI)^{-1}E_{jj}(A+tI)^{-1}$. Both properties hold
trivially at $s\in\{0,1\}$, where $(A^{2})_{ii}=\sum_kA_{ik}^{2}$. Finally,
$\partial(A^{p})_{ii}/\partial A_{ii}=\sum_{k,\ell}u_{ik}^{2}u_{i\ell}^{2}(x^{p})^{[1]}(\lambda_k,\lambda_\ell)>0$,
and $(A^{p})_{ii}\ge A_{ii}^{p}$ by Jensen's inequality, since $(A^{p})_{ii}=\sum_ku_{ik}^{2}\lambda_k^{p}$
with $\sum_ku_{ik}^{2}\lambda_k=A_{ii}$.

Now let $\gamma_1,\gamma_2\in\mathcal{R}_\alpha$ with associated $A_1,A_2$, let
$t\in(0,1)$, $\bar\gamma=t\gamma_1+(1-t)\gamma_2$ and $\bar A=tA_1+(1-t)A_2\succ0$, so that
$\vecl(\bar A)=\alpha\bar\gamma$ and, by convexity of the diagonal entries,
$\diag(\bar A^{p})\le t\iota+(1-t)\iota=\iota$ componentwise. Let
$\varphi(u)=\diag\{(\bar A+\Diag(u))^{p}\}$ for $u\ge0$ and
$S=\{u\ge0:\varphi(u)\le\iota\}$. The set $S$ contains $0$, is closed, and is bounded because
$\varphi_i(u)\ge(\bar A_{ii}+u_i)^{p}$ forces $u_i\le1-\bar A_{ii}$. Let $u^{*}$ maximise
$\iota'u$ over $S$. If $\varphi_i(u^{*})<1$ for some $i$, then for small $\varepsilon>0$ the
point $u^{*}+\varepsilon e_i$ still satisfies $\varphi_i\le1$ by continuity and
$\varphi_j\le\varphi_j(u^{*})\le1$ for $j\ne i$ by the sign of the cross derivative, so it lies
in $S$ with a larger coordinate sum, a contradiction. Hence $\varphi(u^{*})=\iota$,
$C=(\bar A+\Diag(u^{*}))^{p}$ is a correlation matrix with $\vecl(C^{\alpha})=\alpha\bar\gamma$,
and $\bar\gamma\in\mathcal{R}_\alpha$.

For the explicit inverse, at $\alpha=\tfrac12$ we have $p=2$, and $A=C^{1/2}\succ0$ must have off diagonal
$\gamma/2$ and unit row norms, $\sum_kA_{ik}^{2}=1$, which determines its diagonal up to sign.
The positive root is forced because a positive definite matrix has positive diagonal. So
$A=B(\gamma)$, and conversely if $B(\gamma)\succ0$ then $B(\gamma)^{2}$ is a correlation matrix
with $\Phi_{1/2}(B(\gamma)^{2})=\gamma$.

(ii) Let $C_r$ be equicorrelated with parameter $r\in(0,1)$ on the index set $\{1,2,3\}$ and equal to
the identity elsewhere, and let $D=\diag(-1,1,\dots,1)$. By (\ref{eq:equi}) and block
diagonality, $\Phi_\alpha(C_r)$ has $\gamma_{12}=\gamma_{13}=\gamma_{23}=a_{\alpha,3}(r)$ and
all other coordinates zero, while $\Phi_\alpha(DC_rD)$ agrees except that $\gamma_{12}$ and
$\gamma_{13}$ change sign. Their midpoint is the point $m$ with $m_{23}=a_{\alpha,3}(r)$ and all
other coordinates zero.

Suppose $m=\Phi_\alpha(C)$ for some $C\in\Cn$. For each $i\notin\{2,3\}$ let $D_i$ be the
signature matrix that flips coordinate $i$. Every coordinate $\gamma_{ij}$, $j\ne i$, of $m$
vanishes, so $m$ is fixed by the corresponding sign action, whence
$\Phi_\alpha(D_iCD_i)=\Phi_\alpha(C)$ and injectivity gives $D_iCD_i=C$, that is $C_{ij}=0$ for
all $j\ne i$. Applying this for every $i\notin\{2,3\}$ leaves $C$ block diagonal with a single
$2\times2$ block on $\{2,3\}$, so its transformed $(2,3)$ coordinate is the bivariate coordinate
(\ref{eq:n2}) and therefore $m_{23}<2^{\alpha-1}/\alpha$. But
$a_{\alpha,3}(r)\to3^{\alpha-1}/\alpha$ as $r\uparrow1$, and $3^{\alpha-1}>2^{\alpha-1}$
precisely when $\alpha>1$, so choosing $r$ close enough to one makes
$m_{23}\ge2^{\alpha-1}/\alpha$, a contradiction. Hence $m\notin\mathcal{R}_\alpha$ while both
$\Phi_\alpha(C_r)$ and $\Phi_\alpha(DC_rD)$ lie in it.

(iii) Let $n=4$ first. Let $C=VV'$, where the rows of $V$ are the unit vectors
$v_1=(1,0)$, $v_2=(-\tfrac{\sqrt3}{2},\tfrac12)$, $v_3=(-1,0)$ and $v_4=(0,1)$, at the
angles $0$, $5\pi/6$, $\pi$ and $\pi/2$. Let $C_\varepsilon=(1-\varepsilon)C+\varepsilon I$
for $\varepsilon\in(0,1)$, which is a correlation matrix, and let $D=\diag(1,1,1,-1)$.
Both $\Phi_\alpha(C_\varepsilon)$ and $\Phi_\alpha(DC_\varepsilon D)$ lie in $\mathcal{R}_\alpha$.

The matrix $M=V'V$ has the entries $M_{11}=\tfrac{11}{4}$, $M_{12}=-\tfrac{\sqrt3}{4}$,
$M_{22}=\tfrac54$, eigenvalues $\lambda_\pm=2\pm\tfrac{\sqrt3}{2}$ and spectral projections
$\Pi_\pm$ with $(\Pi_\pm)_{11}=\tfrac12\pm\tfrac{\sqrt3}{4}$ and $(\Pi_\pm)_{12}=\mp\tfrac14$.
If $Mx=\lambda x$ then $C_\varepsilon Vx=\{(1-\varepsilon)\lambda+\varepsilon\}Vx$, and
$C_\varepsilon$ equals $\varepsilon I$ on the orthogonal complement of the range of $V$.
Therefore
\[
  C_\varepsilon^{1/4}=\varepsilon^{1/4}I+VH_\varepsilon V',\qquad
  H_\varepsilon=h_\varepsilon(\lambda_+)\Pi_++h_\varepsilon(\lambda_-)\Pi_-,\qquad
  h_\varepsilon(\lambda)=\frac{\{(1-\varepsilon)\lambda+\varepsilon\}^{1/4}-\varepsilon^{1/4}}{\lambda}.
\]
The off diagonal entries of $C_\varepsilon^{1/4}$ are $v_i'H_\varepsilon v_j$. Since
$v_3=-v_1$, the entries on the pairs $(1,2)$, $(1,3)$ and $(2,3)$ are $-p_\varepsilon$,
$-q_\varepsilon$ and $p_\varepsilon$, with
\[
  q_\varepsilon=(H_\varepsilon)_{11}
  =\bigl(\tfrac12+\tfrac{\sqrt3}{4}\bigr)h_\varepsilon(\lambda_+)
  +\bigl(\tfrac12-\tfrac{\sqrt3}{4}\bigr)h_\varepsilon(\lambda_-),\qquad
  p_\varepsilon=\tfrac{\sqrt3}{2}\,q_\varepsilon
  +\tfrac18\bigl\{h_\varepsilon(\lambda_+)-h_\varepsilon(\lambda_-)\bigr\}.
\]
As $\varepsilon\to0$, $h_\varepsilon(\lambda)\to\lambda^{-3/4}$, and the limits are
$p_0=0.362613\ldots$ and $q_0=0.484532\ldots$, so that the corresponding coordinates of
$\Phi_\alpha(C)$ are $4(-p_0,-q_0,p_0)=(-1.4505,-1.9381,1.4505)$.

The midpoint $m_\varepsilon$ of $\Phi_\alpha(C_\varepsilon)$ and $\Phi_\alpha(DC_\varepsilon D)$
has zero coordinates on the pairs that involve the fourth variable, and its other coordinates
are $4(-p_\varepsilon,-q_\varepsilon,p_\varepsilon)$. Suppose that
$m_\varepsilon=\Phi_\alpha(Y)$ with $Y\in\mathcal{C}_4$. As in part (ii), $m_\varepsilon$ is fixed by the sign action of $D$, so
injectivity gives $DYD=Y$, and $Y=Y_3\oplus1$ with $Y_3\in\mathcal{C}_3$. Then
$A=Y_3^{1/4}$ is positive definite, has $A_{12}=-p_\varepsilon$, $A_{13}=-q_\varepsilon$ and
$A_{23}=p_\varepsilon$, and satisfies $\diag(A^{4})=\diag(Y_3)=\iota$.

Lemma~\ref{lem:quarter} rules this out once (\ref{eq:qa})--(\ref{eq:qc}) hold at
$(p_\varepsilon,q_\varepsilon)$ for some $\bar v$. Let $\bar v=0.59$. At $\varepsilon=0$
the left sides of (\ref{eq:qa}), (\ref{eq:qb}) and (\ref{eq:qc}) are $1.8119$, $1.0126$ and
$0.0295$. All three inequalities are strict, so they continue to hold at
$(p_\varepsilon,q_\varepsilon)$ for all sufficiently small $\varepsilon>0$. For such
$\varepsilon$ the midpoint $m_\varepsilon$ is not in $\mathcal{R}_\alpha$, while both
endpoints are, and $\mathcal{R}_\alpha$ is not convex.

The smallness can be made explicit. For $\lambda\ge1$ and $\varepsilon\in[0,\varepsilon_0]$,
\[
  (1-\varepsilon_0)\lambda^{-3/4}-\varepsilon_0^{1/4}/\lambda\le h_\varepsilon(\lambda)
  \le\lambda^{-3/4},
\]
and $\lambda_->1$. With $\varepsilon_0=10^{-12}$, so that $\varepsilon_0^{1/4}=10^{-3}$,
exact rational interval arithmetic gives $p_\varepsilon\in[0.362287,0.362673]$ and
$q_\varepsilon\in[0.484146,0.484532]$ for every $\varepsilon\in[0,10^{-12}]$, and over these
intervals the left sides of (\ref{eq:qa})--(\ref{eq:qc}) are at least $1.8068$, $1.0102$ and
$0.0268$. The midpoint is therefore outside $\mathcal{R}_\alpha$ for every
$\varepsilon\in(0,10^{-12}]$.

The same certificate gives an open interval of $\alpha$. Fix $\varepsilon\in(0,10^{-12}]$ and keep
$C_\varepsilon$ and $DC_\varepsilon D$ fixed as $\alpha$ varies. Suppose that for a sequence
$\alpha_k\to\tfrac14$ the midpoint of $\Phi_{\alpha_k}(C_\varepsilon)$ and
$\Phi_{\alpha_k}(DC_\varepsilon D)$ lies in $\mathcal{R}_{\alpha_k}$. As above, the midpoint is then
$\Phi_{\alpha_k}(Y_k\oplus1)$ with $Y_k\in\mathcal{C}_3$, and $A_k=Y_k^{\alpha_k}$ has the same off
diagonal entries as $C_\varepsilon^{\alpha_k}$ on the pairs $(1,2)$, $(1,3)$ and $(2,3)$. A subsequence
of $Y_k$ converges to some $Y$ in the closed $3\times3$ elliptope. The map $(\alpha,Y)\mapsto Y^{\alpha}$
is jointly continuous on positive semidefinite matrices for $\alpha$ near $\tfrac14$, so
$A=Y^{1/4}$ is positive semidefinite, has $A_{12}=-p_\varepsilon$, $A_{13}=-q_\varepsilon$ and
$A_{23}=p_\varepsilon$, and satisfies $\diag(A^{4})=\diag(Y)=\iota$. Lemma~\ref{lem:quarter} rules this
out. Hence the midpoint lies outside $\mathcal{R}_\alpha$ for every $\alpha$ in an open interval
around $\tfrac14$, while both endpoints lie in it.

For $n>4$ let $\tilde C_\varepsilon=C_\varepsilon\oplus I_{n-4}$ and
$\tilde D=\diag(1,1,1,-1,1,\dots,1)$. Since $\Phi_\alpha$ maps a block diagonal matrix to a
vector that is zero across blocks, and $f_\alpha(1)=0$, the midpoint of
$\Phi_\alpha(\tilde C_\varepsilon)$ and $\Phi_\alpha(\tilde D\tilde C_\varepsilon\tilde D)$ is
zero on every pair that involves a variable $i\ge4$. If it equals $\Phi_\alpha(Y)$, then for
each $i\ge4$ the sign action that flips variable $i$ fixes it, and injectivity gives
$Y_{ij}=0$ for $j\ne i$. Hence $Y=Y_3\oplus I_{n-3}$, and the argument for $n=4$ applies
without change.
\end{proof}

The only numerical step is the evaluation of the explicit algebraic numbers $p_\varepsilon$
and $q_\varepsilon$ and of three polynomials, and the margins are large enough to be checked
by hand to four digits. The replication files verify the step in exact rational arithmetic,
with rational brackets for $\sqrt3$ and for $\lambda_\pm^{-3/4}$ that are checked by
comparing integer powers. They also verify the algebraic identities in the proof of
Lemma~\ref{lem:quarter} symbolically. As an independent check that does not use the
symmetry reduction, a branch and bound over $d\in[0,1]^{3}$ in exact interval arithmetic,
using only $\diag(A^{4})=\iota$ and the fact that $A$ is not positive semidefinite when
$A+\Diag(t)$ is not for some $t\ge0$, excludes the whole cube with about $100$ boxes. The same
search shows that no point within sup-norm distance $0.015$ of
$4(-p_0,-q_0,p_0)$ lies in the closure of the image of $\mathcal{C}_3$.

\begin{proof}[Proof of Theorem~\ref{thm:cond}]
By the proof of Lemma~\ref{lem:jac}, $J_\alpha(C)$ is the principal submatrix of the positive
definite matrix $M_\alpha(C)$ for the off diagonal coordinates. The eigenvalues of $M_\alpha(C)$
are the divided differences $f_\alpha^{[1]}(\lambda_k,\lambda_\ell)$, with eigenvectors given by
the coordinates of $U(e_ke_\ell'+e_\ell e_k')U'$. Since $f_\alpha'(x)=x^{\alpha-1}$ is monotone on $(0,\infty)$,
the divided differences of $f_\alpha$ are monotone in each argument, so they are extremised at
the extreme eigenvalues, giving
$\kappa(M_\alpha(C))=(\lambda_n/\lambda_1)^{\alpha-1}$ or its reciprocal, that is
$\kappa(C)^{|1-\alpha|}$. The eigenvalues of a principal submatrix of a positive definite
matrix lie between the extreme eigenvalues of the matrix, because its Rayleigh quotient takes a
subset of the values of the Rayleigh quotient of the matrix. Hence
$\kappa(J_\alpha(C))\le\kappa(M_\alpha(C))$, which gives (\ref{eq:cond}).

For sharpness, write $n=2m+r$ with $m\ge2$ and $r\in\{0,1\}$, let
$A=\bigl(\begin{smallmatrix}1&\rho\\ \rho&1\end{smallmatrix}\bigr)$ with $\rho\in(0,1)$, and let
$C$ be block diagonal with $m$ copies of $A$ and, if $r=1$, a final diagonal entry equal to one.
Then $\kappa(C)=(1+\rho)/(1-\rho)$, which takes every value $\kappa_0>1$ as $\rho$ ranges over
$(0,1)$. The matrix $A$ has eigenvalues $1\pm\rho$ and eigenvectors $w_\pm=(1,\pm1)'/\sqrt2$. In $C$,
the vectors $u=(w_-,0,\dots)'$ and $v=(0,w_-,0,\dots)'$ supported on two different blocks are
both eigenvectors for $\lambda_{\min}=1-\rho$, and $uv'+vu'$ has zero diagonal because $u$ and
$v$ have disjoint supports. Hence $uv'+vu'\in\Sn^{0}$ is an eigenvector of $Df_\alpha(C)$ with zero
diagonal, so its off diagonal coordinates form an eigenvector of $J_\alpha(C)$ with eigenvalue $f_\alpha^{[1]}(1-\rho,1-\rho)=(1-\rho)^{\alpha-1}$. The same
construction with $w_+$ gives a hollow eigenvector with eigenvalue $(1+\rho)^{\alpha-1}$. Both
extreme eigenvalues of $M_\alpha(C)$ are therefore eigenvalues of $J_\alpha(C)$, so
(\ref{eq:cond}) holds with equality. Two blocks are needed, whence $n\ge4$.
\end{proof}

\begin{proof}[Proof of Theorem~\ref{thm:varstab}]
For the untransformed correlations, the Gaussian asymptotic covariance of $\hat\rho_{ij}$ and
$\hat\rho_{kl}$ is, see \citet{Nel:1985} and \citet{NeudeckerWesselman:1990},
\[
  \rho_{ik}\rho_{jl}+\rho_{il}\rho_{jk}
  -\rho_{ij}(\rho_{ik}\rho_{il}+\rho_{jk}\rho_{jl})
  -\rho_{kl}(\rho_{ik}\rho_{jk}+\rho_{il}\rho_{jl})
  +\tfrac12\rho_{ij}\rho_{kl}(\rho_{ik}^{2}+\rho_{il}^{2}+\rho_{jk}^{2}+\rho_{jl}^{2}),
\]
a polynomial in the entries of $C$ whose only terms of degree at most one at $C=I$ come from
$\rho_{ik}\rho_{jl}+\rho_{il}\rho_{jk}$. If the pairs coincide this contributes
$1+O(\|\Delta\|^{2})$, if they are disjoint it contributes $O(\|\Delta\|^{2})$, and if they
overlap in a single index $p$ it contributes $\Delta_{uv}+O(\|\Delta\|^{2})$ with $u,v$ the
non-shared indices. Hence $V(I+\Delta)=I_d+B_\Delta+O(\|\Delta\|^{2})$.

For the Jacobian, $f_\alpha(1+\delta)=\delta+\tfrac{\alpha-1}{2}\delta^{2}+O(\delta^{3})$
gives $f_\alpha(I+\Delta)=\Delta+\tfrac{\alpha-1}{2}\Delta^{2}+O(\|\Delta\|^{3})$ and hence
$Df_\alpha(I+\Delta)[H]=H+\tfrac{\alpha-1}{2}(\Delta H+H\Delta)+O(\|\Delta\|^{2})$. For hollow
$H$ and $i\ne j$,
\[
  (\Delta H+H\Delta)_{ij}=\sum_{k\ne i,j}\bigl(\Delta_{ik}H_{kj}+H_{ik}\Delta_{kj}\bigr),
\]
so the coefficient of the free variable $H_{il}$ is $\Delta_{lj}$ and the coefficient of
$H_{kj}$ is $\Delta_{ik}$. Both are entries of the overlap pattern (\ref{eq:overlap}), so the
matrix of $H\mapsto\vecl(\Delta H+H\Delta)$ is exactly $B_\Delta$, which is symmetric. Therefore
$J_\alpha(I+\Delta)=I_d+\tfrac{\alpha-1}{2}B_\Delta+O(\|\Delta\|^{2})$ and
\[
  V_\alpha=J_\alpha VJ_\alpha'
  =\Bigl(I_d+\tfrac{\alpha-1}{2}B_\Delta\Bigr)\bigl(I_d+B_\Delta\bigr)
   \Bigl(I_d+\tfrac{\alpha-1}{2}B_\Delta\Bigr)+O(\|\Delta\|^{2})
  =I_d+\alpha B_\Delta+O(\|\Delta\|^{2}).
\]
The first order term vanishes identically in $\Delta$ if and only if $\alpha=0$.
\end{proof}

\begin{proof}[Proof of Proposition~\ref{prop:equivar}]
Write $C_\rho=\lambda_1P+\lambda_2Q$ with $P=\iota\iota'/n$ and $Q=I-P$. By the formula in the proof of Theorem~\ref{thm:varstab}, $V(C_\rho)$ has the entry $(1-\rho^{2})^{2}$ for equal pairs, $b=\rho(1-\rho)^{2}(1+\tfrac32\rho)$ for pairs that share one index, and $c=2\rho^{2}(1-\rho)^{2}$ for disjoint pairs. Hence $V(C_\rho)=(1-\rho^{2})^{2}I_d+bA_1+cA_2$, where $A_1$ and $A_2$ are the adjacency matrices of the pairs that share one index and of the disjoint pairs. The common, node and pair-specific subspaces are the eigenspaces of $A_1$, with eigenvalues $2(n-2)$, $n-4$ and $-2$, and of $A_2$, with eigenvalues $\tfrac12(n-2)(n-3)$, $-(n-3)$ and $1$. This gives the eigenvalues $v_0=\lambda_1^{2}\lambda_2^{2}$, $v_1=\lambda_2^{2}\{1+(n-2)\rho(1-\rho/2)\}$ and $v_2=\lambda_2^{2}$ of $V(C_\rho)$.

For the Jacobian, the Daleckii--Krein formula gives $Df_\alpha(C_\rho)[X]=\lambda_1^{\alpha-1}PXP+\lambda_2^{\alpha-1}QXQ+f_\alpha^{[1]}(\lambda_1,\lambda_2)(PXQ+QXP)$. For $X=\iota\iota'-I=(n-1)P-Q$ this is $(n-1)\lambda_1^{\alpha-1}P-\lambda_2^{\alpha-1}Q$, whose off diagonal entries are $j_0=\{(n-1)\lambda_1^{\alpha-1}+\lambda_2^{\alpha-1}\}/n$. For hollow $X$ with $X\iota=0$ we have $X=QXQ$, so the image is $j_2X$ with $j_2=\lambda_2^{\alpha-1}$. For $X=a\iota'+\iota a'-2\Diag(a)$ with $\iota'a=0$, whose off diagonal entries are $a_i+a_j$, we have $PXP=0$, $PXQ+QXP=(1-\tfrac2n)(\iota a'+a\iota')$ and $QXQ=-2Q\Diag(a)Q$, whose $(i,j)$ entry for $i\ne j$ is $\tfrac2n(a_i+a_j)$. The off diagonal of the image is therefore $j_1$ times that of $X$, with $j_1=(1-\tfrac2n)f_\alpha^{[1]}(\lambda_1,\lambda_2)+\tfrac2n\lambda_2^{\alpha-1}$. So $J_\alpha(C_\rho)$ acts on the three subspaces as $j_0$, $j_1$ and $j_2$, and $V_\alpha=J_\alpha VJ_\alpha'$ gives $\omega_k=j_k^{2}v_k$, which are the stated expressions. At $\alpha=0$, $j_0^{2}v_0=\{(n-1)\lambda_2+\lambda_1\}^{2}/n^{2}=1$ and $j_2^{2}v_2=1$.

For uniqueness, $\omega_2=(1-\rho)^{2\alpha}$, and the pair-specific subspace is nonzero for $n\ge4$. Further, $\omega_0^{1/2}$ has derivative $\alpha\{(n-1)^{2}-1\}/n=\alpha(n-2)$ at $\rho=0$, which is not zero for $\alpha\ne0$ and $n\ge3$.

For the statement on blocks, let $X$ be hollow, supported on block $k$ and with zero row sums. Then $CX=XC=(1-\rho_k)X$, so $Df_\alpha(C)[X]=(1-\rho_k)^{\alpha-1}X$. The permutations of the variables in block $k$ leave $C$ and the law of $\hat C$ unchanged, so they commute with $V(C)$. The pair-specific subspace of block $k$ carries an irreducible representation of these permutations that does not occur elsewhere in $\R^{d}$, since the pairs with one or no index in block $k$ carry only the trivial and the standard representations. By Schur's lemma $V(C)$ acts on it as a scalar, which, for $\|\vecl(X)\|=1$, equals the asymptotic variance of $\sqrt T\vecl(X)'\vecl(\hat C)$. This depends only on the variables in block $k$, whose correlation matrix is equicorrelated with parameter $\rho_k$, so the scalar is $(1-\rho_k)^{2}$, and the eigenvalue of $V_\alpha(C)$ is $(1-\rho_k)^{2\alpha}$.
\end{proof}

\begin{proof}[Proof of Proposition~\ref{prop:skew}]
(i) Let $X$ be the $T\times n$ data matrix. If the mean is known, set $m=T$ and $Z=X$. If the mean is
estimated, set $m=T-1$ and $Z=Q'X$, where the columns of the $T\times(T-1)$ matrix $Q$ form an
orthonormal basis of the orthogonal complement of $\iota$. In both cases the columns
$z_1,\dots,z_n$ of $Z$ are independent $N(0,I_m)$ vectors, and $\hat C_{ij}=u_i'u_j$ with
$u_v=z_v/\|z_v\|$. The $u_v$ are independent and uniform on the unit sphere of $\R^{m}$, and $u_v$
is independent of $\|z_v\|$. Write $\hat C=I+K$, so that $K$ is hollow with $K_{ij}=u_i'u_j$ for
$i\ne j$.

\emph{Symmetry.} Let $D=\diag(s_1,\dots,s_n)$ with each $s_v\in\{-1,1\}$. Replacing $u_v$ by
$s_vu_v$ maps $\hat C$ to $D\hat CD$ and leaves its distribution unchanged. Since $D$ is
orthogonal, $f_\alpha(D\hat CD)=Df_\alpha(\hat C)D$, so $\hat\gamma_{ij}$ becomes
$s_is_j\hat\gamma_{ij}$. A product of coordinates over the pairs $e_1,\dots,e_p$ is therefore
multiplied by $\prod_vs_v^{d_v}$, where $d_v$ is the number of pairs that contain $v$. If some
$d_v$ is odd, changing the sign of $u_v$ alone changes the sign of the product, so its expectation
is zero. With $p=1$ this gives $\mathbb{E}\hat\gamma_{ij}=0$, and with zero means every joint third
cumulant equals the corresponding third moment. Let $p=3$. If a pair occurs twice, the third pair
either equals it, which gives degree three, or misses one of its indices, which gives degree one.
So the three pairs are distinct. Three distinct pairs with even degrees form a cycle, and a cycle
of three pairs is a triangle. Hence all other third cumulants vanish for every $T$. Permutations of
the variables show that all triangles give the same value, and it remains to compute
$\mathbb{E}(\hat\gamma_{12}\hat\gamma_{13}\hat\gamma_{23})$.

\emph{Products of correlations.} Let $G$ be a list of $p$ pairs $\{i,j\}$ with $i\ne j$, with
repetitions allowed, and let every degree $d_v$ be even. Since $u_v$ is independent of $\|z_v\|$,
\[
  \mathbb{E}\prod_{\{i,j\}\in G}K_{ij}
  =\frac{E\prod_{\{i,j\}\in G}z_i'z_j}{\prod_vE\|z_v\|^{d_v}},
  \qquad \mathbb{E}\|z_v\|^{2k}=m(m+2)\cdots(m+2k-2).
\]
Expand each $z_i'z_j$ as a sum over coordinates and apply Isserlis' theorem to each $z_v$
separately. Each term corresponds to a pairing of the $d_v$ occurrences of each index $v$, and the
sum over coordinates gives $m^{c}$, where $c$ is the number of closed cycles that the pairing
forms. A cycle uses at least two pairs because no pair has equal indices, so $c\le\lfloor
p/2\rfloor$. The denominator is a polynomial in $m$ of degree $p$ with leading coefficient one.
Together with the symmetry argument for odd degrees this gives
\begin{equation}\label{eq:wickbound}
  \mathbb{E}\prod_{\{i,j\}\in G}K_{ij}=O\bigl(m^{-\lceil p/2\rceil}\bigr)
\end{equation}
for every list of $p$ pairs. Three exact values are needed. Since $\mathbb{E}(u_vu_v')=m^{-1}I$ and the
$u_v$ are independent,
\[
  \mathbb{E}(K_{12}K_{23}K_{31})=\mathbb{E}\tr(u_1u_1'u_2u_2'u_3u_3')=\tr(m^{-3}I)=m^{-2},
\]
and in the same way $\mathbb{E}(K_{1k}K_{k2}K_{23}K_{31})=\tr(m^{-4}I)=m^{-3}$ for $k\ge4$. Given $u_3$,
the correlations $K_{13}$ and $K_{23}$ are independent with conditional second moment
$u_3'(m^{-1}I)u_3=m^{-1}$, so $\mathbb{E}(K_{13}^{2}K_{23}^{2})=m^{-2}$.

\emph{Expansion.} Let $a_2$ and $a_3$ be as in the proof of Proposition~\ref{prop:bias}, and let
$A$ be the event $\|K\|\le\tfrac12$ in the spectral norm. On $A$ the eigenvalues of $\hat C$ lie in
$[\tfrac12,\tfrac32]$, where $f_\alpha$ is analytic. Taylor's theorem for $f_\alpha(1+x)$ on
$|x|\le\tfrac12$ and the spectral calculus give constants $c_1$ and $c_2$, which depend only on
$\alpha$, such that on $A$
\[
  f_\alpha(\hat C)=P+R,\qquad P=K+a_2K^{2}+a_3K^{3},\qquad \|R\|\le c_2\|K\|^{4},
\]
with $|\hat\gamma_{ij}|\le c_1\|K\|$ and $|P_{ij}|\le c_1\|K\|$. Hence on $A$
\[
  \bigl|\hat\gamma_{12}\hat\gamma_{13}\hat\gamma_{23}-P_{12}P_{13}P_{23}\bigr|
  \le3c_1^{2}c_2\|K\|^{6}.
\]
Since $\|K\|^{2}\le\sum_{i\ne j}K_{ij}^{2}$ and $\mathbb{E}(K_{ij}^{2q})=(2q-1)!!/\{m(m+2)\cdots(m+2q-2)\}$,
we have $\mathbb{E}\|K\|^{6}=O(m^{-3})$. Outside $A$ some $|K_{ij}|$ exceeds $1/(2n)$, and Markov's
inequality with $q=6$ gives $\Pr(A^{c})=O(m^{-6})$.

The entries of $P$ are bounded because $|K_{ij}|\le1$. For $\alpha>0$ the function $f_\alpha$ is
bounded on $[0,n]$, which contains the spectrum of $\hat C$, so the coordinates of $\hat\gamma$ are
bounded. For $\alpha\le0$ let $s=\max(1,-\alpha)$. Then $|f_\alpha(x)|\le c_3(1+x^{-s})$ on
$(0,n]$, so $|\hat\gamma_{ij}|\le c_3\{1+\lambda_{\min}(\hat C)^{-s}\}$. Further
$\lambda_{\min}(\hat C)^{-1}\le\tr(\hat C^{-1})=\sum_v(1-R_v^{2})^{-1}$, where $R_v^{2}$ is the
squared length of the projection of $u_v$ on the span of the other $u$ vectors. Given the others,
$u_v$ is uniform on the sphere, so $1-R_v^{2}$ has the beta distribution with parameters
$(m-n+1)/2$ and $(n-1)/2$. Its moment of order $-6s$ is finite, and bounded in $m$, when
$m>n-1+12s$, which the condition on $T$ ensures. In both cases
$\mathbb{E}(\hat\gamma_{12}\hat\gamma_{13}\hat\gamma_{23}-P_{12}P_{13}P_{23})^{2}$ is bounded in $m$, and the
Cauchy--Schwarz inequality gives
\[
  \mathbb{E}\bigl|(\hat\gamma_{12}\hat\gamma_{13}\hat\gamma_{23}-P_{12}P_{13}P_{23})1_{A^{c}}\bigr|
  =O\bigl\{\Pr(A^{c})^{1/2}\bigr\}=O(m^{-3}).
\]
With the bound on $A$ this gives
$\mathbb{E}(\hat\gamma_{12}\hat\gamma_{13}\hat\gamma_{23})=\mathbb{E}(P_{12}P_{13}P_{23})+O(m^{-3})$.

\emph{The polynomial part.} $\mathbb{E}(P_{12}P_{13}P_{23})$ is a finite sum of expectations of products of
entries of $K$. By (\ref{eq:wickbound}) the products with five or more factors are $O(m^{-3})$.
The only product with three factors is $K_{12}K_{13}K_{23}$, with expectation $m^{-2}$. The
products with four factors come from one factor $a_2(K^{2})_{ij}$ and two factors from $K$. Since
$K$ is hollow, $(K^{2})_{12}=\sum_{k\ne1,2}K_{1k}K_{k2}$, and the exact values above give
\[
  \mathbb{E}\{(K^{2})_{12}K_{13}K_{23}\}
  =\mathbb{E}(K_{13}^{2}K_{23}^{2})+\sum_{k\ge4}\mathbb{E}(K_{1k}K_{k2}K_{23}K_{31})
  =m^{-2}+(n-3)m^{-3}.
\]
The other two terms with four factors equal this one by symmetry. Therefore
\[
  \mathbb{E}(\hat\gamma_{12}\hat\gamma_{13}\hat\gamma_{23})
  =\frac{1+3a_2}{m^{2}}+O(m^{-3})
  =\frac{3\alpha-1}{2m^{2}}+O(m^{-3}).
\]
Since $m=T$ or $m=T-1$, we have $m^{-2}=T^{-2}+O(T^{-3})$, which proves (i).

(ii) Let $k\notin\{i,j\}$ and let $D$ change the sign of variable $k$. Then $f_\alpha(D\hat CD)=Df_\alpha(\hat C)D$ leaves $\hat\gamma_{ij}$ unchanged, and $D\hat CD$ has the law of the sample correlation matrix at $DCD$, in which every $\Delta_{kl}$ has changed sign. The first order term of the third cumulant of $\hat\gamma_{ij}$ therefore involves $\Delta_{ij}$ only, and its coefficient can be computed at $C=I+\Delta_{ij}(e_ie_j'+e_je_i')$. This matrix is block diagonal with the blocks $\{i,j\}$ and the remaining variables. By the Daleckii--Krein formula the derivative of $\gamma_{ij}$ with respect to every other free entry of $C$ is zero there, and by the formula in the proof of Theorem~\ref{thm:varstab} the asymptotic covariance of $\hat\rho_{ij}$ with every other sample correlation is zero. In the delta method expansion of the third cumulant of $\hat\gamma_{ij}$, which involves the first derivatives through the third cumulants of the sample correlations and the first and second derivatives through their covariances, only the terms in $\hat\rho_{ij}$ itself remain. Since $\hat\rho_{ij}$ depends only on variables $i$ and $j$, and $\gamma_{ij}=\Phi_\alpha(\rho_{ij})$ with the bivariate map (\ref{eq:n2}), the leading third cumulant is the $n=2$ expression given before Theorem~\ref{thm:varstab}, evaluated at $\rho=\Delta_{ij}$. There $\Phi_\alpha'(\rho)=1+O(\rho^{2})$ and $\Phi_\alpha''(\rho)=(\alpha-1)(\alpha-2)\rho+O(\rho^{3})$, so the third cumulant is $\{3(\alpha-1)(\alpha-2)-6\}\Delta_{ij}/T^{2}=3\alpha(\alpha-3)\Delta_{ij}/T^{2}$ to first order.
\end{proof}

\begin{proof}[Proof of Proposition~\ref{prop:normal}]
Since $H$ is hollow, $\diag(e^{H})=\iota+\tfrac12\diag(H^{2})+O(\|H\|^{3})$, so the normalising
factor is $\Diag\{\diag(e^{H})\}^{-1/2}=I-\tfrac14\Diag\{\diag(H^{2})\}+O(\|H\|^{3})$ and
\[
  \Cor(e^{H})=I+H+\tfrac12\bigl[H^{2}-\Diag\{\diag(H^{2})\}\bigr]+O(\|H\|^{3}).
\]
Writing $E$ for the perturbation and using $f_\alpha(I+E)=E+\tfrac{\alpha-1}{2}E^{2}+O(\|E\|^{3})$
with $E^{2}=H^{2}+O(\|H\|^{3})$,
\[
  f_\alpha\{\Cor(e^{H})\}=H+\frac{\alpha}{2}H^{2}
  -\tfrac12\Diag\{\diag(H^{2})\}+O(\|H\|^{3}).
\]
The diagonal term is annihilated by $\vecl$, which gives (\ref{eq:normal}). For $i\ne j$ we have
$(H^{2})_{ij}=\sum_{k\ne i,j}H_{ik}H_{kj}$, an empty sum when $n=2$ and generically nonzero when
$n\ge3$.

For the remainder at $\alpha=0$, let $X=-\tfrac12\log\Diag\{\diag(e^{H})\}$, which is diagonal
and $O(\|H\|^{2})$. Then $\Cor(e^{H})=e^{X}e^{H}e^{X}$, and the symmetric
Baker--Campbell--Hausdorff expansion has no single-commutator term, so
\[
  \log\Cor(e^{H})=H+2X+\tfrac16[H,[H,X]]+O(\|H\|^{5}).
\]
The term $2X$ is diagonal and is annihilated by $\vecl$, and the double commutator is
$O(\|H\|^{4})$, so $\Phi_0\{\Cor(e^{H})\}=\vecl(H)+O(\|H\|^{4})$.

It remains to identify the coordinates, which in geometric terms are the normal coordinates of
the quotient at $I$. The path $t\mapsto e^{tH}$ is the geodesic of the covariance matrices
through $I$ with initial velocity $H$, which is orthogonal to the nuisance directions at $I$. A
geodesic that starts orthogonally to the nuisance directions stays so and projects to a geodesic
of the quotient, a general property of Riemannian submersions used in this setting by
\citet{ThanwerdasPennec:2021}. Hence $t\mapsto\Cor(e^{tH})$ is the quotient geodesic
through $I$ with initial velocity $H$, and $\vecl(H)$ is the normal coordinate of
$\Cor(e^{H})$. By the definition of normal coordinates, the metric is the identity at
the base point and its first derivatives vanish there.
\end{proof}

\begin{proof}[Proof of Proposition~\ref{prop:kron}]
Let $C=C_A\otimes C_B$. By (\ref{eq:kron}) at $\alpha=0$, $\log C=\log C_A\otimes I_q+I_p\otimes\log C_B$, whose entry in position $((a,b),(a',b'))$ is $(\log C_A)_{aa'}\delta_{bb'}+\delta_{aa'}(\log C_B)_{bb'}$. This is zero when $a\ne a'$ and $b\ne b'$, it equals $(\log C_A)_{aa'}$ when $a\ne a'$ and $b=b'$, and it equals $(\log C_B)_{bb'}$ when $a=a'$ and $b\ne b'$. Hence $\Phi_0(C)\in S_{p,q}$, and the common values are the coordinates of $\Phi_0(C_A)$ and $\Phi_0(C_B)$. Conversely, let $\gamma\in S_{p,q}$, and let $\gamma_A\in\R^{p(p-1)/2}$ and $\gamma_B\in\R^{q(q-1)/2}$ collect its common values. By Theorem~\ref{thm:main}(ii) there are $C_A\in\mathcal{C}_p$ and $C_B\in\mathcal{C}_q$ with $\Phi_0(C_A)=\gamma_A$ and $\Phi_0(C_B)=\gamma_B$, and by the first part $\Phi_0(C_A\otimes C_B)=\gamma$. Since $C_A\otimes C_B$ is a positive definite correlation matrix and $\Phi_0$ is injective, $\Phi_0^{-1}(\gamma)=C_A\otimes C_B$. Finally, an element of $S_{p,q}$ is determined by its common values, which can be chosen freely, so $S_{p,q}$ has dimension $p(p-1)/2+q(q-1)/2$.
\end{proof}

\begin{proof}[Proof of Theorem~\ref{thm:scaled}]
(i) By (\ref{eq:pgrad}), $\nabla P_\alpha(x)=2\operatorname{sgn}(\alpha)\{\diag A(x)^{\alpha}-\iota\}$
for $\alpha\ne0$ and $\nabla p_0(x)=2\diag\log A(x)$, so the stationary points are exactly the
scalings in (\ref{eq:scaled}), and by Lemma~\ref{lem:congconv} there is at most one. For
existence it suffices that the function is coercive. Let $\alpha>0$, let the base matrix $C$ be
any positive definite matrix, not necessarily with unit diagonal, write $P_\alpha(x;C)$ for the
function defined with $A(x)=e^{X}Ce^{X}$, and let $m=\max_ix_i$. Since
$\tr A^{\alpha}\ge\lambda_{\max}(A)^{\alpha}\ge(A_{ii})^{\alpha}=C_{ii}^{\alpha}e^{2\alpha x_i}$
for every $i$, and $-2\iota'x\ge-2nm$, we get $P_\alpha(x;C)\ge c\,e^{2\alpha m}/\alpha-2nm$ with
$c=\min_iC_{ii}^{\alpha}>0$, which tends to infinity with $m$. If $m$ stays bounded above while
$\|x\|\to\infty$, then $\min_ix_i\to-\infty$, so $-2\iota'x\to\infty$ while $\tr A^{\alpha}>0$.
For $\alpha<0$ note that $A(x)^{-1}=e^{-X}Ke^{-X}$ with $K=C^{-1}$, so
$P_\alpha(x;C)=P_{|\alpha|}(-x;K)$, and the case just treated applies with the positive definite
base matrix $K$. For $\alpha=0$, $p_0$ is strongly convex by Lemma~\ref{lem:congconv}, hence
coercive.

(ii) If $G=f_\alpha(D_\alpha CD_\alpha)$ then $G$ is hollow and
$I+\alpha G=(D_\alpha CD_\alpha)^{\alpha}\succ0$, so $\tilde\Phi_\alpha(C)\in\mathcal{H}_\alpha$.
Conversely let $G\in\mathcal{H}_\alpha$, set $M=e_\alpha(G)\succ0$ and $C=\Cor(M)$.
Then $M=D'CD'$ with $D'=\Diag(M)^{1/2}$, and $f_\alpha(D'CD')=G$ is hollow, so by uniqueness
$D_\alpha(C)=D'$ and $\tilde\Phi_\alpha(C)=G$. Hence $\tilde\Phi_\alpha$ maps onto
$\mathcal{H}_\alpha$ and (\ref{eq:scaledinv}) is a two sided inverse. The inverse is real
analytic on the open set $\mathcal{H}_\alpha$, being a composition of the spectral calculus of
an analytic function with the normalisation. For the forward map, $D_\alpha(C)$ is the
solution of $\nabla P_\alpha=0$, a real analytic equation in $(x,C)$ whose Jacobian in $x$ is
the positive definite Hessian of Lemma~\ref{lem:congconv}, so $C\mapsto D_\alpha(C)$ is real analytic
by the analytic implicit function theorem, and $\tilde\Phi_\alpha$ is a composition of analytic
maps.

(iii) $\mathcal{H}_\alpha$ is the intersection of the linear space $\Sn^{0}$ with the
preimage of the open convex cone under an affine map, hence open and convex. Let
$G\in\Sn^{0}$ have eigenvalues $\lambda_1\ge\dots\ge\lambda_n$, which sum to $\tr G=0$. If
$\alpha>0$ then $I+\alpha G\succ0$ means $\lambda_n>-1/\alpha$, and then
$\lambda_1=-\sum_{k\ge2}\lambda_k<(n-1)/\alpha$, so $\|G\|_2<(n-1)/\alpha$. The case
$\alpha<0$ is symmetric. At $\alpha=0$ every hollow $G$ qualifies since $e^{G}\succ0$.
\end{proof}

\section{Supplementary results}\label{app:supp}

This appendix collects results that support the main text, grouped by topic, together with their
proofs.

\subsection{The diagonal problem}\label{app:diag}

\begin{lemma}\label{lem:jac}
For every $\alpha\in\R$ the map $\Phi_\alpha$ is real analytic on $\Cn$, and its Jacobian
$J_\alpha(C)$ with respect to the free off diagonal entries of $C$ is symmetric positive definite
at every $C\in\Cn$. Hence $\Phi_\alpha$ is a real analytic diffeomorphism onto
its image $\mathcal{R}_\alpha=\Phi_\alpha(\Cn)$, which is an open subset of $\R^{d}$, and
$\Phi_\alpha^{-1}$ is real analytic.
\end{lemma}

\begin{proof}[Proof of Lemma~\ref{lem:jac}]
For analyticity, $x\mapsto x^{\alpha}$ and $x\mapsto\log x$ are holomorphic on a neighbourhood
of $(0,\infty)$ in the right half plane, so on the positive definite cone
\[
  f_\alpha(C)=\frac{1}{2\pi i}\oint_{\Sigma}f_\alpha(z)(zI-C)^{-1}dz
\]
with $\Sigma$ a contour enclosing the spectrum. The integrand is rational, hence real analytic,
in the entries of $C$, and $\Sigma$ may be held fixed under small perturbations, so
$C\mapsto f_\alpha(C)$ is real analytic.

The tangent space of $\Cn$ is the space $\Sn^{0}$ of hollow symmetric matrices, of dimension
$d$. The differential of $C\mapsto f_\alpha(C)$ is $Df_\alpha(C)$, a self adjoint operator on
$\Sn$ with quadratic form
$\sum_{k\ell}f_\alpha^{[1]}(\lambda_k,\lambda_\ell)[(U'HU)_{k\ell}]^{2}$, positive definite
because $f_\alpha$ is strictly increasing on $(0,\infty)$ for every $\alpha$. For hollow $H$ and $\tilde H$ with
$x=\vecl(H)$ and $y=\vecl(\tilde H)$ we have $\langle\tilde H,Df_\alpha(C)[H]\rangle=2y'J_\alpha(C)x$,
because the trace inner product counts each off diagonal entry twice. The left side is symmetric in
$H$ and $\tilde H$ and positive for $H=\tilde H\ne0$, so $J_\alpha(C)$ is symmetric positive
definite, and its condition number $\kappa(J_\alpha)$ is the ratio of its extreme eigenvalues.
Let $M_\alpha(C)$ be the matrix of $Df_\alpha(C)$ in the orthonormal basis of $\Sn$ formed by
the matrices $e_ie_i'$ and $(e_ie_j'+e_je_i')/\sqrt2$, $i>j$, a positive definite matrix of
dimension $n(n+1)/2$. The coordinates of $H$ in this basis are its diagonal entries and
$\sqrt2H_{ij}$, so $J_\alpha(C)$ is the $d\times d$ principal submatrix of $M_\alpha(C)$ for the
off diagonal coordinates, that is, the derivative of the off diagonal entries of $f_\alpha(C)$
with respect to those of $C$ with the unit diagonal held fixed. The
analytic inverse function theorem then makes $\Phi_\alpha$ a local real analytic diffeomorphism
at every point, and a local diffeomorphism is an open map, so the image is open.
\end{proof}

\begin{proposition}\label{prop:sign}
Let $C\in\Cn$ and $\delta^{*}=\diag f_\alpha(C)$. Then componentwise
$\delta^{*}\le0$ for $\alpha<1$, $\delta^{*}=0$ for $\alpha=1$ and $\delta^{*}\ge0$ for
$\alpha>1$. For $\alpha\ne1$, $\delta^{*}_i=0$ if and only if $C_{ij}=0$ for every $j\ne i$.
\end{proposition}

\begin{proof}[Proof of Proposition~\ref{prop:sign}]
Write $C=U\diag(\lambda)U'$. Since $C_{ii}=\sum_ku_{ik}^{2}\lambda_k=1$ and the weights
$u_{ik}^{2}$ sum to one, $\delta^{*}_i=\sum_ku_{ik}^{2}f_\alpha(\lambda_k)$ is the value of
$f_\alpha$ averaged over a distribution with mean one, and $f_\alpha(1)=0$. As
$f_\alpha''(x)=(\alpha-1)x^{\alpha-2}$, Jensen's inequality gives the sign. For $\alpha\ne1$
equality holds exactly when every eigenvalue carrying positive weight in coordinate $i$ equals
one, which makes $e_i$ an eigenvector of $C$ with eigenvalue one, that is $C_{ij}=0$ for
$j\ne i$, and the converse is immediate.
\end{proof}

\begin{proposition}\label{prop:dual}
Let $\alpha>0$ and $\gamma\in\R^{d}$. Then $\hat L_\alpha$ is coercive and attains its
infimum over $\R^{n}$. If $\gamma\in\mathcal{R}_\alpha$ the minimiser is the stationary point
$\delta^{*}\in\mathcal{D}_\alpha$ of Proposition~\ref{prop:convex}. If
$\gamma\notin\mathcal{R}_\alpha$ every minimiser lies outside $\mathcal{D}_\alpha$,
$\inf_{\mathcal{D}_\alpha}L_\alpha$ is not attained, and it can strictly exceed
$\min_{\R^{n}}\hat L_\alpha$.
\end{proposition}

\begin{proof}[Proof of Proposition~\ref{prop:dual}]
For convex $\phi$ the trace inequality $\tr\phi(G)\ge\sum_i\phi(G_{ii})$ gives
$\hat L_\alpha(\delta)\ge\sum_i\{\hat\Psi_\alpha(\delta_i)-\delta_i\}$, which tends to
$+\infty$ as $\|\delta\|\to\infty$ because $\hat\Psi_\alpha(y)-y$ grows superlinearly as
$y\to\infty$ and linearly as $y\to-\infty$. So $\hat L_\alpha$ is coercive and continuous and
its infimum is attained. If $\gamma\in\mathcal{R}_\alpha$, the stationary point
$\delta^{*}\in\mathcal{D}_\alpha$ of $L_\alpha$ is a stationary point of the convex function
$\hat L_\alpha$, hence a global minimiser, and by strict convexity of $L_\alpha$ on
$\mathcal{D}_\alpha$ no other minimiser lies in $\mathcal{D}_\alpha$. None lies outside
either, since $\hat L_\alpha$ is convex and a second minimiser would make the whole segment
between the two minimisers consist of minimisers, part of which lies in the open set
$\mathcal{D}_\alpha$. If $\gamma\notin\mathcal{R}_\alpha$, a minimiser in $\mathcal{D}_\alpha$
would be a stationary point of $L_\alpha$ and $e_\alpha(G(\hat\delta))$ would be a correlation
matrix mapping to $\gamma$, and for the same reason $\inf_{\mathcal{D}_\alpha}L_\alpha$ is not
attained. It can exceed the minimum of $\hat L_\alpha$. Take $n=2$, $\alpha=1$ and $\gamma=2$,
so $\Psi_1(y)=(1+y)^{2}/2$ and $G(\delta)$ has off diagonal $2$. Writing $a_i=1+\delta_i$, the
domain is $\{a_1,a_2>0,\;a_1a_2>4\}$ and
$L_1(\delta)=\tfrac12(a_1-1)^{2}+\tfrac12(a_2-1)^{2}+5$, whose infimum over the domain is $6$,
approached as $a_1=a_2\downarrow2$, whereas at $\delta=(-1,-1)$ the matrix $I+G$ has eigenvalues
$\pm2$ and $\hat L_1(\delta)=\hat\Psi_1(1)+\hat\Psi_1(-3)+2=4$.
\end{proof}

\subsection{The image near $\alpha=\tfrac14$}\label{app:shape}

Part (iii) of Theorem~\ref{thm:shape} is proved with the correlation matrix $C$ of the unit vectors in
$\R^{2}$ at the angles $0$, $5\pi/6$, $\pi$ and $\pi/2$,
\[
  C=\begin{pmatrix}
  1&-\tfrac{\sqrt3}{2}&-1&0\\
  -\tfrac{\sqrt3}{2}&1&\tfrac{\sqrt3}{2}&\tfrac12\\
  -1&\tfrac{\sqrt3}{2}&1&0\\
  0&\tfrac12&0&1
  \end{pmatrix},
\]
which has rank two, together with $D=\diag(1,1,1,-1)$ and $C_\varepsilon=(1-\varepsilon)C+\varepsilon I$.
The midpoint of $\Phi_\alpha(C_\varepsilon)$ and $\Phi_\alpha(DC_\varepsilon D)$ has zero coordinates
for the pairs that involve the fourth variable. By the argument for Theorem~\ref{thm:shape}(ii), the midpoint
therefore lies in $\mathcal{R}_\alpha$ if and only if its remaining three coordinates lie in the
image of the $3\times3$ correlation matrices. At $\alpha=\tfrac14$ this is a polynomial question,
because a vector lies in the closure of that image if and only if there is a positive
semidefinite $A$ with the prescribed off diagonal and $\diag(A^{4})=\iota$. Three polynomial
inequalities then show that the midpoint is outside $\mathcal{R}_{1/4}$ for every
$\varepsilon\in(0,10^{-12}]$, and a compactness argument extends this to an open interval of
$\alpha$ around $\tfrac14$. The required $\varepsilon$ is small, and for $\varepsilon=10^{-6}$ the
midpoint is in $\mathcal{R}_{1/4}$. The coordinates themselves give no sign of the problem,
because by (\ref{eq:bound}) every coordinate obeys the bivariate bound for $0<\alpha\le1$. The
obstruction comes from which coordinates can occur together.

For other values of $\alpha$ in $(0,\tfrac12)$ the evidence is numerical. For the same witness and
$\alpha=0.1$ and $0.4$, the smallest distances found from the limiting coordinates to the closure of the
$3\times3$ image were about $0.14$ and $0.0014$, against $0.03$ at $\alpha=\tfrac14$, where the distance is positive by Theorem~\ref{thm:shape}(iii), and zero at
$\alpha=\tfrac12$, so the margin shrinks as $\alpha\uparrow\tfrac12$. Appendix~\ref{app:num}
describes the computation. At $\alpha=1/k$, with $k$ an integer, the same reduction gives the
polynomial condition $\diag(A^{k})=\iota$, so these values can be settled in the same way. For
$n=3$ an extensive search found no violation, and since every coordinate obeys the bivariate bound
when $\alpha<1$, the construction in the proof of Theorem~\ref{thm:shape}(ii) is not available there. We therefore
conjecture the following.

\begin{conjecture}\label{conj:convex}
Let $\alpha>0$. For $n\ge4$ the image $\mathcal{R}_\alpha$ is convex if and only if
$\alpha\in[\tfrac12,1]$. For $n=3$ it is convex for every $\alpha\in(0,1]$.
\end{conjecture}

\subsection{Conditioning for $n=3$}\label{app:cond}

\begin{proposition}\label{prop:n3}
Let $n=3$.
\begin{enumerate}
\item[(i)] Fix $r\in(0,1)$ and let $C_\varepsilon$ be the correlation matrix with off
diagonal entries $(1-\varepsilon,r,r)$. Then $\kappa(C_\varepsilon)=\Theta(\varepsilon^{-1})$
and, for every $\alpha<1$, $\log\kappa\{J_\alpha(C_\varepsilon)\}/\log\kappa(C_\varepsilon)\to
|1-\alpha|$ as $\varepsilon\downarrow0$. The exponent in (\ref{eq:cond}) is therefore
asymptotically sharp at $n=3$ on the whole surjective range.
\item[(ii)] For $\alpha\ne1$, equality holds in (\ref{eq:cond}) only at $C=I$. At $\alpha=1$
the Jacobian is the identity and equality holds for every $C$, since both sides equal one.
\item[(iii)] For $\alpha>1$, $\sup_{C\in\mathcal{C}_3}\kappa\{J_\alpha(C)\}<\infty$.
\end{enumerate}
\end{proposition}

Proposition~\ref{prop:n3} settles the case $n=3$. For $\alpha<1$ the
exponent is asymptotically sharp, but equality away from $C=I$ is impossible. At $\alpha=1$ the Jacobian is the identity, and for $\alpha>1$ the condition number of $J_\alpha$ stays bounded however ill conditioned $C$ is. For $n\ge4$ the two-block construction in the proof of Theorem~\ref{thm:cond} attains the exponent on both sides of one. The proof of part (iii) shows where $n=3$ enters. For $\alpha>1$ the Jacobian $J_\alpha(C)$ extends to the closed elliptope and remains positive definite there, because in dimension three no nonzero hollow symmetric matrix is supported on the kernel of a singular correlation matrix. At $\alpha=2$ direct differentiation of the expression for
$\Phi_2$ given after Theorem~\ref{thm:shape} gives $J_2(C)=\tfrac12(I_3+PCP')$, where $P$ is the
permutation matrix that reverses the order of the three variables, so that
$\kappa\{J_2(C)\}=\{1+\lambda_{\max}(C)\}/\{1+\lambda_{\min}(C)\}<4$ for every $C\in\mathcal{C}_3$,
however large $\kappa(C)$ is. For $\alpha<1$, how closely the
bound is approached depends on the way $C$ becomes singular, which in statistical terms is the
form of the near multicollinearity. Along $C_\varepsilon$ two of the three variables become nearly
collinear while the third stays apart, the eigenvector $u=(1,-1,0)'/\sqrt2$ of the small
eigenvalue has a zero component, and for $\alpha<1$ the exponent tends to $|1-\alpha|$. The convergence is
slow, since the proof of Proposition~\ref{prop:n3}(i) gives $\log\kappa(J_\alpha)=|1-\alpha|\log\kappa(C)+O(1)$, so the ratio
$\log\kappa(J_\alpha)/\log\kappa(C)$ approaches $|1-\alpha|$ only at the rate $1/\log\kappa(C)$.

When the near collinearity involves all three variables, the exponent is lower for $\alpha<0$,
and at $\alpha=0$ it is still one but with an additional logarithmic factor. Along the
equicorrelation path $C_\rho$ with $\rho\downarrow-\tfrac12$ it is the sum of the three variables
whose variance vanishes, the eigenvector $u=\iota/\sqrt3$ has no zero component, and no nonzero
hollow direction avoids the small eigenvalue, because for $n=3$ a hollow symmetric matrix with
zero row sums must vanish. Writing $\varepsilon=1+2\rho$ and $\lambda_2=1-\rho$ for the two
eigenvalues of $C_\rho$, the Jacobian $J_\alpha(C_\rho)$ has the
simple eigenvalue $\tfrac23\varepsilon^{\alpha-1}+\tfrac13\lambda_2^{\alpha-1}$ on
$\iota\iota'-I$ and the double eigenvalue
$\tfrac13f_\alpha^{[1]}(\varepsilon,\lambda_2)+\tfrac23\lambda_2^{\alpha-1}$ on its hollow
complement, so that
\[
  \kappa\{J_\alpha(C_\rho)\}
  =\frac{2\varepsilon^{\alpha-1}+\lambda_2^{\alpha-1}}{f_\alpha^{[1]}(\varepsilon,\lambda_2)+2\lambda_2^{\alpha-1}}
  \sim
  \begin{cases}
    2|\alpha|\kappa(C_\rho), & \alpha<0,\\[2pt]
    2\kappa(C_\rho)/\log\kappa(C_\rho), & \alpha=0,
  \end{cases}
  \qquad\varepsilon\downarrow0,
\]
since $f_\alpha^{[1]}(\varepsilon,\lambda_2)\sim\varepsilon^{\alpha}/(|\alpha|\lambda_2)$ for
$\alpha<0$ and $\sim\log(1/\varepsilon)/\lambda_2$ at $\alpha=0$, while
$\kappa(C_\rho)=\lambda_2/\varepsilon$. Along this path the exponent is one for every $\alpha\le0$,
and the extra power $|\alpha|$ in the bound is never seen, because for $\alpha<0$ the smallest
eigenvalue of $J_\alpha(C_\rho)$ is itself of order $\varepsilon^{\alpha}$, while at $\alpha=0$ it
grows only logarithmically. At $\alpha=-1$, for instance, the formula reduces to
$\kappa(J_{-1})=\{2\kappa(C)^{2}+1\}/\{\kappa(C)+2\}$. The two paths illustrate different
mechanisms and do not bound the behaviour of a general $C$, as the random draws reported in
Appendix~\ref{app:num} show. For $n\ge4$ and $\alpha<1$ the
equicorrelation path with $\rho\downarrow-1/(n-1)$ also attains the worst-case exponent
$|1-\alpha|$ asymptotically, since hollow matrices with zero row sums then exist.

\begin{proof}[Proof of Proposition~\ref{prop:n3}]
(i) The vector $u=(1,-1,0)'/\sqrt2$ satisfies $C_\varepsilon u=\varepsilon u$, while the two
remaining eigenvalues converge to those of the limiting rank two matrix and stay bounded above
and away from zero, so $\kappa(C_\varepsilon)=\Theta(\varepsilon^{-1})$. For the lower bound on
$\lambda_{\max}$, take the hollow direction $H$ with $H_{12}=H_{21}=1$ and all other entries
zero. Its component along $uu'$ is $\langle H,uu'\rangle=-1\ne0$, so its Rayleigh quotient under
$Df_\alpha(C_\varepsilon)$ is at least $f_\alpha'(\varepsilon)\langle H,uu'\rangle^{2}/\|H\|_F^{2}
=\Omega(\varepsilon^{\alpha-1})$. For the upper bound on $\lambda_{\min}$, take the hollow $K$
with $K_{13}=K_{31}=K_{23}=K_{32}=1$ and $K_{12}=0$. Then $Ku=0$, so in the eigenbasis
$(U'KU)_{1\ell}=u'Ku_\ell=0$ for every $\ell$ and the quadratic form at $K$ involves no divided
difference containing $\lambda_{\min}$, hence is bounded as $\varepsilon\downarrow0$. Combined
with $\lambda_{\min}(J_\alpha)\ge f_\alpha'(\lambda_{\max})=\Theta(1)$ this gives
$\lambda_{\min}(J_\alpha)=\Theta(1)$, and dividing the two bounds gives (i).

(ii) The case $\alpha=1$ is immediate from $J_1=I$. Let $\alpha\ne1$, and write $E_{\min}$ and
$E_{\max}$ for the eigenspaces of $\lambda_{\min}$ and $\lambda_{\max}$. Since $f_\alpha'$ is
monotone, the extreme divided differences $f_\alpha^{[1]}(\lambda_k,\lambda_\ell)$ occur at
$k=\ell$ with $\lambda_k$ extreme, so the extreme Rayleigh quotients of $Df_\alpha(C)$ are
attained exactly on the symmetric matrices supported in $E_{\min}$ and in $E_{\max}$
respectively. Equality in (\ref{eq:cond}) therefore requires a nonzero hollow symmetric matrix in
each of $\mathrm{Sym}(E_{\min})$ and $\mathrm{Sym}(E_{\max})$.

A nonzero hollow symmetric matrix has zero trace, and a symmetric matrix of rank one is
$\pm ww'$ with trace $\pm\|w\|^{2}$, so a nonzero hollow symmetric matrix has rank at least two.
A symmetric matrix supported in a subspace $E$ has rank at most $\dim E$, so both $\dim E_{\min}$
and $\dim E_{\max}$ are at least two. Two eigenspaces of distinct eigenvalues are orthogonal, so
in $\R^{3}$ they cannot both have dimension two. Hence $\lambda_{\min}=\lambda_{\max}$, $C$ is a
multiple of the identity, and $\diag C=\iota$ forces $C=I$.

(iii) Let $\alpha>1$. The divided difference $f_\alpha^{[1]}(x,y)$, which is $(x^{\alpha}-y^{\alpha})/\{\alpha(x-y)\}$ for $x\ne y$ and $x^{\alpha-1}$ for $x=y$, is continuous on $[0,\infty)^{2}$ and positive unless $x=y=0$. For $C=U\diag(\lambda)U'$ in the closed elliptope $\bar{\mathcal{C}}_3$ and hollow symmetric $H$ let
\[
  Q_C(H)=\sum_{k,\ell}f_\alpha^{[1]}(\lambda_k,\lambda_\ell)\{(U'HU)_{k\ell}\}^{2}.
\]
This does not depend on the choice of $U$, it is the quadratic form of $Df_\alpha(C)$ when $C\succ0$, and by the continuity of $f_\alpha^{[1]}$ it is continuous in $(C,H)$ \citep[Ch.~V]{Bhatia:1997}. By Lemma~\ref{lem:jac}, $\kappa\{J_\alpha(C)\}$ is the ratio of the maximum and the minimum of $Q_C(H)$ over hollow $H$ with $\|H\|_F=1$. The maximum is bounded on the compact set $\bar{\mathcal{C}}_3$, so it suffices to show that $Q_C(H)>0$ for every $C\in\bar{\mathcal{C}}_3$ and every nonzero hollow $H$, since the minimum is then bounded away from zero by compactness.

Suppose that $Q_C(H)=0$. Then $(U'HU)_{k\ell}=0$ unless $\lambda_k=\lambda_\ell=0$, so $H=PHP$ with $P$ the orthogonal projection onto $\ker C$. If $C\succ0$, then $P=0$ and $H=0$. If $C$ has rank two, then $\ker C$ is spanned by a unit vector $w$ and $H=cww'$, so $\diag H=0$ gives $cw_i^{2}=0$ for every $i$ and hence $c=0$. If $C$ has rank one, then $C=ss'$ with $s_i=\pm1$, because $\diag C=\iota$, and $Hs=PHPs=0$. With $S=\diag(s)$ the hollow matrix $K=SHS$ satisfies $K\iota=0$, that is $K_{12}+K_{13}=K_{12}+K_{23}=K_{13}+K_{23}=0$, so $K=0$ and $H=0$.
\end{proof}

\subsection{Second order covariance, the node eigenvalue and bias}\label{app:sampling}

At $\alpha=0$ the first order term in (\ref{eq:varstab}) vanishes, and the second order term has a closed form. For the pair $p=(i,j)$ let $h_p=(e_ie_j'+e_je_i')/\sqrt2$, so that the $h_p$ form an orthonormal basis of $\Sn^{0}$, write $[A,B]=AB-BA$, and let $\Gamma_\Delta$ be the $d\times d$ Gram matrix with entries $[\Gamma_\Delta]_{pq}=\langle[\Delta,h_p],[\Delta,h_q]\rangle$, where $\langle A,B\rangle=\tr(A'B)$. The transpose matters, because the commutators are skew symmetric. In the next result $V_0(C)$ denotes the asymptotic covariance of $\sqrt T\{\Phi_0(\hat C)-\Phi_0(C)\}$ under elliptical sampling.

\begin{proposition}\label{prop:secondorder}
Let $\hat C$ be the sample correlation matrix of $T$ independent draws from an elliptical distribution with finite fourth moments, correlation matrix $I+\Delta$ and kurtosis parameter $\kappa$. As $\Delta\to0$ through hollow symmetric matrices,
\begin{equation}\label{eq:secondorder}
  V_0(I+\Delta)=(1+\kappa)\bigl(I_d-\tfrac{1}{12}\Gamma_\Delta\bigr)+O(\|\Delta\|^{3}).
\end{equation}
In particular,
\[
  \operatorname{avar}(\hat\gamma_{ij})=(1+\kappa)\Bigl\{1-\frac{1}{12}\sum_{m\notin\{i,j\}}(\Delta_{im}^{2}+\Delta_{jm}^{2})\Bigr\}+O(\|\Delta\|^{3}),
\]
whose quadratic term does not depend on $\Delta_{ij}$. For distinct indices $i,j,k,l$ the off diagonal entries of $\Gamma_\Delta$ are $[\Gamma_\Delta]_{(ij),(ik)}=\sum_{m\notin\{i,j,k\}}\Delta_{jm}\Delta_{km}-\Delta_{ij}\Delta_{ik}$ and $[\Gamma_\Delta]_{(ij),(kl)}=-2(\Delta_{ik}\Delta_{jl}+\Delta_{il}\Delta_{jk})$.
\end{proposition}

To second order, each GFT variance falls below $1+\kappa$ by $(1+\kappa)/12$ times the sum of the squared correlations of variables $i$ and $j$ with the other variables, and the correlation $\Delta_{ij}$ of the pair itself does not enter. The matrix $\Gamma_\Delta$ is a Gram matrix, so the quadratic correction $-\tfrac{1}{12}(1+\kappa)\Gamma_\Delta$ is negative semidefinite. This does not mean that the exact covariance lies below $(1+\kappa)I_d$. Since $[\Delta,\Delta]=0$ we have $\Gamma_\Delta\vecl(\Delta)=0$, so the correction vanishes in the direction $\vecl(\Delta)$, and coupling with other directions can still produce an eigenvalue above $1+\kappa$. Under Gaussian sampling with $C_{12}=C_{13}=0.1$ and $C_{23}=0$, for instance, the largest eigenvalue of $V_0(C)$ is about $1.0000086$. For $n=2$ the matrix $\Gamma_\Delta$ is zero, in line with the exact variance stabilisation of Fisher's $z$.

\begin{proof}[Proof of Proposition~\ref{prop:secondorder}]
The sample correlation matrix does not depend on the scale of the variables, so we may take the covariance matrix of the draws to be $C=I+\Delta$. Let $\hat\Sigma$ be the sample covariance matrix. Under elliptical sampling $\sqrt T(\hat\Sigma-C)$ converges to a centred normal matrix with covariance $(1+\kappa)(I+K_n)(C\otimes C)+\kappa\operatorname{vec}(C)\operatorname{vec}(C)'$, where $K_n$ is the commutation matrix \citep{BrowneShapiro:1986}. The first part is the covariance of $(1+\kappa)^{1/2}H_0$ with $H_0=C^{1/2}ZC^{1/2}$, where $Z$ is symmetric with independent entries $Z_{aa}\sim N(0,2)$ and $Z_{ab}\sim N(0,1)$ for $a<b$. The second part is a rank one term in the direction of $C$. The derivative of $\Sigma\mapsto\Cor(\Sigma)$ at $C$ maps a perturbation $E$ to $E-\tfrac12\{\Diag(E)C+C\Diag(E)\}$, which is zero for $E=C$. The second part therefore does not contribute, as in the proof of Theorem~1 of \citet{ArchakovHansen:2026}, and $\sqrt T(\hat C-C)$ converges to $(1+\kappa)^{1/2}[H_0-\tfrac12\{\Diag(H_0)C+C\Diag(H_0)\}]$.

Let $C=U\diag(\lambda)U'$ and let $M_{ab}$ be the logarithmic mean of $\lambda_a$ and $\lambda_b$, so that $\log^{[1]}(\lambda_a,\lambda_b)=1/M_{ab}$, see \citet[Lemma~A.2]{ArchakovHansen:2026}. Since $U'H_0U=\diag(\lambda)^{1/2}(U'ZU)\diag(\lambda)^{1/2}$, and $U'\{\Diag(H_0)C+C\Diag(H_0)\}U$ has entries $(\lambda_a+\lambda_b)(U'\Diag(H_0)U)_{ab}$, the Daleckii--Krein formula gives
\[
  \sqrt T(\hat\gamma-\gamma)\Rightarrow(1+\kappa)^{1/2}\vecl\bigl\{U[(U'ZU)\circ c]U'-U[(U'\Diag(H_0)U)\circ\ell]U'\bigr\},
\]
with $c_{ab}=(\lambda_a\lambda_b)^{1/2}/M_{ab}$ and $\ell_{ab}=(\lambda_a+\lambda_b)/(2M_{ab})-1$, where the remaining term $-\Diag(H_0)$ is diagonal and is annihilated by $\vecl$. With $\eta_{ab}=\tfrac12(\log\lambda_a-\log\lambda_b)$ we have $c_{ab}=\eta_{ab}/\sinh\eta_{ab}$ and $\ell_{ab}=\eta_{ab}\coth\eta_{ab}-1$.

At $C=I+\Delta$ let $\Delta=U\diag(\omega)U'$, so that $\lambda_a=1+\omega_a$. Then $c_{ab}=1-(\omega_a-\omega_b)^{2}/24+O(\|\Delta\|^{3})$ and $\ell_{ab}=(\omega_a-\omega_b)^{2}/12+O(\|\Delta\|^{3})$. Since $(U'[\Delta,[\Delta,X]]U)_{ab}=(\omega_a-\omega_b)^{2}(U'XU)_{ab}$, the Schur multiplier $(\omega_a-\omega_b)^{2}$ is the map $X\mapsto[\Delta,[\Delta,X]]$. Moreover $\Diag(H_0)=\Diag(Z)+O(\|\Delta\|\|Z\|)$, and it is multiplied by the second order $\ell$, so the limit equals $(1+\kappa)^{1/2}\vecl(Y)$ with
\[
  Y=Z-\tfrac{1}{24}[\Delta,[\Delta,Z]]-\tfrac{1}{12}[\Delta,[\Delta,\Diag(Z)]]+R,\qquad \mathbb{E}\|R\|^{2}=O(\|\Delta\|^{6}).
\]

The off diagonal entries of $Z$ are independent of its diagonal, and $\vecl(Z)$ has covariance $I_d$. The third term of $Y$ is therefore uncorrelated with $\vecl(Z)$, so the normalisation does not contribute at second order. The off diagonal part of $Z$ is $\sum_q\sqrt2Z_qh_q$ with $Z_q=\vecl(Z)_q$, and $\vecl(X)_p=\langle h_p,X\rangle/\sqrt2$ for symmetric $X$. Hence $\mathrm{Cov}(Z_q,\vecl([\Delta,[\Delta,Z]])_p)=\langle h_p,[\Delta,[\Delta,h_q]]\rangle=\langle[\Delta,h_p],[\Delta,h_q]\rangle=[\Gamma_\Delta]_{pq}$, because $\langle X,[\Delta,Y]\rangle=\langle[\Delta,X],Y\rangle$ for symmetric $\Delta$. The covariance of the product of two correction terms is $O(\|\Delta\|^{4})$, and the remainder contributes $O(\|\Delta\|^{3})$ by the Cauchy--Schwarz inequality. Therefore $V_0(I+\Delta)=(1+\kappa)\{I_d-\tfrac{1}{24}(\Gamma_\Delta+\Gamma_\Delta')\}+O(\|\Delta\|^{3})$, which is (\ref{eq:secondorder}) since $\Gamma_\Delta$ is symmetric.

For the entries, let $a_m=\Delta e_m$ and $S(x,y)=xy'-yx'$. Then $\sqrt2[\Delta,h_p]=S(a_i,e_j)+S(a_j,e_i)$ for $p=(i,j)$, and $\langle S(x,y),S(z,w)\rangle=2\{(x'z)(y'w)-(x'w)(y'z)\}$. Expanding the inner products and using $\Delta_{mm}=0$ gives the three expressions in the proposition.
\end{proof}

The node eigenvalue in Proposition~\ref{prop:equivar}, with $\lambda_1=1+(n-1)\rho$ and $\lambda_2=1-\rho$ the eigenvalues of $C_\rho$, is
\begin{equation}\label{eq:omega1}
  \omega_1=\Bigl\{\Bigl(1-\frac{2}{n}\Bigr)f_\alpha^{[1]}(\lambda_1,\lambda_2)+\frac{2}{n}\lambda_2^{\alpha-1}\Bigr\}^{2}\lambda_2^{2}\Bigl\{1+(n-2)\rho\Bigl(1-\frac{\rho}{2}\Bigr)\Bigr\}.
\end{equation}

We now turn to the bias under Gaussian sampling. Write $c$ for
the free entries $\vecl(C)$ and let
\begin{equation}\label{eq:beta}
  \beta_\alpha(C)=J_\alpha(C)b(C)+\frac12\sum_{p,q}V_{pq}(C)\,
  \frac{\partial^{2}\Phi_\alpha(C)}{\partial c_p\,\partial c_q},
  \qquad b_{ij}(C)=-\tfrac12C_{ij}(1-C_{ij}^{2}),
\end{equation}
where $T^{-1}b(C)$ is the classical leading bias of the sample correlations
\citep{OlkinPratt:1958}. Then $T^{-1}\beta_\alpha(C)$ is the leading bias of $\Phi_\alpha(\hat C)$
given by the second order delta method.

\begin{proposition}\label{prop:bias}
As $\Delta\to0$ through hollow symmetric matrices,
\[
  \beta_\alpha(I+\Delta)=c_n(\alpha)\vecl(\Delta)+O(\|\Delta\|^{2}),
  \qquad
  c_n(\alpha)=\frac{(2n-1)\alpha^{2}-3(n+1)\alpha+(n+1)}{6}.
\]
\end{proposition}

Since $J_\alpha(I)=I_d$ for every $\alpha$, all members agree to first order at independence, so
the coefficients $c_n(\alpha)$ can be compared directly. For $n=2$,
$c_2(\alpha)=(\alpha^{2}-3\alpha+1)/2$, which is $\tfrac12$ for Fisher's $z$, whose bias is
$\rho/(2T)$, and $-\tfrac12$ for $\hat\rho$. On the surjective range $c_n(\alpha)$ is positive and
decreasing in $\alpha$, so every member with $\alpha\le0$ is biased away from independence, and
the GFT is the least biased of them, with $c_n(0)=(n+1)/6$ against $c_n(-1)=n+\tfrac12$. However, the GFT
is not the least biased member of the Box--Cox family. The coefficient has the two
positive roots
\[
  \alpha_{n,\pm}=\frac{3(n+1)\pm\sqrt{(n+1)(n+13)}}{2(2n-1)}.
\]
The smaller, $\alpha^{*}_n=\alpha_{n,-}$, is the only root in $[0,1]$, and it increases from
$(3-\sqrt5)/2\approx0.38$ at $n=2$ to $\tfrac12$ as $n\to\infty$. The larger exceeds one and tends
to one from above, and at $n=2$ it is $(3+\sqrt5)/2\approx2.62$. The coefficients of the two members
with an explicit inverse do not depend on the dimension, $c_n(\tfrac12)=-\tfrac18$ and
$c_n(1)=-\tfrac12$, whereas that of the GFT grows linearly with $n$.

\begin{proof}[Proof of Proposition~\ref{prop:bias}]
Write $\hat C=C+K$ with $C=I+\Delta$, where $K$ is hollow. For $\|X\|<1$ the spectral calculus
gives
\[
  f_\alpha(I+X)=X+a_2X^{2}+a_3X^{3}+O(\|X\|^{4}),
  \qquad a_2=\frac{\alpha-1}{2},\quad a_3=\frac{(\alpha-1)(\alpha-2)}{6},
\]
from the Taylor expansion of $f_\alpha(1+x)$. With $X=\Delta+K$, the terms of $f_\alpha(I+\Delta+K)$
that are linear in $K$ are $K+a_2(\Delta K+K\Delta)+O(\|\Delta\|^{2}\|K\|)$, and the terms that are
quadratic in $K$ are $a_2K^{2}+a_3(\Delta K^{2}+K\Delta K+K^{2}\Delta)+O(\|\Delta\|^{2}\|K\|^{2})$. In
(\ref{eq:beta}) the linear terms are evaluated at $TE(K)=b(C)+O(T^{-1})$ and the quadratic terms at
the covariances $T\,\mathrm{Cov}(K_{ab},K_{cd})=V_{(a,b),(c,d)}(C)+O(T^{-1})$.

Since $b(I+\Delta)=-\tfrac12\vecl(\Delta)+O(\|\Delta\|^{3})$, the linear terms contribute
$-\tfrac12\vecl(\Delta)+O(\|\Delta\|^{2})$. For the quadratic terms, the proof of
Theorem~\ref{thm:varstab} shows that $V_{(a,b),(c,d)}(I+\Delta)$ equals $1+O(\|\Delta\|^{2})$ when the
pairs coincide, $\Delta_{uv}+O(\|\Delta\|^{2})$ when they share one index and $u,v$ are the other
two, and $O(\|\Delta\|^{2})$ when they are disjoint. For $i\ne j$, hollowness of $K$ gives
$(K^{2})_{ij}=\sum_{k\ne i,j}K_{ik}K_{kj}$, where each term involves two pairs that share the index
$k$, so $T\,\mathbb{E}(K^{2})_{ij}=(n-2)\Delta_{ij}+O(\|\Delta\|^{2})$. In the terms that already contain
$\Delta$ the covariances can be evaluated at $C=I$, where $T\,\mathbb{E}(K_{ab}K_{cd})$ is one if
$\{a,b\}=\{c,d\}$ and zero otherwise. Then $T\,\mathbb{E}(K^{2})_{\ell j}=(n-1)\delta_{\ell j}$, so that
$T\,\mathbb{E}(\Delta K^{2})_{ij}=T\,\mathbb{E}(K^{2}\Delta)_{ij}=(n-1)\Delta_{ij}$, and
$T\,\mathbb{E}(K\Delta K)_{ij}=\sum_{a,b}\Delta_{ab}\,T\,\mathbb{E}(K_{ia}K_{bj})=\Delta_{ji}$, because only $a=j$ and
$b=i$ contribute. Collecting terms,
\[
  \beta_\alpha(I+\Delta)=\Bigl\{-\tfrac12+(n-2)a_2+(2n-1)a_3\Bigr\}\vecl(\Delta)+O(\|\Delta\|^{2}),
\]
and $-\tfrac12+(n-2)(\alpha-1)/2+(2n-1)(\alpha-1)(\alpha-2)/6=c_n(\alpha)$.
\end{proof}

The bias of the scaled coordinates near independence can be computed in the same way as in
Proposition~\ref{prop:bias}. Let $\tilde\beta_\alpha$ be defined by (\ref{eq:beta}) with
$\tilde\Phi_\alpha$ in place of $\Phi_\alpha$.

\begin{proposition}\label{prop:scaledbias}
Let $n\ge2$. As $\Delta\to0$ through hollow symmetric matrices,
\[
  \tilde\beta_\alpha(I+\Delta)=\tilde c_n(\alpha)\vecl(\Delta)+O(\|\Delta\|^{2}),
  \qquad
  \tilde c_n(\alpha)=\frac{(n+1)-(n+4)\alpha^{2}}{6}.
\]
\end{proposition}

The coefficient is even in $\alpha$, and
$\tilde c_n(\alpha)=c_n(\alpha)-\tfrac12\alpha(\alpha-1)(n+1)$. The difference is the
contribution of the scaling. It vanishes at $\alpha=1$, where $D_1=I$, and at $\alpha=0$, where
$\tilde c_n(0)=c_n(0)=(n+1)/6$, in line with the agreement of $\tilde\Phi_0$ and $\Phi_0$
through third order at $I$. At $\alpha=\pm1$ the coefficient is $-\tfrac12$ for every $n$. For
$\alpha=1$ this is the classical bias of $\hat\rho$. For $\alpha=-1$ it agrees with the result of
\citet{Fisher:1924} that, for Gaussian data, the sample partial correlation of two variables given
the other $n-2$ is distributed as a sample correlation computed from $n-2$ fewer observations,
so that its bias is $-\pi_{ij}(1-\pi_{ij}^{2})/(2T)+O(T^{-2})$. At $\alpha=-1$ the scaling therefore
takes the coefficient from $c_n(-1)=n+\tfrac12$ to $\tilde c_n(-1)=-\tfrac12$. For fixed $\alpha\notin\{-1,1\}$ its
absolute value still grows with $n$, since $\tilde c_n(\alpha)=\{n(1-\alpha^{2})+1-4\alpha^{2}\}/6$, and for
example $\tilde c_n(-2)=-(n+5)/2$. The coefficient vanishes at $\alpha=\pm\{(n+1)/(n+4)\}^{1/2}$, and for
$n\ge3$ its absolute value on $[-1,1]$ is largest at $\alpha=0$, so near independence the scaled
GFT is the most biased member of the scaled family in this range.

\begin{proof}[Proof of Proposition~\ref{prop:scaledbias}]
Let $X$ be hollow and small, and let $D=e^{Y}$, with $Y$ diagonal, be the hollowing scaling of
$I+X$. By Theorem~\ref{thm:scaled}, $Y$ is a real analytic function of $X$ with $Y=0$ at $X=0$.
Write $e^{Y}(I+X)e^{Y}=I+E$, so that
\[
  E=X+2Y+(XY+YX)+2Y^{2}+O(\|X\|\|Y\|^{2}+\|Y\|^{3}),
\]
and $f_\alpha(I+E)=E+a_2E^{2}+a_3E^{3}+O(\|E\|^{4})$ with $a_2$ and $a_3$ as in the proof of
Proposition~\ref{prop:bias}. To first order in $(X,Y)$, $\diag f_\alpha(I+E)$ equals
$\diag(X+2Y)=2\diag(Y)$, because $X$ is hollow. The linear part of the analytic map $X\mapsto Y$
therefore vanishes, so $Y=O(\|X\|^{2})$. Then $E=X+2Y+O(\|X\|^{3})$ and
$\diag(E^{2})=\diag(X^{2})+O(\|X\|^{3})$. The condition $\diag f_\alpha(I+E)=0$ then gives
\[
  Y=-\frac{a_2}{2}\,S+O(\|X\|^{3}),\qquad S=\Diag\{\diag(X^{2})\}.
\]

Now let $i\ne j$. The diagonal matrices $2Y$ and $2Y^{2}$ do not contribute to the $(i,j)$ entry,
so $E_{ij}=X_{ij}+(XY+YX)_{ij}+O(\|X\|^{4})$. Since $(X+2Y)^{2}=X^{2}+2(XY+YX)+4Y^{2}$, also
$(E^{2})_{ij}=(X^{2})_{ij}+2(XY+YX)_{ij}+O(\|X\|^{4})$, and $(E^{3})_{ij}=(X^{3})_{ij}+O(\|X\|^{4})$.
With $1+2a_2=\alpha$ this gives
\[
  f_\alpha(D(I+X)D)_{ij}
  =\Bigl\{X+a_2X^{2}+a_3X^{3}-\frac{\alpha a_2}{2}(SX+XS)\Bigr\}_{ij}+O(\|X\|^{4}).
\]
The third order term of $Y$ enters the $(i,j)$ entry only through $XY+YX$, so it contributes
at fourth order. The remainder is a convergent power series in $X$ without terms of degree below
four. With $X=\Delta+K$ its part that is linear in $K$ is therefore $O(\|\Delta\|^{3}\|K\|)$, and its
part that is quadratic in $K$ is $O(\|\Delta\|^{2}\|K\|^{2})$, so it does not affect the
coefficient of $\vecl(\Delta)$.

The terms $X+a_2X^{2}+a_3X^{3}$ are those of the proof of Proposition~\ref{prop:bias}, and they
contribute $c_n(\alpha)\vecl(\Delta)+O(\|\Delta\|^{2})$. It remains to evaluate the scaling term
$SX+XS$ with $X=\Delta+K$. Its part that is linear in $K$ is quadratic in $\Delta$, and since
$T\,\mathbb{E}(K)=b(C)+O(T^{-1})=O(\|\Delta\|)$ it contributes $O(\|\Delta\|^{3})$. Its part that is
quadratic in $K$ is
\[
  \Diag\{\diag(K^{2})\}\Delta+\Delta\Diag\{\diag(K^{2})\}
  +\Diag\{\diag(\Delta K+K\Delta)\}K+K\Diag\{\diag(\Delta K+K\Delta)\}.
\]
Each term already contains $\Delta$, so the covariances can be evaluated at $C=I$, where
$T\,\mathbb{E}(K_{ab}K_{cd})$ is one if $\{a,b\}=\{c,d\}$ and zero otherwise. Then
$T\,\mathbb{E}(K^{2})_{\ell\ell}=n-1$, so the first two terms contribute $(n-1)\Delta_{ij}$ each. Since
$\diag(\Delta K+K\Delta)_{i}=2\sum_{a}\Delta_{ia}K_{ai}$, the $(i,j)$ entry of the third term is
$2\sum_{a}\Delta_{ia}K_{ai}K_{ij}$. Only $a=j$ contributes to its expectation, which gives
$T\,\mathbb{E}=2\Delta_{ij}$. The fourth term gives $2\Delta_{ji}=2\Delta_{ij}$ in the same way. The
scaling term therefore contributes
$-\tfrac12\alpha a_2\{2(n-1)+4\}\Delta_{ij}=-\alpha a_2(n+1)\Delta_{ij}$, and
\[
  \tilde\beta_\alpha(I+\Delta)
  =\bigl\{c_n(\alpha)-\alpha a_2(n+1)\bigr\}\vecl(\Delta)+O(\|\Delta\|^{2}).
\]
Finally $6\{c_n(\alpha)-\alpha a_2(n+1)\}
=(2n-1)\alpha^{2}-3(n+1)\alpha+(n+1)-3(n+1)\alpha^{2}+3(n+1)\alpha
=(n+1)-(n+4)\alpha^{2}$.
\end{proof}

\subsection{Information geometry}\label{app:geom}

This subsection gives the geometric background of Section~\ref{sec:geom}.

For the centred Gaussian model $N(0,\Sigma)$ the information matrix for $\Sigma$,
written as a quadratic form on perturbations $U$ and $W$ of $\Sigma$, is
\begin{equation}\label{eq:fr}
  g_\Sigma(U,W)=\tfrac12\tr(\Sigma^{-1}U\Sigma^{-1}W),
\end{equation}
see \citet{Skovgaard:1984} and \citet{SmithCRB:2005}. Read as an inner product on each tangent
space, (\ref{eq:fr}) is the Fisher--Rao metric of the covariance matrices. Its shortest paths
from $I$, the geodesics, are the power paths $t\mapsto A^{t}=e^{t\log A}$, and the length of such
a path, the information distance, is
\begin{equation}\label{eq:dcov}
  d_{\mathrm{cov}}(I,C)^{2}=\tfrac12\|\log C\|_F^{2},
\end{equation}
which is the quadratic form of the Wald statistic for $\Sigma=I$ in the coordinates $\log\Sigma$.

The quadratic form (\ref{eq:fr}) is invariant under $\Sigma\mapsto D\Sigma D$ for positive
diagonal $D$. Rescaling leaves the correlations unchanged, so at each $\Sigma$ the perturbations
split into the ones that only rescale, which are nuisance directions, and the ones orthogonal to
them in (\ref{eq:fr}), which change the correlations. Restricting (\ref{eq:fr}) to the latter is
exactly what concentrating the variances out of the information matrix does, so the restriction
is the efficient information for the correlations, the inverse of the asymptotic covariance
$V(C)$ of Section~\ref{sec:varstab}. At $\Sigma=I$ the nuisance directions are the diagonal matrices and the
remaining directions are the hollow ones. In the geometric literature this restriction is the
quotient of the Fisher--Rao geometry by diagonal rescaling, introduced by \citet{DavidGu:2019},
whose paths, connection and curvature are developed by \citet{ThanwerdasPennec:2021}, see also
\citet{ThanwerdasPennec:2022}. The orthogonal directions are called horizontal and the nuisance
directions vertical.

It matters which of the two information distances one uses, and (\ref{eq:dcov}) is not the
distance for correlations. With the variances concentrated out the distance is
\begin{equation}\label{eq:dcor}
  d_{\mathrm{cor}}(I,C)^{2}=\inf_{D\succ0\text{ diagonal}}\tfrac12\|\log(DCD)\|_F^{2}
  \;\le\;d_{\mathrm{cov}}(I,C)^{2},
\end{equation}
and the infimum is attained at the scaling that makes $\log(DCD)$ hollow, since that is the
stationarity condition in $\log D$. The inequality in (\ref{eq:dcor}) is therefore strict unless
$\log C$ already has zero diagonal, which for a correlation matrix happens only at $C=I$. Any
statement about rates of divergence at the boundary that is read off (\ref{eq:dcov}) is a
statement about covariance matrices, where an eigenvalue approaching zero recedes at
exactly logarithmic order, matching the $\alpha=0$ line of (\ref{eq:sing}), whereas for correlations
the distance is no larger.

\subsection{Convexity along diagonal congruences}\label{app:congconv}

\begin{lemma}\label{lem:congconv}
Let $C\succ0$, and let $A(x)=e^{X}Ce^{X}$ with $X=\Diag(x)$ for $x\in\R^{n}$. For $\mu\ne0$ let $p_\mu(x)=\tr A(x)^{\mu}/|\mu|$, and let
$p_0(x)=\tfrac12\|\log A(x)\|_F^{2}$. Then $p_\mu$ is strictly convex on $\R^{n}$ for every
$\mu\in\R$, with
\begin{equation}\label{eq:pgrad}
  \nabla p_\mu(x)=2\operatorname{sgn}(\mu)\diag\bigl(A(x)^{\mu}\bigr)\ (\mu\ne0),
  \qquad
  \nabla p_0(x)=2\diag\bigl(\log A(x)\bigr),
\end{equation}
and, writing $A(x)=U\diag(\lambda)U'$,
\begin{equation}\label{eq:phess}
  \bigl[\nabla^{2}p_\mu(x)\bigr]_{ij}
  =2\sum_{k,\ell}c_\mu(\lambda_k,\lambda_\ell)u_{ik}u_{jk}u_{i\ell}u_{j\ell},
\end{equation}
where
\[
  c_\mu(a,b)=\operatorname{sgn}(\mu)(a+b)\frac{a^{\mu}-b^{\mu}}{a-b}\ (\mu\ne0),
  \qquad
  c_0(a,b)=(a+b)\frac{\log a-\log b}{a-b},
\]
read by continuity on the diagonal, where $c_\mu(a,a)=2|\mu|a^{\mu}$ and $c_0(a,a)=2$. All
$c_\mu$ are positive, and $c_0\ge2$, so that $\nabla^{2}p_0(x)\succeq4I$ for every $x$.
\end{lemma}

\begin{proof}[Proof of Lemma~\ref{lem:congconv}]
Fix $v\in\R^{n}$, let $V=\Diag(v)$ and $M(t)=A(x+tv)=e^{tV}A(x)e^{tV}$, so that
$M'(t)=VM+MV$. For a smooth $h$ on $(0,\infty)$,
$\frac{d}{dt}\tr h(M)=\tr\{h'(M)M'\}=2\tr\{h'(M)MV\}$. With $h_\mu(\lambda)=\lambda^{\mu}$,
which is not the Box--Cox function $f_\mu$, this is $2\mu\tr(M^{\mu}V)$, and differentiating
once more, $\frac{d^{2}}{dt^{2}}\tr M^{\mu}=2\mu\tr\{V\,Dh_\mu(M)[VM+MV]\}$. In the eigenbasis of $M$, with $W=U'VU$,
the matrix $U'(VM+MV)U$ has entries $W_{k\ell}(\lambda_k+\lambda_\ell)$, and the
Daleckii--Krein formula gives
\[
  \frac{d^{2}}{dt^{2}}\tr M^{\mu}=2\mu\sum_{k,\ell}h_\mu^{[1]}(\lambda_k,\lambda_\ell)
  (\lambda_k+\lambda_\ell)W_{k\ell}^{2}.
\]
Since $h_\mu$ is strictly increasing for $\mu>0$ and strictly decreasing for $\mu<0$, every
coefficient $\mu h_\mu^{[1]}(\lambda_k+\lambda_\ell)$ is positive, so the sum is nonnegative and
vanishes only if $W=0$, that is $V=0$. Dividing by $|\mu|$ gives the Hessian form with
$c_\mu$, after expanding $W_{k\ell}=\sum_iv_iu_{ik}u_{i\ell}$. For $\mu=0$,
$\frac{d}{dt}\tfrac12\tr(\log M)^{2}=\tr\{\log M\cdot D\log_M[M']\}=\tr\{D\log_M[\log M]M'\}$
because the Fr\'echet derivative is self adjoint, and $D\log_M[\log M]=M^{-1}\log M$ because
$\log M$ commutes with $M$. Hence the derivative is $2\tr(V\log M)$ and the second derivative is
$2\sum_{k,\ell}\log^{[1]}(\lambda_k,\lambda_\ell)(\lambda_k+\lambda_\ell)W_{k\ell}^{2}$. The
gradients follow from the first derivatives, and the bound $c_0\ge2$ is the inequality between
the logarithmic and arithmetic means, (\ref{eq:logmean}) and (\ref{eq:meanineq}) at
$\alpha=0$. Finally $\sum_{k,\ell}u_{ik}u_{jk}u_{i\ell}u_{j\ell}=\delta_{ij}$, so
$c_0\ge2$ gives $\nabla^{2}p_0\succeq4I$.
\end{proof}

\section{Algorithms for the inverse}\label{sec:algo}

This appendix collects algorithms for the inverse map for general $\alpha$. At $\alpha=0$ the
fixed point and Newton method of \citet{ArchakovHansen:Variational} is the method of choice,
and most of what follows extends it to $\alpha\ne0$.

Let $\gamma\in\mathcal{R}_\alpha$, which by Theorem~\ref{thm:main} is automatic when
$\alpha\le0$ but a genuine restriction when $\alpha>0$. At $\alpha=1$ nothing is to be done,
and at $\alpha=\tfrac12$ Theorem~\ref{thm:shape}(i) gives the inverse in closed form,
including the test for membership. Elsewhere, Proposition~\ref{prop:convex} gives a
Newton method, which iterates on $\delta$ with the exact Hessian $H_{ij}=w_{ij}'\Lambda^{[1]}w_{ij}$ and
backtracks on both feasibility of $I+\alpha G(\delta)\succ0$ and sufficient decrease of
$L_\alpha$. The start is $\delta=0$ whenever that is feasible, which is always the case at
$\alpha=0$ and typically for small $|\alpha|$. Otherwise the Gershgorin point
$\delta_i=-1/\alpha\pm(\sum_{j\ne i}|G_{ij}|+1)$, with the sign of $\alpha$, is feasible.
In double precision this converges in $3$ to $19$ steps for $|\alpha|\le2$ and all $n$ we
tested, with recovery error at the level of the residual tolerance.

The dual (\ref{eq:maxent}) gives a second Newton method, in the off diagonal
entries of $C$ rather than in $\delta$. Its stationarity condition is
$\Phi_\alpha(C)=\gamma$, the Newton direction is $J_\alpha(C)^{-1}\{\gamma-\Phi_\alpha(C)\}$
with $J_\alpha$ the Jacobian of Lemma~\ref{lem:jac}, and the line search keeps $C$
positive definite and increases the dual objective. It works in $d$ rather than $n$ variables,
but it has one advantage on $\alpha<-1$. For such $\alpha$ the exponent $(1+\alpha)/\alpha$ lies
in $(0,1)$, so $\Psi_\alpha$ stays finite at the boundary of $\mathcal{D}_\alpha$, and only its
slope diverges. The minimiser is still interior, but the line search no longer keeps the iterates
away from the boundary, where the curvature $e_\alpha'$ is unbounded and the Newton steps can
become very short. A fraction-to-boundary rule prevents this. Each step is restricted to at most
a fixed fraction, say $0.99$, of the largest step that keeps $I+\alpha G(\delta)$ positive definite,
which costs one symmetric eigenvalue problem. Alternatively, the dual objective in
(\ref{eq:maxent}) diverges at the boundary of the elliptope exactly when $\alpha\le-1$, so the dual
Newton method has the barrier that the primal problem lacks. The companion code uses the primal
method and falls back on the dual one.

For $\alpha>0$ and $\gamma\notin\mathcal{R}_\alpha$ the primal iterates are driven towards
the boundary of $\mathcal{D}_\alpha$ and the method must report failure. Alternatively one
can minimise the extended objective $\hat L_\alpha$ of (\ref{eq:Lhat}) without any feasibility
safeguard, in which case membership $\gamma\in\mathcal{R}_\alpha$ is decided by whether the
minimiser lands in $\mathcal{D}_\alpha$, by Proposition~\ref{prop:dual}. This is how the
non-membership claims of Theorem~\ref{thm:shape}(ii) are checked numerically in the
companion code.

The Archakov--Hansen fixed point iteration has an exact analogue,
\begin{equation}\label{eq:fp}
  \delta\longleftarrow\delta-f_\alpha\bigl(\diag C(\delta)\bigr),
\end{equation}
which reduces to $\delta\leftarrow\delta-\log\diag(\exp G)$ at $\alpha=0$. Because
$f_\alpha'(1)=1$ for every $\alpha$, the local convergence rate at a solution is governed by
the spectral radius of $I-H$, so the iteration is locally contractive exactly when the spectrum
of $H$ lies in $(0,2)$. The following identity says where to expect this.

\begin{lemma}\label{lem:rowsum}
At a solution, $H\iota=\diag(C^{1-\alpha})$. Consequently the row sums of $H$ are at most one
for $\alpha\in[0,1]$ and at least one for $\alpha\notin[0,1]$.
\end{lemma}
\begin{proof}
With $G=U\diag(a)U'$ and $\lambda_k=e_\alpha(a_k)$ the eigenvalues of $C$, the diagonal divided
difference is $\Lambda^{[1]}_{kk}=e_\alpha'(a_k)=\lambda_k^{1-\alpha}$, so
\[
  \sum_jH_{ij}=\sum_{k,\ell}\Lambda^{[1]}_{k\ell}u_{ik}u_{i\ell}\sum_ju_{jk}u_{j\ell}
  =\sum_k\Lambda^{[1]}_{kk}u_{ik}^{2}=\sum_k\lambda_k^{1-\alpha}u_{ik}^{2}=(C^{1-\alpha})_{ii}.
\]
For $\alpha\in[0,1]$ the exponent lies in $[0,1]$ and Jensen's inequality applied to the
weights $u_{ik}^{2}$, whose mean is $C_{ii}=1$, gives $(C^{1-\alpha})_{ii}\le1$. For
$\alpha\notin[0,1]$ the function is convex and the inequality reverses.
\end{proof}

The $n\times n$ Hessian $H$ of the diagonal problem should be distinguished from the
$d\times d$ Jacobian $J_\alpha$ of Theorem~\ref{thm:cond}. Both are principal submatrices of the
matrix of a Fr\'echet derivative, $H$ the diagonal block for $De_\alpha(G)$ and $J_\alpha$ the
off diagonal block for $Df_\alpha(C)$. The same argument bounds both.
At every feasible $\delta$, with $C=C(\delta)$, we have $\kappa(H)\le\kappa(C)^{|1-\alpha|}$, because the
eigenvalues of $De_\alpha(G)$ are the divided differences of $e_\alpha$ and therefore lie
between the smallest and the largest of the values $e_\alpha'(a_k)=\lambda_k^{1-\alpha}$, and
the eigenvalues of a principal submatrix of a positive definite matrix lie in the same interval. At
$\alpha=0$ this is the bound $\kappa(H)\le\kappa(C)$ in \citet{ArchakovHansen:Variational},
and Lemma~\ref{lem:rowsum} generalises the row sum identity $H\iota=\iota$ given there. Hessian--vector products
require no explicit $H$, since for $v\in\R^{n}$
\[
  Hv=\diag\Bigl[U\bigl\{\Lambda^{[1]}\circ(U'\Diag(v)U)\bigr\}U'\Bigr],
\]
which costs matrix multiplications once $G$ has been eigendecomposed, so that Newton
iterations can be run matrix free with $\Diag\{\diag(C^{1-\alpha})\}$ as the natural diagonal
preconditioner.

For $\alpha\in[0,1]$ the row sums in Lemma~\ref{lem:rowsum} are at most one, which suggests
that $H\preceq I$, and the following result shows that this holds.

\begin{lemma}\label{lem:contract}
For every $\alpha\in[0,1]$ and every $C\in\Cn$ the Hessian satisfies
$0\prec H\preceq I$, so $\rho(I-H)<1$ and the iteration (\ref{eq:fp}) is locally contractive.
\end{lemma}

\begin{proof}
The entries of $\Lambda^{[1]}$ are the divided differences of $e_\alpha$, which in terms of the
eigenvalues $\lambda$ of $C$ are
\[
  \Lambda^{[1]}_{k\ell}=K_\alpha(\lambda_k,\lambda_\ell),
  \qquad
  K_\alpha(a,b)=\frac{\alpha(a-b)}{a^{\alpha}-b^{\alpha}}
  =\Bigl\{\int_0^{1}\bigl[(1-t)a+tb\bigr]^{\alpha-1}dt\Bigr\}^{-1},
\]
the integral form coming from $a^{\alpha}-b^{\alpha}=\alpha\int_b^{a}x^{\alpha-1}dx$. It is
read by continuity where the quotient is indeterminate, so that at $\alpha=0$ it is the logarithmic
mean
\begin{equation}\label{eq:logmean}
  K_0(a,b)=\frac{a-b}{\log a-\log b},
\end{equation}
and at $a=b$ it is $a^{1-\alpha}$. For $\alpha\in[0,1]$ the function $x\mapsto x^{\alpha-1}$
is convex, so Jensen's inequality applied to the uniform distribution on $[0,1]$ gives
\begin{equation}\label{eq:meanineq}
  K_\alpha(a,b)\le\Bigl(\frac{a+b}{2}\Bigr)^{1-\alpha},
  \qquad a,b>0,\;\alpha\in[0,1],
\end{equation}
with equality if and only if $a=b$ or $\alpha=1$. At $\alpha=0$ this is the inequality
between the logarithmic and arithmetic means.

Now let $\|v\|=1$, write $M=U'\Diag(v)U$ and $w_{k\ell}=M_{k\ell}^{2}$. These weights satisfy
$\sum_{k,\ell}w_{k\ell}=\|M\|_F^{2}=1$ and
\[
  \sum_{k,\ell}w_{k\ell}\frac{\lambda_k+\lambda_\ell}{2}
  =\sum_{k,\ell}\lambda_kM_{k\ell}^{2}=\tr\{\diag(\lambda)M^{2}\}
  =\tr\{C\Diag(v)^{2}\}=\sum_iv_i^{2}C_{ii}=1.
\]
Since $x\mapsto x^{1-\alpha}$ is concave for $\alpha\in[0,1]$, (\ref{eq:meanineq}) and
Jensen's inequality give
\[
  v'Hv=\sum_{k,\ell}w_{k\ell}K_\alpha(\lambda_k,\lambda_\ell)
  \le\sum_{k,\ell}w_{k\ell}\Bigl(\frac{\lambda_k+\lambda_\ell}{2}\Bigr)^{1-\alpha}
  \le\Bigl\{\sum_{k,\ell}w_{k\ell}\frac{\lambda_k+\lambda_\ell}{2}\Bigr\}^{1-\alpha}=1.
\]
Positivity of $H$ is Proposition~\ref{prop:convex}.
\end{proof}

Outside $[0,1]$ contractivity can fail. In our draws $\lambda_{\max}(H)$ exceeded $2$ in every case at $\alpha=\pm2$ and reached $10^{5}$ at
$\alpha=3$. This is a statement about typical draws and not a universal one, since at $C=I$ we
have $H=I$ for every $\alpha$ and the iteration converges in one step. In practice we
recommend a hybrid, which uses (\ref{eq:fp}) while it decreases the residual and switches to
damped Newton otherwise. For $|\alpha|>1$ we recommend damped Newton alone.

Finally, (\ref{eq:cond}) also predicts where numerical deterioration can be expected, although
as an upper bound it guarantees nothing. We drew $150$ unrestricted Wishart correlation matrices
with $n\in\{4,5,6,8\}$ at each of $\alpha=-3$ and $\alpha=-4$ and required the round trip
to be accurate to $10^{-6}$. The primal Newton method alone solved $112$ and $67$ of them, and with the dual
fallback the counts rose to $140$ and $113$. At $\alpha=-3$ the median of
$\kappa(C)^{|1-\alpha|}$ was $5.8\cdot10^{6}$ among the draws the full method solved and
$1.5\cdot10^{12}$ among those it did not, and at $\alpha=-4$ the medians were
$2.7\cdot10^{7}$ and $4.9\cdot10^{12}$. The separation is clear, but the threshold is
empirical and depends on the tolerance and on the algorithm.

\section{Computational details}\label{app:num}

Except for the simulations of Section~\ref{sec:tests}, the computations use the companion script \texttt{gft\_boxcox.py} together with
\texttt{verify\_claims.py}, which checks every quantitative statement in the paper.

Unless stated otherwise, correlation matrices are drawn as $C=D^{-1/2}SD^{-1/2}$ with
$S\sim W_n(\nu,I_n)$ a standard Wishart with $\nu$ between $n$ and $n+4$ degrees of freedom and
$D=\diag(S)$. Writing $S=D^{1/2}CD^{1/2}$, the Wishart density factorises into a function of $D$
and a function of $C$, and the marginal density of $C$ is proportional to
$\det(C)^{(\nu-n-1)/2}$, with no residual dependence on $D$. The determinant exponent runs from
$-1/2$ at $\nu=n$ to $3/2$ at $\nu=n+4$, so the draws sit closer to the boundary for the smaller
$\nu$ and closer to the identity for the larger. Where ill-conditioned matrices are required, a
spectrum $\lambda_k\propto\exp\{(k-1)\log\kappa_0/(n-1)\}$ with $\sum_k\lambda_k=n$ is rotated by
a Haar orthogonal matrix and the result is normalised to unit diagonal. The normalisation
changes the spectrum, so $\kappa_0$ only steers the draw towards the required conditioning. The
condition number reported, and used in all selection rules, is the realised $\kappa(C)$ of the
correlation matrix itself. Draws whose normalisation is degenerate are rejected.

The 400 matrices in Figures~\ref{fig:image} and \ref{fig:hull} are drawn by rejection sampling, with the three off diagonal entries uniform on $[-1,1]^{3}$ and retained when $\det C>0.006$, so they are uniformly distributed on that part of the elliptope. Points outside the plotted region are drawn on its edge in the same direction, and for $\alpha>0$ the grey surface in Figure~\ref{fig:image} is the image of the singular correlation matrices under the continuous extension of $\Phi_\alpha$.

The rejection rates in Table~\ref{tab:sim} and the power figures in Section~\ref{sec:tests} are computed with the Julia script \texttt{kron\_sim.jl}, which uses only the standard libraries. Each replication has its own random number generator, seeded from the base seed 19191018, the birthday of George E.~P.~Box, so the results do not depend on the number of threads. The covariance $V_0(\hat C)$ is computed as $J_0(\hat C)V(\hat C)J_0(\hat C)'$, with the Daleckii--Krein Jacobian and the Gaussian covariance of the sample correlations, and the inverse GFT by the fixed point of \citet{ArchakovHansen:Variational}. The bootstrap rejection rates use the warp-speed method of \citet{GiacominiPolitisWhite:2013}. In each replication one bootstrap sample is drawn under the restricted estimate, and the test rejects when the statistic exceeds the $95\%$ quantile of the bootstrap statistics across replications. The conventional bootstrap tests are computed with \texttt{nested\_boot.jl}, with $2{,}000$ replications for the $4\times5$ design and $1{,}000$ for the $5\times8$ design, $B=199$ bootstrap samples for each replication and the bootstrap $p$-value $(1+\#\{b:W_b^{*}\ge W\})/(B+1)$. The random number streams are those of \texttt{kron\_sim.jl}, so the data and the first bootstrap sample of each replication are the ones behind Table~\ref{tab:sim}. For the power comparison $E=GG'/n$ with $G$ a fixed $n\times n$ matrix of independent standard normal entries, and the size-adjusted critical values are the $95\%$ quantiles of the statistics in separate replications under the null hypothesis, with $5{,}000$ replications for each value of $\tau$.

The distances to the closure of the $3\times3$ image quoted in Appendix~\ref{app:shape} are
Euclidean distances in the $\Phi_\alpha$ coordinates. The closed elliptope is parametrised by the
Gram matrices of three unit vectors in $\R^{3}$, with three angles, and
$\|\Phi_\alpha(VV')-s\|$ is minimised from 1{,}000 random starting angles by the Nelder--Mead
method, stopped when the change in the angles is below $10^{-12}$ and the change in the objective
below $10^{-16}$, or after 4{,}000 iterations, followed by Powell's method. The reported distance
is the smallest value found, so it is an upper bound for the distance. At $\alpha=\tfrac14$ it is
$0.0295$, and Theorem~\ref{thm:shape}(iii) certifies that the distance is positive.

Spectral functions and divided differences are computed from \texttt{eigh}. The solver for the diagonal problem is
a damped Newton method on $L_\alpha$ with exact Daleckii--Krein Hessian and a feasibility plus
Armijo backtracking line search, with tolerance $10^{-10}$ on $\|\diag C-\iota\|_\infty$, the
dual Newton method of Appendix~\ref{sec:algo} serving as a fallback. Round trips
$C\mapsto\Phi_\alpha(C)\mapsto C$ are checked for $n\in\{3,5,8\}$ and
$\alpha\in\{-3,-2,-1,-\tfrac12,0,\tfrac14,\tfrac12,1,2,3\}$ on draws with
$\kappa(C)^{|\alpha|}\le10^{7}$. This is a selection rule for the experiment and not a
boundary. It keeps the entries of $f_\alpha(C)$, which scale like $\kappa(C)^{|\alpha|}$ for
$\alpha<0$, within a range where the tolerance is meaningful, while for $\alpha>0$ the
coordinates are bounded by (\ref{eq:bound}) whatever $\kappa(C)$ is. Every retained case is
solved, and every failure would be counted, not skipped. The unrestricted draws at
$\alpha=-3,-4$ reported in Appendix~\ref{sec:algo} are a separate experiment. For
Theorem~\ref{thm:shape}(i) the script checks the closed form at $\alpha=\tfrac12$, the
sign of the cross derivatives of $A\mapsto\diag(A^{p})$ on either side of $p=2$, and membership
of random convex combinations of image points, and for the equicorrelation path in
Section~\ref{sec:special} it checks the
closed form for $\kappa\{J_\alpha(C_\rho)\}$ and reproduces the following figures. Along
$C_\varepsilon$ at $\alpha=0$ the ratio $\log\kappa(J_\alpha)/\log\kappa(C)$ is at most $0.889$ for
$\kappa(C_\varepsilon)\in[10^{3},10^{4}]$, and with $r=0.3$ it is still only $0.92$ at
$\kappa(C_\varepsilon)\approx2\cdot10^{6}$. Among $3\times3$ matrices with
$\kappa(C)\in[10^{3},10^{4}]$, two thousand draws at $\alpha=-1$ gave ratios between $0.51$ and
$1.82$ with median $1.22$, against a bound of two and against $1.075$ to $1.10$ along the
equicorrelation path, and at $\alpha=0$ a search of $6\cdot10^{5}$ draws reached $0.887$, close to
the value along $C_\varepsilon$ at that conditioning. For
Section~\ref{sec:scaled} the script checks the Hessian formula (\ref{eq:phess}) against finite
differences, the bound $\nabla^{2}p_0\succeq4I$, uniqueness of the scaling from random
starting points, and the explicit inverse (\ref{eq:scaledinv}). The
claims inside theorem statements are proved analytically. Reported percentages and search results, such as those just quoted, refer to the sampling laws above, are descriptive, and support no claim in
a theorem.

\setlength{\bibsep}{0pt plus 0.1ex}
{\footnotesize
\bibliographystyle{apalike}
\bibliography{prh}
}

\end{document}